\documentclass[12pt]{iopart}

\usepackage{iopams}
\usepackage{cite}
\usepackage{alltt}
\usepackage[colorlinks=true,linkcolor=blue,citecolor=blue,urlcolor=blue]{hyperref}
\begin{document}
\nocite{*}
\title[ Fermion as a non-local particle-hole excitation  ]{Fermion as a non-local particle-hole excitation}

\author{Alok Kushwaha$^1$, Rishi Paresh Joshi$^2$  
and Girish Sampath Setlur$^1$}
\address{$^1$ Department of Physics, Indian Institute of Technology Guwahati, Assam, India}
\address{$^2$ Department of Physics, Homi Bhabha National Institute,
Anushaktinagar, Mumbai, India}
\ead{gsetlur@iitg.ac.in}

\begin{abstract}
       We show that the fermion, in the context of a system that comprises many such entities - which, by virtue of the Pauli exclusion principle, possesses a Fermi surface at zero temperature - may itself be thought of as a collection of non-local particle-hole excitations across this Fermi surface. This result is purely kinematical and completely general - not being restricted to any specific dimension, applicable to both continuum and lattice systems. There is also no implication that it is applicable only to low-energy phenomena close to the Fermi surface. We are able to derive the full single-particle dynamical Green function of this fermion at finite temperature by viewing it as a collection of these non-local particle-hole excitations. The Green function of the fermion then manifests itself as a solution to a first-order differential equation in a parameter that controls the number of particle-hole pairs across the Fermi surface, and this equation itself reveals variable coefficients that may be identified with a Bose-Einstein distribution - implying that there is a sense in which the non-local particle-hole excitations have bosonic qualities while not being exact bosons at the level of operators. We also recall the definition of the non-local particle-hole operator that may be used to diagonalize the kinetic energy of free fermions of the sort mentioned above. Number-conserving products of creation and annihilation operators of fermions are expressible as a (rather complicated) combination of these non-local particle-hole operators.
\end{abstract}
\section{Introduction}

The subject of bosonization in one spatial dimension is well known and has been the subject of numerous review articles (\cite{Gia03,JanSch98,Sén04,Voi95}). The inspiration behind this starts with S. Tomonaga\cite{Tom50} in 1950, and in 1963, J.M. Luttinger\cite{Lut63} reformulated the theory in terms of Bloch sound waves. Through this, an exactly soluble model of a one‐dimensional many‐fermion system was proposed. In 1965, D.C. Mattis and E. H. Lieb\cite{MatLieb65} observed that the charge density $\rho({p})$ obeys bosonic commutation rules known now as the chiral anomaly. They then used this observation to solve for and obtain the exact (and nontrivial) energy spectrum, free energy, dielectric constant, and the single-particle Green function of the interacting fermions. In higher dimensions, the attempts of Luther\cite{Luth79} and, most notably, Haldane\cite{Hal92HPA,Hal92} that try to mimic the one-dimensional Tomonaga-Luttinger model in each radial direction (known as the ``tomographic" approach) are noteworthy. Other authors have tried to calculate the properties of interacting systems using this idea\cite{HoughMarsRSha94,KopHerSch95,NetFra94,NetFra94PRLB,Kop97}. Haldane's idea is to write a particle-hole operator that is pinned to the Fermi surface and symmetrically creates particles above the Fermi surface and holes below it. 
All these works, however, suffer from the same malady, viz., the operators they regard as bosons and use to diagonalize the kinetic energy of free fermions are not bosons at all. No proof is offered by any of these authors that they are bosons. Furthermore, the energy of the particle-hole pair vanishes when the momentum transfer is parallel to the Fermi surface, leading to mathematical singularities in practical calculations. Some hand-waving explanations are given that are not convincing.  One of the present authors' own early work nicely summarizes the state of the art in the field that existed at the time and suggests that the idea may be generalized to include particle-hole excitations not necessarily close to the Fermi surface\cite{GSS98}. But even here, an unproven assumption that the operators in question are bosons was made without proof. Nevertheless, a hack enabled the derivation of the correct Luttinger exponents by this author, and the higher-dimensional quasi-particle residue and so on were similarly derived\cite{GSS04,GSS13}. The state of affairs until now has, therefore, been rather unsatisfactory.

The present work shows how it is actually done. Non-local operators that correspond to particle-hole excitations across the Fermi surface are introduced, which are then used to diagonalize the kinetic energy of free fermions\cite{GSS13}. The number-conserving products of a pair of creation and annihilation operators of these fermions (local particle-hole excitation) is expressible as a (rather complicated) combination of these non-local particle-hole operators\cite{GSS13}.

Later, ``bosonization" - the idea that properties of fermions may be parametrized in terms of objects that possess properties that we normally associate with bosons - is shown to be a kind of inverse of Wick's theorem. While Wick's theorem involves writing higher-order correlations as products of lower-order ones, this method (which we continue to call ``bosonization" for lack of a better word) involves writing the one-particle Green function in terms of the two-particle Green function tempered by an operator that controls for the number of particle-hole pairs across the Fermi surface. These ideas are then used to write down an equation for the single-particle Green function of free fermions at finite temperature as a first-order ordinary differential equation (ODE) in a parameter that controls for the number of particle-hole pairs across the Fermi surface. The non-constant coefficients in this ODE are seen to possess a form that may be identified with a Bose-Einstein distribution. The solution to this ODE involves integrating with respect to this parameter (which we call $ \lambda $), thereby justifying the use of the adjective ``non-local" as the operation of integration is decidedly a non-local one. The non-constant coefficients that appear in the equation are Bose-Einstein distributions involving this parameter $ \lambda $. We conclude by comparing the solution obtained by this method with the straightforward method involving quantum statistical analysis of free fermions that obey fermion (anti) commutation rules.

 This work lends further credence to the suggestion made by one of the present authors that the term ``bosonization" is a misnomer since the fermion is expressible as a boson in general only at the level of correlation functions rather than at the level of operators \footnote{... except perhaps for chiral fermions in one spatial dimension, where the only convincing way to show that this is valid at the operator level is to prove that the bosonized form of the fermion properly (anti) commutes with the conventional fermion - a demonstration missing from the literature (except as a private communication to one of the authors by F.D.M. Haldane) }.

\section{ Non-local particle-hole operators }

 In the present context what we have in mind are fermions described by annihilation operators $ c_{ {\bf{p}} } $ and creation operators $ c^{\dagger}_{ {\bf{p}} } $ together with the understanding that at zero temperature all states labeled by $ {\bf{p}} $ (which is typically momentum or crystal momentum in case of lattice systems) are occupied when $ \epsilon_{ {\bf{p}} }< E_F $. The parameter $ E_F $ is, of course, the Fermi energy. The zero temperature momentum distribution is now identified with $ <c^{\dagger}_{ {\bf{p}} }c_{ {\bf{p}} }> \equiv n_F({\bf{p}}) \equiv
\theta(E_F-\epsilon_{ {\bf{p}} }) $. Where $ \theta(X > 0) = 1, \theta(X< 0 ) = 0, \theta(X=0) = 1/2 $ is the Heaviside step function.
 We now introduce the notations $ c_{ {\bf{p}}, < } \equiv c_{ {\bf{p}} }n_F({\bf{p}}) $ and  $ c_{ {\bf{p}}, > } \equiv c_{ {\bf{p}} }(1-n_F({\bf{p}})) $.  The main idea of the present section is the definition of an operator that corresponds to particle-hole excitation across the Fermi surface. But this is not the usual particle-hole excitation, a total number conserving Fermi bilinear. Rather, it is a non-local version of this. This non-local version is the one that is needed to ensure that the kinetic energy of fermions would be diagonal in these operators, just as it was in the original fermion description. First, define an operator that corresponds to the number of particle-hole pairs. 
 \begin{eqnarray}
N_{>} = \sum_{ {\bf{p}} }c_{ {\bf{p}}, < }c^{\dagger}_{ {\bf{p}}, < } 
\end{eqnarray}
We now define
 \begin{eqnarray}
A_{ {\bf{k}} }({\bf{q}}) =  c^{\dagger}_{ {\bf{k}} - {\bf{q}}/2, < }\frac{1}{\sqrt{N_{>}}}c_{ {\bf{k}} + {\bf{q}}/2, > } =  c^{\dagger}_{ {\bf{k}} - {\bf{q}}/2, < }c_{ {\bf{k}} + {\bf{q}}/2, > }\frac{1}{\sqrt{N_{>}}}
\end{eqnarray}
This operator makes sense when it acts on all states in the $ N^0 = \sum_{ {\bf{p}} }n_F({\bf{p}}) $ particle subspace of the Fock space except on the special state that contains no particle-hole pairs viz. on the state $ |FS> $ such that $ N_{>}|FS> = 0 $. In such a case, we assert by the fact that,
 \begin{eqnarray}
A_{ {\bf{k}} }({\bf{q}}) |FS> = 0 
\end{eqnarray}
From this, we may conclude that,
 \begin{eqnarray}
c^{\dagger}_{ {\bf{k}} + {\bf{q}}/2, < }c_{ {\bf{k}} - {\bf{q}}/2, > }  =  A_{ {\bf{k}} }(-{\bf{q}})\mbox{  }\sqrt{N_{>}}  
\end{eqnarray}
This is a form of inversion of the definition of the non-local particle-hole operator $ A_{ {\bf{k}} }({\bf{q}}) $ to obtain the usual Fermi-bilinear. The adjoint of the above formula gives,
 \begin{eqnarray}
c^{\dagger}_{ {\bf{k}} + {\bf{q}}/2, > }c_{ {\bf{k}} - {\bf{q}}/2, < }  = \sqrt{N_{>}}  \mbox{  }  A^{\dagger}_{ {\bf{k}} }({\bf{q}})
\end{eqnarray}
Obtaining the remaining bilinears viz. $ c^{\dagger}_{ {\bf{k}} + {\bf{q}}/2, > }c_{ {\bf{k}} - {\bf{q}}/2, > }   $ and $  c^{\dagger}_{ {\bf{k}} + {\bf{q}}/2, < }c_{ {\bf{k}} - {\bf{q}}/2, < }   $ requires a clever juxtaposition of these two formulas as follows. First, note that,
\begin{eqnarray}
  A_{ {\bf{k}} - {\bf{q}}_1/2 }( {\bf{q}}_1-{\bf{q}}) = 
  c^{\dagger}_{ {\bf{k}} + {\bf{q}}/2 - {\bf{q}}_1, < }\frac{1}{\sqrt{N_{>}}}c_{ {\bf{k}} - {\bf{q}}/2, > }
\end{eqnarray}
and
\begin{eqnarray}
  A^{\dagger}_{ {\bf{k}} + {\bf{q}}/2 -{\bf{q}}_1/2 }( {\bf{q}}_1) = c^{\dagger}_{ {\bf{k}} + {\bf{q}}/2, > }\frac{1}{\sqrt{N_{>}}}  c_{ {\bf{k}} -{\bf{q}}_1  + {\bf{q}}/2, < }
\end{eqnarray}
This means,
\begin{eqnarray}
  c^{\dagger}_{ {\bf{k}} + {\bf{q}}/2, > }c_{ {\bf{k}} - {\bf{q}}/2, > } =  \sum_{ {\bf{q}}_1 } \mbox{  }A^{\dagger}_{ {\bf{k}} + {\bf{q}}/2 -{\bf{q}}_1/2 }( {\bf{q}}_1) A_{ {\bf{k}} - {\bf{q}}_1/2 }( {\bf{q}}_1-{\bf{q}})
\end{eqnarray}
Similarly, 
\begin{eqnarray}
  A_{ {\bf{k}} + {\bf{q}}_1/2 }( {\bf{q}}_1-{\bf{q}}) =  c^{\dagger}_{ {\bf{k}} + {\bf{q}}/2, < }c_{ {\bf{k}} + {\bf{q}}_1   -{\bf{q}}/2, > }\frac{1}{\sqrt{N_{>}}}
\end{eqnarray}
and
\begin{eqnarray}
  A^{\dagger}_{ {\bf{k}} - {\bf{q}}/2 +{\bf{q}}_1/2 }( {\bf{q}}_1) = \frac{1}{\sqrt{N_{>}}}   c^{\dagger}_{{\bf{k}} - {\bf{q}}/2 +{\bf{q}}_1 , > }c_{ {\bf{k}} - {\bf{q}}/2  , < }
\end{eqnarray}
hence,
\begin{eqnarray}
 \sum_{ {\bf{q}}_1 } \mbox{      }  A^{\dagger}_{ {\bf{k}} - {\bf{q}}/2 +{\bf{q}}_1/2 }( {\bf{q}}_1)A_{ {\bf{k}} + {\bf{q}}_1/2 }( {\bf{q}}_1-{\bf{q}}) =   \frac{\sqrt{N^{'}_{>}}}{\sqrt{N_{>}}}  c_{ {\bf{k}} - {\bf{q}}/2  , < }c^{\dagger}_{ {\bf{k}} + {\bf{q}}/2, < }\frac{\sqrt{N^{'}_{>}}}{\sqrt{N_{>}}}
 \label{adaga}
\end{eqnarray}
where,
\begin{eqnarray}
N^{'}_{>} = \sum_{ {\bf{p}} }c^{\dagger}_{ {\bf{p}}, > }c_{ {\bf{p}}, > }
\end{eqnarray}
These two notions of the number of particle-hole pairs are linked through the identity,
\begin{eqnarray}
N^{'}_{>}-N_{>} = N-N^0
\end{eqnarray}
where $ N = \sum_{ {\bf{p}} }c^{\dagger}_{ {\bf{p}} }c_{ {\bf{p}} }$. So long as we are only interested in matrix elements of equation (\ref{adaga}) with respect to states that all contain $ N^0 $ number of particles, then $ N^{'}_{>} = N_{>} $ and we may write,
\begin{eqnarray}
&\fl \sum_{ {\bf{q}}_1 } \mbox{      }  A^{\dagger}_{ {\bf{k}} - {\bf{q}}/2 +{\bf{q}}_1/2 }( {\bf{q}}_1)A_{ {\bf{k}} + {\bf{q}}_1/2 }( {\bf{q}}_1-{\bf{q}}) &=   c_{ {\bf{k}} - {\bf{q}}/2  , < }c^{\dagger}_{ {\bf{k}} + {\bf{q}}/2, < } \\&\mbox{ } & = n_F({\bf{k}})\mbox{  }\delta_{ {\bf{q}}, 0 }
  - c^{\dagger}_{ {\bf{k}} + {\bf{q}}/2, < } c_{ {\bf{k}} - {\bf{q}}/2  , < }
\end{eqnarray}
This means the kinetic energy of free fermions may be written in the language of the A-operators as follows (we assume, for illustration, $ \epsilon_{ {\bf{k}} } = \frac{k^2}{2m} $).
\begin{eqnarray}
& K = \sum_{ {\bf{k}} }\epsilon_{ {\bf{k}} }c^{\dagger}_{ {\bf{k}} }c_{ {\bf{k}} } = \sum_{ {\bf{k}} }\epsilon_{ {\bf{k}} }n_F({\bf{k}})
 + \sum_{ {\bf{k}} } \frac{ {\bf{k.q}} }{m}\mbox{  }A^{\dagger}_{ {\bf{k}} }({\bf{q}})A_{ {\bf{k}} }({\bf{q}}) 
\end{eqnarray}
Thus, the kinetic energy of free fermions is diagonal in both the original Fermi language and in this new language of the A-operators. However, contrary to the assumptions made in earlier works (Haldane \cite{Hal92}, Castro Neto and Fradkin\cite{NetFra94}), these A-operators are not bosons but obey complicated commutation rules, making this approach unwieldy and practically useless. This is why we turn to the next section, where we show how to recast the ideas in the present section to make it practically useful.

\section{ Inverse of Wick's theorem }

The central idea of the present work is the realization that it is fruitful to regard the one-particle Green function as a peculiar version of the two-particle correlation function (also known as the 4-point function). In order to do this, we invoke the venerated tool used in many areas of theoretical physics, viz., resolution of the identity. Colloquially, it involves writing unity in unrecognizable ways and surreptitiously inserting it in places we feel are going to lead to something worthwhile.
 The number of particle-hole pairs may be written either as,
\begin{eqnarray}
N_{>} = \sum_{ {\bf{p}} }c_{ {\bf{p}}, < }c^{\dagger}_{ {\bf{p}}, < }  \mbox{               } ; \qquad
N^{'}_{>} = \sum_{ {\bf{p}} }c^{\dagger}_{ {\bf{p}}, > }c_{ {\bf{p}}, > }
\end{eqnarray}
The two notions are nearly the same but have subtle differences as we have already seen.  The resolution of identity we have in mind is one of two assertions.
\begin{eqnarray}
{\bf{1}}\mbox{               }  \equiv    \frac{1}{N_{>}} \mbox{  }\sum_{ {\bf{p}} }c_{ {\bf{p}}, < }c^{\dagger}_{ {\bf{p}}, < }  \mbox{            };\qquad   {\bf{1}}   \equiv \mbox{            } \frac{1}{ N^{'}_{>} }   \sum_{ {\bf{p}} }c^{\dagger}_{ {\bf{p}}, > }c_{ {\bf{p}}, > }
\end{eqnarray}
The troublesome operators in the denominator are reinterpreted as,
\begin{eqnarray}
 \fl {\bf{1}}\mbox{               }  \equiv   \int^{\infty}_0d\lambda \mbox{      } \rme^{ - \lambda N_{>}} \mbox{  }\sum_{ {\bf{p}} }c_{ {\bf{p}}, < }c^{\dagger}_{ {\bf{p}}, < }  \mbox{            };    \qquad {\bf{1}}   \equiv   \int^{\infty}_0d\lambda^{'} \mbox{      }\rme^{ - \lambda^{'} N^{'}_{>} }   \sum_{ {\bf{p}} }c^{\dagger}_{ {\bf{p}}, > }c_{ {\bf{p}}, > }
\end{eqnarray}
This then enables us to view the two-point function as a kind of 4-point function as follows:
\begin{eqnarray}
\fl <Tc_{ {\bf{k}} }(t) c^{\dagger}_{ {\bf{k}} }(t^{'})>\mbox{ }\equiv \mbox{ }\int^{\infty}_0d\lambda\sum_{ {\bf{p}} }
<T \rme^{ - \lambda N_{>}(t_1) }c_{ {\bf{p}}, < }(t_1)c^{\dagger}_{ {\bf{p}}, < }(t_1)c_{ {\bf{k}} }(t) c^{\dagger}_{ {\bf{k}} }(t^{'})>
\end{eqnarray}
or,
\begin{eqnarray}
\fl <Tc_{ {\bf{k}} }(t) c^{\dagger}_{ {\bf{k}} }(t^{'})> \mbox{ }\equiv \mbox{ }\int^{\infty}_0d\lambda^{'} \sum_{ {\bf{p}} }<\!T \rme^{ - \lambda^{'} N^{'}_{>}(t_1) }c^{\dagger}_{ {\bf{p}}, > }(t_1) c_{ {\bf{p}}, > }(t_1)c_{ {\bf{k}} }(t) c^{\dagger}_{ {\bf{k}} }(t^{'})\!>
\end{eqnarray}
The rest of this work involves deriving closed (but coupled) first-order differential equations in $ \lambda, \lambda^{'} $ for the generalized two-point function of free fermions viz.
\begin{eqnarray}
& G_{ {\bf{k}},> }(\lambda, \lambda^{'}; t,t^{'}) \mbox{       } \equiv \mbox{       } 
<T  \rme^{ - \lambda N_{>}(t) } \rme^{ - \lambda^{'} N^{'}_{>}(t) }  c_{ {\bf{k}},> }(t) c^{\dagger}_{ {\bf{k}},> }(t^{'}) > ; \nonumber\\&  \mbox{  }G_{ {\bf{k}},< }(\lambda, \lambda^{'}; t,t^{'}) \mbox{       } \equiv \mbox{       }<T  \rme^{ - \lambda N_{>}(t) } \rme^{ - \lambda^{'} N^{'}_{>}(t) }  c_{ {\bf{k}}, < }(t) c^{\dagger}_{ {\bf{k}},<  }(t^{'}) >
\end{eqnarray}
These are then solved using the method of separation of variables suitably adapted to the present case. We now present a brief outline of the derivation of the coupled equations that we alluded to. The detailed derivation is left to the supplementary material attached to this article. 
\section{Equations of motion}
We define the usual particle-hole operators, 
\begin{equation}
    a_{\bf k}({\bf q}) = c^\dagger_{\mathbf{k}-\frac{\mathbf{q}}{2},<}c_{\mathbf{k}+\frac{\mathbf{q}}{2},>};\qquad 
    a^\dagger_{\bf k}({\bf q}) = c^\dagger_{\mathbf{k}+\frac{\mathbf{q}}{2},>}c_{\mathbf{k}-\frac{\mathbf{q}}{2},<} 
\end{equation}
Let us consider a four-point function (apart from the additional factors that control the number of particle-hole pairs) which is defined as
\begin{eqnarray}
  F_{\bf k, <}({\bf q}; t_1)\mbox{ } &= \mbox{ }< T \rme^{-\lambda N_>(t) -\lambda' N'_>(t)} \mbox{ } c_{\bf k+q, >}(t) \mbox{ } a^\dagger_{{\bf k} + \frac{{\bf q}} {2} }({\bf q}, t_1 )\mbox{ }  c^\dagger_{\bf k,<}(t')  >\\
  F_{\bf k, >}({\bf q}; t_1) \mbox{ }&= \mbox{ }< T \rme^{-\lambda N_>(t) -\lambda' N'_>(t)} \mbox{ } c_{\bf k-q, <}(t) \mbox{ }a_{{\bf k} - \frac{{\bf q}} {2} }({\bf q}, t_1 ) \mbox{ }  c^\dagger_{\bf k,>}(t')  >
\end{eqnarray}
  The equation for time-dependence of $ F_{\bf k <}({\bf q}; t_1)$ is given by 
\begin{eqnarray}
\fl  \rmi \partial_{t_1}  F_{\bf k, <}({\bf q}; t_1) \mbox{ }&= \mbox{ } (\epsilon_{\bf k} - \epsilon_{\bf k +q}) \,F_{\bf k, <}({\bf q}; t_1)+(\rme^{\lambda + \lambda'} - 1) F_{\bf k, <}({\bf q}; t)\mbox{ } \rmi\delta(t_1-t) \nonumber\\ 
    &\quad  -(1 - n_F({\bf k + q}))  \, \rme^{\lambda + \lambda'}\,G_{{\bf k,<}}(\lambda, \lambda^{'} ; t, t') \mbox{ } \rmi\delta(t_1-t) \nonumber \\
    &\quad + n_F({\bf k}) G_{{\bf k + q, >}}(\lambda, \lambda^{'} ; t, t') \mbox{ } \rmi\delta(t_1-t') \label{eq1}
\end{eqnarray} 
Here we have taken $ \hbar = 1$. In the same way, equation for time-dependence of $ F_{\bf k >}({\bf q}; t_1)$ is given by
\begin{eqnarray}
\fl    \rmi \partial_{t_1}  F_{\bf k, >}({\bf q}; t_1)&= (\epsilon_{\bf k} - \epsilon_{\bf k - q}) \,F_{\bf k, >}({\bf q}; t_1)+(\rme^{ - \lambda - \lambda'} - 1) F_{\bf k, >}({\bf q}; t) \mbox{ } \rmi\delta(t_1-t) \nonumber\\ 
    &\quad  - n_F({\bf k - q}) \, \rme^{ - \lambda - \lambda'}  \,G_{{\bf k,>}}(\lambda, \lambda^{'} ; t, t') \mbox{ } \rmi\delta(t_1-t) \nonumber\\
    &\quad + (1 - n_F({\bf k})) G_{{\bf k - q, <}}(\lambda, \lambda^{'} ; t, t') \mbox{ } \rmi \delta(t_1-t') \label{eq2}
\end{eqnarray} 
We can see clearly that equations (\ref{eq1}) and (\ref{eq2}) are linear differential equations in $t_1$, which are solved easily and the constants of integration are determined by the Kubo-Martin-Schwinger (KMS) boundary condition at $t_1 = 0$ and $t_1 = -\rmi\beta$. The solution to $F_{\bf k, <}({\bf q}; t_1)$ and $F_{\bf k, >}({\bf q}; t_1)$ are given by  
\begin{eqnarray}
  \fl F_{\bf k, <}({\bf q}; t_1) \mbox{ }=& \mbox{ }
  \rme^{-\rmi (t_1 - t)(\epsilon_{\bf k} - \epsilon_{\bf k +q})} \bigg[  (\rme^{\lambda + \lambda'} - 1) F_{\bf k, <}({\bf q}; t)  \nonumber  \\ 
  & -(1 - n_F({\bf k + q})) \, \rme^{\lambda + \lambda'}  \,G_{{\bf k,<}}(\lambda, \lambda^{'} ; t, t') \bigg]
   \left[  \theta(t_1- t) - \frac{1}{1 - \rme^{\beta(\epsilon_{\bf k} - \epsilon_{\bf k +q})}} \right]      \nonumber      \\
    &\hspace{-0.2in} + n_F({\bf k})   \rme^{-\rmi (t_1 - t')(\epsilon_{\bf k} - \epsilon_{\bf k +q})}  G_{{\bf k + q, >}}(\lambda, \lambda^{'} ; t, t')  \left[   \theta(t_1- t') - \frac{1}{1 - \rme^{\beta(\epsilon_{\bf k} - \epsilon_{\bf k +q})}}\right]   
\end{eqnarray}
and 
\begin{eqnarray}
  \fl F_{\bf k, >}({\bf q}; t_1) \mbox{ }=&\mbox{ }  \rme^{-\rmi (t_1 - t)(\epsilon_{\bf k} - \epsilon_{\bf k - q})} \bigg[  (\rme^{- \lambda - \lambda'} - 1) F_{\bf k, >}({\bf q}; t) \nonumber \\
  &-n_F({\bf k - q}) \, \rme^{-\lambda - \lambda'}  \,G_{{\bf k,>}}(\lambda, \lambda^{'} ; t, t') \bigg] \left[  \theta(t_1- t) - \frac{1}{1 - \rme^{\beta(\epsilon_{\bf k} - \epsilon_{\bf k - q})}}\right]      \nonumber      \\
    &\hspace{-0.6in} + (1 - n_F({\bf k}))   \rme^{-\rmi (t_1 - t')(\epsilon_{\bf k} - \epsilon_{\bf k - q})}  G_{{\bf k - q, <}}(\lambda, \lambda^{'} ; t, t')  \left[  \theta(t_1- t') - \frac{1}{1 - \rme^{\beta(\epsilon_{\bf k} - \epsilon_{\bf k -q})}}\right]  
\end{eqnarray}
 In order to find equations which involve Green's functions, we must now take limit $ t_1 \rightarrow t_-$ and sum over ${\bf q}$. The details are relegated to the supplementary material. The crucial closed equations for the one-particle Green functions are now writable as,
 \newpage
 \begin{eqnarray}
&\fl -\partial_{ \lambda } \,G_{{\bf k,>}}(\lambda, \lambda'; t, t')&
\nonumber \\
 & \mbox{} &\fl =\sum_{ {\bf{p}} }
  \frac{1}{\rme^{ \beta  (\epsilon_{ {\bf{k}} }-\epsilon_{ {\bf{p}} })  }\rme^{\lambda + \lambda^{'} }-1}
 \bigg[ \rme^{ \lambda+\lambda^{'} }\rme^{ \rmi   (t-t^{'}) (\epsilon_{ {\bf{p}} } - \epsilon_{ {\bf{k}} }) }
  \mbox{          }
   G_{{\bf p,<}}(\lambda, \lambda^{'} ; t, t') (1- n_F({\bf{k}}) )  
    \nonumber \\
   &\mbox{ } &\fl  \quad \times  \left(   \theta (t^{'}-t)+\theta (t-t^{'})\rme^{ \beta (\epsilon_{ {\bf{k}} }-\epsilon_{ {\bf{p}} }) } \right)-  G_{{\bf k,>}}(\lambda, \lambda^{'} ; t, t')   \mbox{             } n_F({\bf{p}}) \bigg]\label{ceq1}
\end{eqnarray}
 and
  \begin{eqnarray}
&\fl -\partial_{ \lambda^{'} } \mbox{             }G_{{\bf k,<}}(\lambda, \lambda'; t, t')
 \mbox{             }& \mbox{             }
 \nonumber \\
& \mbox{} &\fl 
  = \sum_{ {\bf{p}} }\mbox{          }
   \frac{1}
 { \rme^{ \beta (\epsilon_{ {\bf{p}} }-\epsilon_{ {\bf{k}} }) }\rme^{\lambda + \lambda^{'} }-1 }
  \rme^{ \rmi  ( t-t^{'}) (\epsilon_{ {\bf{p}} }-\epsilon_{ {\bf{k}} } )  }
 \mbox{              }G_{{\bf p,>}}(\lambda, \lambda^{'} ; t, t')  \mbox{             }  n_F({\bf{k}})
 \nonumber\\
 & \mbox{} & \fl \quad \times
 \left(  \theta (t-t^{'})+\rme^{ \beta (\epsilon_{ {\bf{p}} }-\epsilon_{ {\bf{k}} }) }\mbox{  } \theta (t^{'}-t)\right) - G_{{\bf k,<}}(\lambda, \lambda^{'}; t, t')\mbox{             }
 (1-  n_F( {\bf{p}} ) ) \label{ceq2}
\end{eqnarray}
 In order to find out $G_{{\bf k,<}}(\lambda, \lambda', t, t')$ and $G_{{\bf k,>}}(\lambda, \lambda', t, t')$, it remains for us to solve equations (\ref{ceq1}) and (\ref{ceq2}).
  \paragraph{}
\section{ The Coupled Equations }
The coupled equations for the free fermion Green functions may be written down suggestively as follows:
\setlength{\arraycolsep}{2pt}
\begin{eqnarray}
 &\fl \nonumber-\partial_{ \lambda }G_{ {\bf{k}},> } (\lambda, \lambda^{'}; t,t^{'})& \\
 \nonumber&\mbox{ } &\fl =\sum_{ {\bf{p}} }( \rme^{ \lambda+\lambda^{'} }  \rme^{ \rmi  (t-t^{'}) (\epsilon_{ {\bf{p}} } - \epsilon_{ {\bf{k}} }) }
G_{ {\bf{p}},< }(\lambda, \lambda^{'}; t,t^{'})   (1- n_F({\bf{k}}) )(\theta (t^{'}-t)+\theta (t-t^{'})e^{ \beta (\epsilon_{ {\bf{k}} }-\epsilon_{ {\bf{p}} }) }  ) \\&\mbox{} &\fl  \qquad-
   G_{ {\bf{k}},> }(\lambda, \lambda^{'}; t,t^{'})     n_F({\bf{p}}) ) \mbox{                   } 
   n_B(\epsilon_{ {\bf{k}} }-\epsilon_{ {\bf{p}} }, - \frac{\lambda + \lambda^{'} }{\beta})
   \label{CPL1}
\end{eqnarray}
and
\begin{eqnarray}
& \fl -\partial_{ \lambda^{'} }  G_{ {\bf{k}},< }(\lambda, \lambda^{'}; t,t^{'})
 &  
\nonumber \\ &\mbox{} &\fl  = \sum_{ {\bf{p}} }\mbox{          }(\rme^{ \rmi ( t-t^{'}) (\epsilon_{ {\bf{p}} }-\epsilon_{ {\bf{k}} } )  }
\mbox{              } G_{ {\bf{p}},> }(\lambda, \lambda^{'}; t,t^{'})    n_F({\bf{k}})
 (  \theta (t-t^{'})+\rme^{ \beta (\epsilon_{ {\bf{p}} }-\epsilon_{ {\bf{k}} }) }\mbox{  } \theta (t^{'}-t) ) \nonumber \\&\mbox{} &\fl \qquad -
 G_{ {\bf{k}},< }(\lambda, \lambda^{'}; t,t^{'}) 
(1-  n_F( {\bf{p}} ) ) )   n_B(\epsilon_{ {\bf{p}} }-\epsilon_{ {\bf{k}} }, - \frac{\lambda + \lambda^{'} }{\beta})
   \label{CPL2}
\end{eqnarray}
where,
\begin{eqnarray}
 n_B(E, \mu) \equiv \frac{1}{\rme^{ \beta (E-\mu) } - 1 }
\end{eqnarray}
is the Bose-Einstein distribution. Thus, the unknown fermion Green functions viz. $ G_{ {\bf{k}},<(>) }(t,t^{'};\lambda, \lambda^{'}) $ obey first-order ordinary differential equations in the parameters $ \lambda, \lambda^{'} $ with non-constant coefficients and the coefficients are nothing but Bose-Einstein distributions of the particle-hole like objects that the fermion is being described in terms of.

\section{ Solution of the coupled equations }

Here, we outline the main steps in determining the solution to these coupled equations. The details may be found in the supplementary files. First, we note that the time dependence of the fermion Green function may be largely eliminated through the substitution ($ a = >, < $):
\begin{eqnarray}
G_{ {\bf{k}}, a }(\lambda, \lambda^{'}; t,t^{'}) = {\tilde{G}}_{ {\bf{k}},a }(\lambda, \lambda^{'}; t,t^{'}) \mbox{  } \rme^{ - \rmi\epsilon_{ {\bf{k}} }(t-t^{'}) }
\end{eqnarray}
The reduced Green function $ {\tilde{G}}_{ {\bf{k}},a }(\lambda, \lambda^{'}; t,t^{'})  $ may be solved through the introduction of a form of separation of variables as shown below. The suggested separation, though seemingly ad-hoc, is well motivated since the left-hand sides of equation (\ref{CPL1}) and equation (\ref{CPL2}) depend only on $ {\bf{k}} $ but the summands on the right depend on both $ {\bf{k}} $ and $ {\bf{p}} $. We want the summand to factor into two parts, one depending only on $ {\bf{k}} $ and the other only on $ {\bf{p}} $. This suggestion is sufficient to fully solve the equations in the previous section reductively without taking help from the known solution obtained from quantum statistics of fermions (i.e. trivial solution of the free fermion Green function). 
\begin{eqnarray}
&\rme^{ \lambda+\lambda^{'} }\mbox{             }
   {\tilde{G}}_{ {\bf{p}},< }(\lambda, \lambda^{'}; t,t^{'}) (1- n_F({\bf{k}}) ) \nonumber  
   \left(   \theta (t^{'}-t)+\theta (t-t^{'})\rme^{ \beta (\epsilon_{ {\bf{k}} }-\epsilon_{ {\bf{p}} }) } \right) \\ & - 
  {\tilde{G}}_{ {\bf{k}},> }(\lambda, \lambda^{'}; t,t^{'})      \mbox{             } n_F({\bf{p}}) \nonumber \\& \qquad =
(\rme^{ \beta  (\epsilon_{ {\bf{k}} }-\epsilon_{ {\bf{p}} })  }\rme^{\lambda + \lambda^{'} }-1)\mbox{               }
L_{>}({\bf{k}}; \lambda, \lambda^{'};t-t^{'}) R_{<}({\bf{p}}; \lambda, \lambda^{'};t-t^{'})
\end{eqnarray}
and
\begin{eqnarray}
& \fl   {\tilde{G}}_{ {\bf{p}},> }(\lambda, \lambda^{'}; t,t^{'})    n_F({\bf{k}})
&\left(  \theta (t-t^{'})+\rme^{ \beta (\epsilon_{ {\bf{p}} }-\epsilon_{ {\bf{k}} }) }\mbox{  } \theta (t^{'}-t)\right) - 
 {\tilde{G}}_{ {\bf{k}},< }(\lambda, \lambda^{'}; t,t^{'})  
(1-  n_F( {\bf{p}} ) )
\nonumber \\& \mbox{} & = 
( \rme^{ \beta (\epsilon_{ {\bf{p}} }-\epsilon_{ {\bf{k}} }) }\rme^{\lambda + \lambda^{'} }-1 )
L_{<}({\bf{k}}; \lambda, \lambda^{'};t-t^{'}) R_{>}({\bf{p}}; \lambda, \lambda^{'};t-t^{'})
\end{eqnarray}
In the above assertions, we have separated the $ {\bf{k}}  $ and $ {\bf{p}} $ dependence through the introduction of four new functions viz. $ L_{>,<} , R_{>,<} $. After some rearrangement, we may write down formulas for the Green functions directly in terms of these reduced functions.
\begin{eqnarray}
&\! \fl {\tilde{G}}_{ {\bf{p}}, < }(\lambda, \lambda^{'}; t,t^{'})
\mbox{      } (1-n_F({\bf{k}}))\mbox{            } &= \mbox{                }
L_{<}({\bf{p}}; \lambda, \lambda^{'};t-t^{'}) R_{>}({\bf{k}}; \lambda, \lambda^{'};t-t^{'}) \nonumber
\\& \mbox{ } &\, +L_{>}({\bf{k}}; \lambda, \lambda^{'};t-t^{'}) R_{<}({\bf{p}}; \lambda, \lambda^{'};t-t^{'})
\mbox{             } \rme^{ \theta(t^{'}-t) \beta (\epsilon_{ {\bf{k}} }- \epsilon_{ {\bf{p}} })  }
\end{eqnarray}
and
\begin{eqnarray}
&\! \fl  {\tilde{G}}_{ {\bf{k}}, > }(\lambda, \lambda^{'}; t,t^{'})
\mbox{      }  n_F({\bf{p}}) \mbox{            } &= \mbox{                }  L_{>}({\bf{k}}; \lambda, \lambda^{'};t-t^{'}) R_{<}({\bf{p}}; \lambda, \lambda^{'};t-t^{'}) \nonumber \\&\mbox{ }&\quad  +L_{<}({\bf{p}}; \lambda, \lambda^{'};t-t^{'}) R_{>}({\bf{k}}; \lambda, \lambda^{'};t-t^{'})
\mbox{           } \rme^{\theta (t-t^{'}) \beta  (\epsilon_{ {\bf{k}} }-\epsilon_{ {\bf{p}} })}  \mbox{          }
 \rme^{\lambda + \lambda^{'} }
\end{eqnarray}
Given that the left-hand side of these expressions depends only on $ {\bf{p}} $ or $ {\bf{k}} $ (apart from the step functions), the right-hand side has to be something special in order for this to be possible. After some contemplation, we conclude that the only way this can be achieved is through the identifications below:
\begin{eqnarray}
& \fl R_{<}({\bf{p}}; \lambda, \lambda^{'};t-t^{'})\mbox{        }  =&  \mbox{     } g_{<}(\lambda,\lambda^{'};t-t^{'})\mbox{      } n_F({\bf{p}}) \nonumber \\&
\mbox{ }& \times ( 1+
 \rme^{-  \beta  \epsilon_{ {\bf{p}} } }  c^{-1}_{<}(\lambda,\lambda^{'};t-t^{'})  c_{>}(\lambda,\lambda^{'};t-t^{'}) \mbox{          }
 \rme^{\lambda + \lambda^{'} } )^{-1}
\end{eqnarray}
and
\begin{eqnarray}
& \fl R_{>}({\bf{k}}; \lambda, \lambda^{'};t-t^{'}) \mbox{             }=& \mbox{     } g_{>}(\lambda,\lambda^{'};t-t^{'})\mbox{      } (1-n_F({\bf{k}}))\nonumber \\&  \mbox{ } &\times ( 1
+   \rme^{  \beta  \epsilon_{ {\bf{k}} } } c^{-1}_{>}(\lambda,\lambda^{'};t-t^{'}) \mbox{          }c_{<}(\lambda,\lambda^{'};t-t^{'}))^{-1}
\end{eqnarray}
and
\begin{eqnarray}
&\! \fl  L_{<}({\bf{p}}; \lambda, \lambda^{'};t-t^{'}) \mbox{                 } &= \mbox{                  }c^{-1}_{<}(\lambda,\lambda^{'};t-t^{'}) \mbox{     } g_{<}(\lambda,\lambda^{'};t-t^{'})\mbox{      } n_F({\bf{p}})\mbox{      }
\nonumber \\& \mbox{ } & \,\, \times ( 1+
 \rme^{-  \beta  \epsilon_{ {\bf{p}} } }  c^{-1}_{<}(\lambda,\lambda^{'};t-t^{'})  c_{>}(\lambda,\lambda^{'};t-t^{'}) \mbox{          }
 \rme^{\lambda + \lambda^{'} } )^{-1}
\mbox{             } \rme^{ -\theta(t^{'}-t) \beta  \epsilon_{ {\bf{p}} } }
\end{eqnarray}
and
\begin{eqnarray}
&\fl  L_{>}({\bf{k}}; \lambda, \lambda^{'};t-t^{'})  \mbox{                 } &= \mbox{                  } c^{-1}_{>}(\lambda,\lambda^{'};t-t^{'}) \mbox{          }\mbox{     } g_{>}(\lambda,\lambda^{'};t-t^{'})\mbox{      } (1-n_F({\bf{k}})) \nonumber \\& \mbox{} & \times ( 1
+   \rme^{  \beta  \epsilon_{ {\bf{k}} } }   \mbox{             } c^{-1}_{>}(\lambda,\lambda^{'};t-t^{'}) \mbox{          }c_{<}(\lambda,\lambda^{'};t-t^{'})
)^{-1}  \rme^{\theta (t-t^{'}) \beta  \epsilon_{ {\bf{k}} } }
\end{eqnarray}
where $ c_{>,<}, g_{>,<} $ etc. are suitable reduced quantities that have to be later determined. The Green functions themselves may be written in terms of these reduced quantities.
\begin{eqnarray}
& \fl {\tilde{G}}_{ {\bf{p}}, < }(\lambda, \lambda^{'}; t,t^{'})
=& c^{-1}_{<}(\lambda,\lambda^{'};t-t^{'}) g_{<}(\lambda,\lambda^{'};t-t^{'})  g_{>}(\lambda,\lambda^{'};t-t^{'})  n_F({\bf{p}}) 
\nonumber \\& \mbox{} & \times \left( 1+
 \rme^{-  \beta  \epsilon_{ {\bf{p}} } }  c^{-1}_{<}(\lambda,\lambda^{'};t-t^{'})  c_{>}(\lambda,\lambda^{'};t-t^{'})  
 \rme^{\lambda + \lambda^{'} } \right)^{-1}
 \rme^{ -\theta(t^{'}-t) \beta  \epsilon_{ {\bf{p}} } }
\end{eqnarray}
and
\begin{eqnarray}
&\fl  {\tilde{G}}_{ {\bf{k}}, > }(\lambda, \lambda^{'}; t,t^{'})
=&
c^{-1}_{>}(\lambda,\lambda^{'};t-t^{'}) \mbox{          }   g_{>}(\lambda,\lambda^{'};t-t^{'})
 g_{<}(\lambda,\lambda^{'};t-t^{'})  (1-n_F({\bf{k}})) \nonumber \\&\mbox{} &  \times ( 1
+   \rme^{  \beta  \epsilon_{ {\bf{k}} } }   \mbox{             } c^{-1}_{>}(\lambda,\lambda^{'};t-t^{'}) \mbox{          }c_{<}(\lambda,\lambda^{'};t-t^{'})
)^{-1} \rme^{\theta (t-t^{'}) \beta  \epsilon_{ {\bf{k}} } }
\end{eqnarray}
The coupled equations of the previous section may be written as follows:
\begin{eqnarray}
-\partial_{ \lambda }   {\tilde{G}}_{ {\bf{k}}, > }(\lambda, \lambda^{'}; t,t^{'})  =  L_{>}({\bf{k}}; \lambda, \lambda^{'};t-t^{'}) \sum_{ {\bf{p}} } 
R_{<}({\bf{p}}; \lambda, \lambda^{'};t-t^{'})
\end{eqnarray}
and
\begin{eqnarray}
-\partial_{ \lambda^{'} }  {\tilde{G}}_{ {\bf{p}}, < }(\lambda, \lambda^{'}; t,t^{'})
 = L_{<}({\bf{p}}; \lambda, \lambda^{'};t-t^{'})  \sum_{ {\bf{k}} }\mbox{          } R_{>}({\bf{k}}; \lambda, \lambda^{'};t-t^{'})
\end{eqnarray}
These may be used to solve for the remaining unknowns. Since the details are quite lengthy, they are relegated to the supplementary documents attached to this paper.
\section{  Final results }
 After following through with the suggestions above, we may derive,
\begin{eqnarray}
& \fl {\tilde{G}}_{ {\bf{p}}, < }(\lambda, \lambda^{'}; t,t^{'}) =&  sgn(t-t^{'})\mbox{                }
  \frac{ n_F({\bf{p}})\mbox{      }
  \rme^{ -\theta(t^{'}-t) \beta  (\epsilon_{ {\bf{p}} }-\mu) } }{\left( 1+
 \rme^{-  \beta  (\epsilon_{ {\bf{p}} }-\mu) }   \mbox{          }
  \rme^{\lambda   }    \right)} \nonumber \\& \mbox{} & \times \rme^{  - \int^{\lambda}_{0} d s\mbox{             }   \sum_{ {\bf{p}} }\mbox{      } \frac{ n_F({\bf{p}})
 }{
\left( 1+
 \rme^{-  \beta  (\epsilon_{ {\bf{p}} }-\mu) }  \mbox{          }
  \rme^{s  } \right) } }\mbox{             }  \rme^{ - \int^{ \lambda^{'} }_{ 0 } ds \sum_{ {\bf{k}} }\mbox{      } \frac{ (1-n_F({\bf{k}}))
 }{ \left( 1
+   \rme^{  \beta  (\epsilon_{ {\bf{k}} }-\mu) }   \mbox{             }  \rme^{  s }
\right) }   }
\end{eqnarray}
and
\begin{eqnarray}
& \fl {\tilde{G}}_{ {\bf{k}}, > }(\lambda, \lambda^{'}; t,t^{'}) =& 
 \rme^{ \lambda^{'} }  \mbox{           }
sgn(t-t^{'})\mbox{                }   \frac{ (1-n_F({\bf{k}})) \rme^{\theta (t-t^{'}) \beta  (\epsilon_{ {\bf{k}} }-\mu) } }{  \left( 1
+   \rme^{  \beta  (\epsilon_{ {\bf{k}} }-\mu) }   \mbox{             }
   \rme^{  \lambda^{'} }
\right) } \nonumber \\& \mbox{} & \times \rme^{  - \int^{\lambda}_{0} d s\mbox{             }   \sum_{ {\bf{p}} }\mbox{      } \frac{ n_F({\bf{p}})
 }{
\left( 1+
 \rme^{-  \beta  (\epsilon_{ {\bf{p}} }-\mu) }  \mbox{          }
  \rme^{s  } \right) } }  \mbox{          } \rme^{ - \int^{ \lambda^{'} }_{ 0 } ds \sum_{ {\bf{k}} }\mbox{      } \frac{ (1-n_F({\bf{k}}))
 }{ \left( 1
+   \rme^{  \beta  (\epsilon_{ {\bf{k}} }-\mu) }   \mbox{             }  \rme^{  s }
\right) }   }
\end{eqnarray}
Note that the times $ 0 < t,t^{'} < -\rmi\beta $ are on the imaginary axis in the interval $ (0,-\rmi\beta) $. Also
$ sgn(X) \equiv \theta(X) - \theta(-X) $. These results are identical to those obtained directly from the quantum statistics of the original fermions by performing trace over fermion Fock spaces in the context of the grand canonical ensemble. Those results (derived again in the supplementary material) are as follows:
\begin{eqnarray}
&\fl  {\tilde{G}}_{ {\bf{p}}, < }(\lambda, \lambda^{'}; t,t^{'}) =&  sgn(t-t^{'})\mbox{                }
  \frac{ n_F({\bf{p}})\mbox{      }
  \rme^{ -\theta(t^{'}-t) \beta  (\epsilon_{ {\bf{p}} }-\mu) } }{\left( 1+
 \rme^{-  \beta  (\epsilon_{ {\bf{p}} }-\mu) }   \mbox{          }
  \rme^{\lambda   }    \right)} \nonumber \\& \mbox{} &\times\mbox{  }
 \prod_{ {\bf{k}}, n_F({\bf{k}}) = 1   } \mbox{  } \frac{ (\rme^{ - \beta  (\epsilon_{ {\bf{k}} }-\mu)  }
  + 
\rme^{ -  \lambda 
  }
)}{ (1 + \rme^{ - \beta (\epsilon_{ {\bf{k}} }-\mu)  })}
  \mbox{        }
\prod_{ {\bf{k}}, n_F({\bf{k}}) = 0   } \mbox{  } \frac{ (\rme^{ - \beta (\epsilon_{ {\bf{k}} }-\mu)  }
\rme^{ -    \lambda^{'}   } + 1
)}{ (1 + \rme^{ - \beta (\epsilon_{ {\bf{k}} }-\mu)  })}
\end{eqnarray}
and
\begin{eqnarray}
& \fl {\tilde{G}}_{ {\bf{k}}, > }(\lambda, \lambda^{'}; t,t^{'}) =& 
 \rme^{ \lambda^{'} }  \mbox{           }
sgn(t-t^{'})\mbox{                }   \frac{ (1-n_F({\bf{k}})) \rme^{\theta (t-t^{'}) \beta  (\epsilon_{ {\bf{k}} }-\mu) } }{  \left( 1
+   \rme^{  \beta  (\epsilon_{ {\bf{k}} }-\mu) }   \mbox{             }
   \rme^{  \lambda^{'} }
\right) } \nonumber \\& \mbox{} & \times \prod_{ {\bf{p}}, n_F({\bf{p}}) = 1   } \mbox{  } \frac{ (\rme^{ - \beta  (\epsilon_{ {\bf{p}} }-\mu)  }
  + 
\rme^{ -  \lambda 
  }
)}{ (1 + \rme^{ - \beta (\epsilon_{ {\bf{p}} }-\mu)  })}
  \mbox{        }
\prod_{ {\bf{p}}, n_F({\bf{p}}) = 0   } \mbox{  } \frac{ (\rme^{ - \beta (\epsilon_{ {\bf{p}} }-\mu)  }
\rme^{ -    \lambda^{'}   } + 1
)}{ (1 + \rme^{ - \beta (\epsilon_{ {\bf{p}} }-\mu)  })}
\end{eqnarray}
These two match giving us confidence that the present approach is likely to be fruitful when applied to more interesting systems where interactions between fermions are present. However, for the study of such systems, approximations will have to be made and, more importantly, justified. This is a difficult next step that will not be attempted here.

\section{Conclusions}

In this work, we have shown how a fermion in the context of a many-body system possessing a Fermi surface may be regarded alternatively as a (collection of) non-local particle-hole excitations across this Fermi surface. The Green function of fermions (more generally, all the N-point functions) has been shown to be a non-local combination of a Bose-Einstein distribution, indicating thereby that the non-local particle-hole excitation obeys statistics resembling bosons with a chemical potential, even though at the operator level, they are not bosons at all. The total number conserving Fermi bilinears have been shown to be expressible in terms of these non-local particle-hole excitations in such a way that the kinetic energy of free fermions is diagonal in both the original Fermi language and also in this new language of non-local particle-hole operators. ChatGPT has suggested the name ``splashon" to describe these objects (presumably because it conjures up an image of the activity seen when a stone is thrown into a pool of water and some water exits the surface, leaving behind a temporary void in the pool beneath, complete with the attendant ripples and so on that it leaves behind). 

{\bf{Author contributions:}} The idea behind this paper is due to Girish, especially from Chap.12 of his textbook \cite{GSS13}. Alok independently did all the derivations in this paper and found more compact versions of the derivations suggested by Girish. Rishi also independently verified all the derivations and wrote the paper. 
\bibliography{main}
\end{document}

\title{ Detailed derivations pertaining to the paper ``Fermion as a non-local particle-hole excitation" }
\date{}
\author{Alok Kushwaha, Rishi Paresh Joshi and Girish S. Setlur}

\maketitle

\section{ Free theory equations of motion }

Fermions are described by annihilation and creation operators $ c_{ {\bf{p}} }, c^{\dagger}_{ {\bf{p}} } $. These fermions have a Fermi surface at zero temperature, described by $ E_F = \epsilon_{ {\bf{p}} } $. At zero temperature they have the momentum distribution $ <c^{\dagger}_{ {\bf{p}} }c_{ {\bf{p}} }> \equiv n_F({\bf{p}})
 \equiv \theta(E_F-\epsilon_{ {\bf{p}} }) $. Here $ \theta(X> 0 ) = 1 , \theta(X<0) = 0 $ and $ \theta(0) = 1/2 $ is the Heaviside step function. We define
$ c_{ {\bf{p}}, < } \equiv n_F({\bf{p}})\mbox{  }c_{ {\bf{p}} } $ and $ c_{ {\bf{p}}, > } \equiv (1-n_F({\bf{p}}))\mbox{  }c_{ {\bf{p}} } $.
Define,
\[
N_{>}(t) = \sum_{ {\bf{k}} }c_{ {\bf{k}}, < }(t) c^{\dagger}_{ {\bf{k}}, < }(t); \mbox{          } \mbox{        }
N^{'}_{>}(t) = \sum_{ {\bf{k}} }c^{\dagger}_{ {\bf{k}}, > }(t) c_{ {\bf{k}}, > }(t)
\]
They represent different version of the number of particle hole pairs. For notational simplicity, in the rest of the description below we assume $ \epsilon_{ {\bf{k}} } \equiv \frac{k^2}{2m} $. This means $ \frac{({\bf{k}}-{\bf{q}}/2)\cdot {\bf{q}} }{m} $ is shorthand for $ \epsilon_{ {\bf{k}} } - \epsilon_{ {\bf{k}}-{\bf{q}} } $  and so on. In other words, the discussion below is completely general and applicable to any $ \epsilon_{ {\bf{k}} } $. Consider the operators 

\[
a_{\bf k}({\bf q})= c^\dagger_{\mathbf{k}-\frac{\mathbf{q}}{2}<}c_{\mathbf{k}+\frac{\mathbf{q}}{2}>} ;
 \quad    a^\dagger_{\bf k}({\bf q}) = c^\dagger_{\mathbf{k}+\frac{\mathbf{q}}{2}>}c_{\mathbf{k}-\frac{\mathbf{q}}{2}<}
\]

---------------------------------------------------------------------------------------------------------------

Let us define four point function

\begin{equation*}
   F_{\bf k <}({\bf -q};t_1) \equiv <T \mbox{   }e^{ - \lambda N_{>}(t)} \mbox{             }e^{ - \lambda^{'} N^{'}_{>}(t)}   \mbox{        } c_{ {\bf{k-q}}, > }(t) \mbox{             }a^{\dagger}_{ {\bf{k}} - {\bf{q}}/2 }(-{\bf{q}},t_1)c^{\dagger}_{ {\bf{k}},< }(t^{'}) >
\end{equation*}
and 

\begin{equation*}
   F_{\bf k >}({\bf q};t_1) \equiv  <T \mbox{   }e^{ - \lambda N_{>}(t)}  \mbox{             }e^{ - \lambda^{'} N^{'}_{>}(t)} \mbox{             } a_{ {\bf{k}} - {\bf{q}}/2 }({\bf{q}},t_1)  \mbox{        }c_{ {\bf{k-q}}, < }(t)
    c^{\dagger}_{ {\bf{k}},> }(t^{'}) >
\end{equation*}

Equations of motion for these functions are given by
\begin{align*}
i\partial_{t_1}<T \mbox{   }e^{ - \lambda N_{>}(t)} \mbox{             }e^{ - \lambda^{'} N^{'}_{>}(t)} \mbox{             }&c_{ {\bf{k-q}}, > }(t) \mbox{ }a^{\dagger}_{ {\bf{k}} - {\bf{q}}/2 }(-{\bf{q}},t_1)  \mbox{        } 
c^{\dagger}_{ {\bf{k}},< }(t^{'}) >  \mbox{             }= \\
&(\epsilon_{ {\bf{k}} } - \epsilon_{ {\bf{k}}-{\bf{q}} })\mbox{  }<T \mbox{   }e^{ - \lambda N_{>}(t)}  \mbox{             }e^{ - \lambda^{'} N^{'}_{>}(t)}    \mbox{        }c_{ {\bf{k-q}}, > }(t) \mbox{             }a^{\dagger}_{ {\bf{k}} - {\bf{q}}/2 }(-{\bf{q}},t_1)
c^{\dagger}_{ {\bf{k}},< }(t^{'}) >\\
&+ (e^{\lambda+\lambda^{'}} -  1)\mbox{   } <T \mbox{            } e^{ - \lambda N_{>}(t)} \mbox{             }e^{ - \lambda^{'} N^{'}_{>}(t)} \mbox{             }
a^{\dagger}_{ {\bf{k}} - {\bf{q}}/2 }(-{\bf{q}},t)
\mbox{        }
 c_{ {\bf{k-q}}, > }(t)
c^{\dagger}_{ {\bf{k}},< }(t^{'}) >   \mbox{             } i\mbox{   }\delta(t_1-t)\\
&- (1-n_F({\bf{k}}-{\bf{q}}))\mbox{       } <T \mbox{            } e^{ - \lambda N_{>}(t)} \mbox{             }e^{ - \lambda^{'} N^{'}_{>}(t)} \mbox{             }
c_{ {\bf{k}}, < }(t)
c^{\dagger}_{ {\bf{k}},< }(t^{'}) >   \mbox{             } i\mbox{   }\delta(t_1-t)\\
 &+ n_F({\bf{k}}) \mbox{  }  <T \mbox{           } e^{ - \lambda N_{>}(t)} \mbox{             }e^{ - \lambda^{'} N^{'}_{>}(t)} \mbox{             } c_{ {\bf{k-q}}, > }(t)
 c^{\dagger}_{ {\bf{k}} - {\bf{q}}, > }(t^{'}) >   \mbox{             }  i\mbox{       }   \delta(t_1-t^{'})
\end{align*}
--------------------------------------------------------------------------------------------------------------

\begin{align*}
i\partial_{t_1}<T \mbox{   }e^{ - \lambda N_{>}(t)}  \mbox{             }e^{ - \lambda^{'} N^{'}_{>}(t)} \mbox{             }&c_{ {\bf{k-q}}, < }(t) \mbox{ } a_{ {\bf{k}} - {\bf{q}}/2 }({\bf{q}},t_1)  \mbox{        }
c^{\dagger}_{ {\bf{k}},> }(t^{'}) >  \mbox{             }=\\
&(\epsilon_{ {\bf{k}} } - \epsilon_{ {\bf{k}}-{\bf{q}} })\mbox{  }<T \mbox{   }e^{ - \lambda N_{>}(t)} \mbox{             }e^{ - \lambda^{'} N^{'}_{>}(t)}   \mbox{        } c_{ {\bf{k-q}}, < }(t) \mbox{             } a_{ {\bf{k}} - {\bf{q}}/2 }({\bf{q}},t_1)
c^{\dagger}_{ {\bf{k}},> }(t^{'}) >
\\
&
+(e^{ - \lambda-\lambda^{'} }  -1) \mbox{       } < T\mbox{                }
 e^{ - \lambda N_{>}(t)}  \mbox{             }e^{ - \lambda^{'} N^{'}_{>}(t)}\mbox{ } a_{ {\bf{k}} - {\bf{q}}/2 }({\bf{q}},t) \mbox{             } c_{ {\bf{k-q}}, < }(t)
\mbox{             }
  c^{\dagger}_{ {\bf{k}},> }(t^{'}) >\mbox{   }i\delta(t_1-t)\\
&
-  n_F({\bf{k}}-{\bf{q}})\mbox{          } < T\mbox{                } e^{ - \lambda N_{>}(t)}  \mbox{             }e^{ - \lambda^{'} N^{'}_{>}(t)} \mbox{             }
c_{ {\bf{k}}, > }(t)
c^{\dagger}_{ {\bf{k}},> }(t^{'}) >\mbox{   }i\delta(t_1-t)
\\
&
+ <T\mbox{          } e^{ - \lambda N_{>}(t)}  \mbox{             }e^{ - \lambda^{'} N^{'}_{>}(t)} \mbox{             } c_{ {\bf{k-q}}, < }(t)\mbox{  }
c^{\dagger}_{ {\bf{k}}-{\bf{q}},< }(t^{'})
 >\mbox{   }(1-n_F({\bf{k}}))\mbox{   } i\delta(t_1-t^{'})
\end{align*}
====================================================================

\begin{align*}
i\partial_{t_1}<T \mbox{   }e^{ - \lambda N_{>}(t)} \mbox{             }e^{ - \lambda^{'} N^{'}_{>}(t)} \mbox{             } &c_{ {\bf{k-q}}, > }(t) a^{\dagger}_{ {\bf{k}} - {\bf{q}}/2 }(-{\bf{q}},t_1)  \mbox{        }
c^{\dagger}_{ {\bf{k}},< }(t^{'}) >  \mbox{             }=\\
&(\epsilon_{ {\bf{k}} } - \epsilon_{ {\bf{k}}-{\bf{q}} })\mbox{  }<T \mbox{   }e^{ - \lambda N_{>}(t)}  \mbox{             }e^{ - \lambda^{'} N^{'}_{>}(t)}   \mbox{        }c_{ {\bf{k-q}}, > }(t)\mbox{             } a^{\dagger}_{ {\bf{k}} - {\bf{q}}/2 }(-{\bf{q}},t_1)
c^{\dagger}_{ {\bf{k}},< }(t^{'}) >
\\
&
+ (e^{\lambda+\lambda^{'}} -  1)\mbox{   } <T \mbox{            } e^{ - \lambda N_{>}(t)} \mbox{             }e^{ - \lambda^{'} N^{'}_{>}(t)} \mbox{             }
 c_{ {\bf{k-q}}, > }(t) \mbox{ } a^{\dagger}_{ {\bf{k}} - {\bf{q}}/2 }(-{\bf{q}},t)\mbox{        }
c^{\dagger}_{ {\bf{k}},< }(t^{'}) >   \mbox{             } i\mbox{   }\delta(t_1-t)
\\
&
- (1-n_F({\bf{k}}-{\bf{q}}))\mbox{       } e^{\lambda+\lambda^{'}} \mbox{ } <T \mbox{            } e^{ - \lambda N_{>}(t)} \mbox{             }e^{ - \lambda^{'} N^{'}_{>}(t)} \mbox{             }
c_{ {\bf{k}}, < }(t)
c^{\dagger}_{ {\bf{k}},< }(t^{'}) >   \mbox{             } i\mbox{   }\delta(t_1-t)
\\
&
 + n_F({\bf{k}}) \mbox{  }  <T \mbox{           } e^{ - \lambda N_{>}(t)} \mbox{             }e^{ - \lambda^{'} N^{'}_{>}(t)} \mbox{             } c_{ {\bf{k-q}}, > }(t)
 c^{\dagger}_{ {\bf{k}} - {\bf{q}}, > }(t^{'}) >   \mbox{             }  i\mbox{       }   \delta(t_1-t^{'})
\end{align*}

--------------------------------------------------------------------------------------------------------------

\begin{align*}
i\partial_{t_1}<T \mbox{   }e^{ - \lambda N_{>}(t)}  \mbox{             }e^{ - \lambda^{'} N^{'}_{>}(t)} \mbox{             } &c_{ {\bf{k-q}}, < }(t)
a_{ {\bf{k}} - {\bf{q}}/2 }({\bf{q}},t_1)  \mbox{        }
c^{\dagger}_{ {\bf{k}},> }(t^{'}) >  \mbox{             }=\\
& 
(\epsilon_{ {\bf{k}} } - \epsilon_{ {\bf{k}}-{\bf{q}} })\mbox{  }<T \mbox{   }e^{ - \lambda N_{>}(t)} \mbox{             }e^{ - \lambda^{'} N^{'}_{>}(t)} \mbox{             } c_{ {\bf{k-q}}, < }(t)
a_{ {\bf{k}} - {\bf{q}}/2 }({\bf{q}},t_1)  \mbox{        }
c^{\dagger}_{ {\bf{k}},> }(t^{'}) >
\\
&
+(e^{ - \lambda-\lambda^{'} }  -1) \mbox{       } < T\mbox{                }
 e^{ - \lambda N_{>}(t)}  \mbox{             }e^{ - \lambda^{'} N^{'}_{>}(t)} \mbox{             }
 \mbox{             } c_{ {\bf{k-q}}, < }(t) a_{ {\bf{k}} - {\bf{q}}/2 }({\bf{q}},t)
c^{\dagger}_{ {\bf{k}},> }(t^{'}) >\mbox{   }i\delta(t_1-t)
\\
&
-  n_F({\bf{k}}-{\bf{q}})\mbox{          }e^{ -\lambda - \lambda^{'}} < T\mbox{                } e^{ - \lambda N_{>}(t)}  \mbox{             }e^{ - \lambda^{'} N^{'}_{>}(t)} \mbox{             }
c_{ {\bf{k}}, > }(t)
c^{\dagger}_{ {\bf{k}},> }(t^{'}) >\mbox{   }i\delta(t_1-t)
\\
&
+ <T\mbox{          } e^{ - \lambda N_{>}(t)}  \mbox{             }e^{ - \lambda^{'} N^{'}_{>}(t)} \mbox{             } c_{ {\bf{k-q}}, < }(t)\mbox{  }
c^{\dagger}_{ {\bf{k}}-{\bf{q}},< }(t^{'})
 >\mbox{   }(1-n_F({\bf{k}}))\mbox{   } i\delta(t_1-t^{'})
\end{align*}

\section{Solutions of equations of motion  }

The above equations are linear differential equation. The solution of these equation contains integration of constant, which is determined by the periodic boundary condition at $t_1 = 0$ and $t_1 = -i \beta$. After substituting these constants in equation we obtain (Mathematica solution)

\begin{align*}
<T \mbox{   }e^{ - \lambda N_{>}(t)} &\mbox{             }e^{ - \lambda^{'} N^{'}_{>}(t)}    \mbox{        } c_{ {\bf{k-q}}, > }(t) a^{\dagger}_{ {\bf{k}} - {\bf{q}}/2 }(-{\bf{q}},t_1) \mbox{             }
c^{\dagger}_{ {\bf{k}},< }(t^{'}) >  \mbox{             } =\\
&
e^{-i(t_1 -t)(\epsilon_{ {\bf{k}} } - \epsilon_{ {\bf{k}}-{\bf{q}} })} \bigg[ (e^{\lambda+\lambda^{'}} -  1)\mbox{   } <T \mbox{            } e^{ - \lambda N_{>}(t)} \mbox{             }e^{ - \lambda^{'} N^{'}_{>}(t)}  c_{ {\bf{k-q}}, > }(t)\mbox{             }
a^{\dagger}_{ {\bf{k}} - {\bf{q}}/2 }(-{\bf{q}},t) \mbox{        }
c^{\dagger}_{ {\bf{k}},< }(t^{'}) >   \mbox{             } 
\\
&
- (1-n_F({\bf{k}}-{\bf{q}}))\mbox{       }e^{\lambda+\lambda^{'}}\mbox{ } <T \mbox{            } e^{ - \lambda N_{>}(t)} \mbox{             }e^{ - \lambda^{'} N^{'}_{>}(t)} \mbox{             }
c_{ {\bf{k}}, < }(t)
c^{\dagger}_{ {\bf{k}},< }(t^{'}) >  \bigg] \mbox{             } \left( \theta(t_1 -t) - \frac{1}{1- e^{\beta(\epsilon_{ {\bf{k}} }  - \epsilon_{ {\bf{k}}-{\bf{q}} })}}      \right)
\\
&
 + n_F({\bf{k}}) e^{-i(t_1 -t')(\epsilon_{ {\bf{k}} } - \epsilon_{ {\bf{k}}-{\bf{q}} })} \mbox{  }  <T \mbox{           } e^{ - \lambda N_{>}(t)} \mbox{             }e^{ - \lambda^{'} N^{'}_{>}(t)} \mbox{             } c_{ {\bf{k-q}}, > }(t)
 c^{\dagger}_{ {\bf{k}} - {\bf{q}}, > }(t^{'}) >   \mbox{             }  \mbox{       }  \left( \theta(t_1 -t') - \frac{1}{1- e^{\beta(\epsilon_{ {\bf{k}} }  - \epsilon_{ {\bf{k}}-{\bf{q}} })}}      \right)
\end{align*}

--------------------------------------------------------------------------------------------------------------

\begin{align*}
<T \mbox{   }e^{ - \lambda N_{>}(t)} & \mbox{             }e^{ - \lambda^{'} N^{'}_{>}(t)} \mbox{             } c_{ {\bf{k-q}}, < }(t)
a_{ {\bf{k}} - {\bf{q}}/2 }({\bf{q}},t_1)  \mbox{        }
c^{\dagger}_{ {\bf{k}},> }(t^{'}) >  \mbox{             }= 
\\
&
e^{-i(t_1 -t)(\epsilon_{ {\bf{k}} } - \epsilon_{ {\bf{k}}-{\bf{q}} })} \bigg[ (e^{ - \lambda-\lambda^{'} }  -1) \mbox{       } < T\mbox{                }
 e^{ - \lambda N_{>}(t)}  \mbox{             }e^{ - \lambda^{'} N^{'}_{>}(t)} \mbox{             }
 \mbox{             } c_{ {\bf{k-q}}, < }(t) a_{ {\bf{k}} - {\bf{q}}/2 }({\bf{q}},t)
c^{\dagger}_{ {\bf{k}},> }(t^{'}) >\mbox{   }
\\
&
-  n_F({\bf{k}}-{\bf{q}})\mbox{          }e^{ -\lambda - \lambda^{'}} < T\mbox{                } e^{ - \lambda N_{>}(t)}  \mbox{             }e^{ - \lambda^{'} N^{'}_{>}(t)} \mbox{             }
c_{ {\bf{k}}, > }(t)
c^{\dagger}_{ {\bf{k}},> }(t^{'}) >\mbox{   }\bigg] \left( \theta(t_1 -t) - \frac{1}{1- e^{\beta(\epsilon_{ {\bf{k}} }  - \epsilon_{ {\bf{k}}-{\bf{q}} })}}      \right)
\\
&
+ e^{-i(t_1 -t')(\epsilon_{ {\bf{k}} } - \epsilon_{ {\bf{k}}-{\bf{q}} })} <T\mbox{          } e^{ - \lambda N_{>}(t)}  \mbox{             }e^{ - \lambda^{'} N^{'}_{>}(t)} \mbox{             } c_{ {\bf{k-q}}, < }(t)\mbox{  }
c^{\dagger}_{ {\bf{k}}-{\bf{q}},< }(t^{'})
 >\mbox{   }(1-n_F({\bf{k}}))\mbox{   } \left( \theta(t_1 -t') - \frac{1}{1- e^{\beta(\epsilon_{ {\bf{k}} }  - \epsilon_{ {\bf{k}}-{\bf{q}} })}}      \right)
\end{align*}

=====================================================

Now taking $t_1 \to t_-$ limit 

\begin{align*}
<T \mbox{   }e^{ - \lambda N_{>}(t)} &\mbox{             }e^{ - \lambda^{'} N^{'}_{>}(t)}\mbox{ } c_{ {\bf{k-q}}, > }(t) \mbox{ } a^{\dagger}_{ {\bf{k}} - {\bf{q}}/2 }(-{\bf{q}},t)
  \mbox{        }
c^{\dagger}_{ {\bf{k}},< }(t^{'}) >  \mbox{             } =
\\
&
 (1-n_F({\bf{k}}-{\bf{q}}))\mbox{       } <T \mbox{            } e^{ - \lambda N_{>}(t)} \mbox{             }e^{ - \lambda^{'} N^{'}_{>}(t)} \mbox{             }
c_{ {\bf{k}}, < }(t)
c^{\dagger}_{ {\bf{k}},< }(t^{'}) >  \mbox{             } \left( \frac{e^{ \lambda + \lambda^{'}}}{e^{ \lambda + \lambda^{'}}- e^{\beta(\epsilon_{ {\bf{k}} }  - \epsilon_{ {\bf{k}}-{\bf{q}} })}}      \right)
\\
&
 - n_F({\bf{k}}) e^{-i(t -t')(\epsilon_{ {\bf{k}} } - \epsilon_{ {\bf{k}}-{\bf{q}} })} \mbox{  }  <T \mbox{           } e^{ - \lambda N_{>}(t)} \mbox{             }e^{ - \lambda^{'} N^{'}_{>}(t)} \mbox{             } c_{ {\bf{k-q}}, > }(t)
 c^{\dagger}_{ {\bf{k}} - {\bf{q}}, > }(t^{'}) >   \mbox{             }  \mbox{       }  \left(  \frac{\theta(t -t')e^{\beta(\epsilon_{ {\bf{k}} }  - \epsilon_{ {\bf{k}}-{\bf{q}} })} + \theta(t'-t)}{e^{ \lambda + \lambda^{'}}- e^{\beta(\epsilon_{ {\bf{k}} }  - \epsilon_{ {\bf{k}}-{\bf{q}} })}}   \right)
\end{align*}
or, 
\begin{align*}
<T \mbox{   }e^{ - \lambda N_{>}(t)} &\mbox{             }e^{ - \lambda^{'} N^{'}_{>}(t)}\mbox{ } a^{\dagger}_{ {\bf{k}} - {\bf{q}}/2 }(-{\bf{q}},t) \mbox{ } c_{ {\bf{k-q}}, > }(t) \mbox{ } 
  \mbox{        }
c^{\dagger}_{ {\bf{k}},< }(t^{'}) >  \mbox{             } =
\\
&
 (1-n_F({\bf{k}}-{\bf{q}}))\mbox{       } <T \mbox{            } e^{ - \lambda N_{>}(t)} \mbox{             }e^{ - \lambda^{'} N^{'}_{>}(t)} \mbox{             }
c_{ {\bf{k}}, < }(t)
c^{\dagger}_{ {\bf{k}},< }(t^{'}) >  \mbox{             } \left( \frac{e^{\beta(\epsilon_{ {\bf{k}} }  - \epsilon_{ {\bf{k}}-{\bf{q}} })}}{e^{ \lambda + \lambda^{'}}- e^{\beta(\epsilon_{ {\bf{k}} }  - \epsilon_{ {\bf{k}}-{\bf{q}} })}}      \right)
\\
&
 - n_F({\bf{k}}) e^{-i(t -t')(\epsilon_{ {\bf{k}} } - \epsilon_{ {\bf{k}}-{\bf{q}} })} \mbox{  }  <T \mbox{           } e^{ - \lambda N_{>}(t)} \mbox{             }e^{ - \lambda^{'} N^{'}_{>}(t)} \mbox{             } c_{ {\bf{k-q}}, > }(t)
 c^{\dagger}_{ {\bf{k}} - {\bf{q}}, > }(t^{'}) >   \mbox{             }  \mbox{       }  \left(  \frac{\theta(t -t')e^{\beta(\epsilon_{ {\bf{k}} }  - \epsilon_{ {\bf{k}}-{\bf{q}} })} + \theta(t'-t)}{e^{ \lambda + \lambda^{'}}- e^{\beta(\epsilon_{ {\bf{k}} }  - \epsilon_{ {\bf{k}}-{\bf{q}} })}}   \right)
\end{align*}
--------------------------------------------------------------------------------------------------------------

\begin{align*}
<T \mbox{   }e^{ - \lambda N_{>}(t)} & \mbox{             }e^{ - \lambda^{'} N^{'}_{>}(t)} \mbox{             } c_{ {\bf{k-q}}, < }(t)
a_{ {\bf{k}} - {\bf{q}}/2 }({\bf{q}},t)  \mbox{        }
c^{\dagger}_{ {\bf{k}},> }(t^{'}) >  \mbox{             }= 
\\
&
  n_F({\bf{k}}-{\bf{q}})\mbox{          } < T\mbox{                } e^{ - \lambda N_{>}(t)}  \mbox{             }e^{ - \lambda^{'} N^{'}_{>}(t)} \mbox{             }
c_{ {\bf{k}}, > }(t)
c^{\dagger}_{ {\bf{k}},> }(t^{'}) >\mbox{   } \left( \frac{e^{ -\lambda - \lambda^{'}}}{e^{ -\lambda - \lambda^{'}}- e^{\beta(\epsilon_{ {\bf{k}} }  - \epsilon_{ {\bf{k}}-{\bf{q}} })}}      \right)
\\
&
- e^{-i(t -t')(\epsilon_{ {\bf{k}} } - \epsilon_{ {\bf{k}}-{\bf{q}} })} <T\mbox{          } e^{ - \lambda N_{>}(t)}  \mbox{             }e^{ - \lambda^{'} N^{'}_{>}(t)} \mbox{             } c_{ {\bf{k-q}}, < }(t)\mbox{  }
c^{\dagger}_{ {\bf{k}}-{\bf{q}},< }(t^{'})
 >\mbox{   }(1-n_F({\bf{k}}))\mbox{   } \left(  \frac{\theta(t -t')e^{\beta(\epsilon_{ {\bf{k}} }  - \epsilon_{ {\bf{k}}-{\bf{q}} })} + \theta(t'-t)}{e^{- \lambda - \lambda^{'}}- e^{\beta(\epsilon_{ {\bf{k}} }  - \epsilon_{ {\bf{k}}-{\bf{q}} })}}   \right)
\end{align*}

=========================================================

LHS can be written as 

\begin{align*}
\sum_q <T \mbox{   }e^{ - \lambda N_{>}(t)} \mbox{             }e^{ - \lambda^{'} N^{'}_{>}(t)} \mbox{           }a^{\dagger}_{ {\bf{k}} - {\bf{q}}/2 }(-{\bf{q}},t)  \mbox{        } c_{ {\bf{k-q}}, > }(t)
&c^{\dagger}_{ {\bf{k}},< }(t^{'}) >  \mbox{             } \\
=\sum_q <T\mbox{   } &e^{ - \lambda N_{>}(t)} \mbox{             }e^{ - \lambda^{'} N^{'}_{>}(t)} \mbox{             } c^{\dagger}_{ {\bf{k-q}}, > }(t)  \mbox{        } c_{ {\bf{k}},< }(t^{})c_{ {\bf{k-q}}, > }(t)
c^{\dagger}_{ {\bf{k}},< }(t^{'}) > 
\\
= - \sum_q <T& \mbox{   }e^{ - \lambda N_{>}(t)} \mbox{             }e^{ - \lambda^{'} N^{'}_{>}(t)} \mbox{             } c^{\dagger}_{ {\bf{k-q}}, > }(t) c_{ {\bf{k-q}}, > }(t) \mbox{        } c_{ {\bf{k}},< }(t^{})
c^{\dagger}_{ {\bf{k}},< }(t^{'}) > \\
\end{align*}

---------------------------------------------------------
\begin{align*}
\sum_q <T \mbox{   }e^{ - \lambda N_{>}(t)}  \mbox{             }e^{ - \lambda^{'} N^{'}_{>}(t)} \mbox{             } c_{ {\bf{k-q}}, < }(t)
a_{ {\bf{k}} - {\bf{q}}/2 }({\bf{q}},t)  \mbox{        }
&c^{\dagger}_{ {\bf{k}},> }(t^{'}) >  \mbox{             } \\
=&\sum_q <T \mbox{   }e^{ - \lambda N_{>}(t)}  \mbox{             }e^{ - \lambda^{'} N^{'}_{>}(t)} \mbox{             } c_{ {\bf{k-q}}, < }(t)
c^{\dagger}_{ {\bf{k-q}}, < }(t)  \mbox{        }
c_{ {\bf{k}},> }(t)c^{\dagger}_{ {\bf{k}},> }(t^{'}) > 
\end{align*}

===================================================================

Taking sum over ${\bf q}$ and replacing ${\bf k -q} \to {\bf p}$, we obtain

\begin{align*}
- \partial_{ \lambda' } <T \mbox{   }e^{ - \lambda N_{>}(t)} &\mbox{             }e^{ - \lambda^{'} N^{'}_{>}(t)} \mbox{             }  \mbox{        } c_{ {\bf{k}},< }(t^{})
c^{\dagger}_{ {\bf{k}},< }(t^{'}) > =
\\
&
- \sum_p (1-n_F({\bf p}))\mbox{       } <T \mbox{            } e^{ - \lambda N_{>}(t)} \mbox{             }e^{ - \lambda^{'} N^{'}_{>}(t)} \mbox{             }
c_{ {\bf{k}}, < }(t)
c^{\dagger}_{ {\bf{k}},< }(t^{'}) >  \mbox{             } \left( \frac{1}{e^{ \lambda + \lambda^{'}}e^{\beta(\epsilon_{ {\bf{p}} }  - \epsilon_{ \bf k })}- 1}      \right)
\\
&
+ \sum_p  n_F({\bf{k}}) e^{i(t -t')(\epsilon_{ {\bf{p}} } - \epsilon_{ \bf k })} \mbox{  }  <T \mbox{           } e^{ - \lambda N_{>}(t)} \mbox{             }e^{ - \lambda^{'} N^{'}_{>}(t)} \mbox{             } c_{ {\bf{p}}, > }(t)
 c^{\dagger}_{ {\bf p}, > }(t^{'}) >   \mbox{             }  \mbox{       }  \left(  \frac{\theta(t -t') + \theta(t'-t)e^{\beta(\epsilon_{ {\bf{p}} }  - \epsilon_{ \bf k })}}{e^{ \lambda + \lambda^{'}}e^{\beta(\epsilon_{ {\bf{p}} }  - \epsilon_{ \bf k })}- 1}   \right)
\end{align*}

--------------------------------------------------------------------------------------------------------------

\begin{align*}
-\partial_{ \lambda  } <T \mbox{   } &e^{ - \lambda N_{>}(t)}  \mbox{             }e^{ - \lambda^{'} N^{'}_{>}(t)} \mbox{             } 
c_{ {\bf{k}},> }(t)c^{\dagger}_{ {\bf{k}},> }(t^{'}) > 
\\
&
= - \sum_p  n_F({\bf p})\mbox{          } < T\mbox{                } e^{ - \lambda N_{>}(t)}  \mbox{             }e^{ - \lambda^{'} N^{'}_{>}(t)} \mbox{             }
c_{ {\bf{k}}, > }(t)
c^{\dagger}_{ {\bf{k}},> }(t^{'}) >\mbox{   } \left( \frac{1}{e^{ \lambda + \lambda^{'}}e^{\beta(\epsilon_{ {\bf{k}} }  - \epsilon_{ \bf p })}- 1}      \right)
\\
&
+ \sum_p  e^{-i(t -t')(\epsilon_{ {\bf{k}} } - \epsilon_{ \bf p })} e^{ \lambda + \lambda^{'}} <T\mbox{          } e^{ - \lambda N_{>}(t)}  \mbox{             }e^{ - \lambda^{'} N^{'}_{>}(t)} \mbox{             } c_{ {\bf{p}}, < }(t)\mbox{  }
c^{\dagger}_{ {\bf p},< }(t^{'})
 >\mbox{   }(1-n_F({\bf{k}}))\mbox{   } \left(  \frac{\theta(t -t')e^{\beta(\epsilon_{ {\bf{k}} }  - \epsilon_{ \bf p})} + \theta(t'-t)}{e^{ \lambda + \lambda^{'}}e^{\beta(\epsilon_{ {\bf{k}} }  - \epsilon_{ \bf p })}- 1}   \right)
\end{align*}

==================================================================================
\begin{align*}
-\partial_{ \lambda^{'} } \mbox{             }<T \mbox{   }e^{ - \lambda N_{>}(t)}  \mbox{             }e^{ - \lambda^{'} N^{'}_{>}(t)}
\mbox{        }   &c_{ {\bf{k}}, < }(t)
c^{\dagger}_{ {\bf{k}},< }(t^{'}) >
\mbox{             }= \mbox{             }
\\
&
  \sum_{ {\bf{p}} }\mbox{          }
  \frac{1 }{ e^{ \beta (\epsilon_{ {\bf{p}} }-\epsilon_{ {\bf{k}} }) }e^{\lambda + \lambda^{'} }-1 }
\bigg[ e^{ i  ( t-t^{'}) (\epsilon_{ {\bf{p}} }-\epsilon_{ {\bf{k}} } )  }
\mbox{              }<T\mbox{   }e^{ - \lambda N_{>}(t)}  \mbox{             }e^{ - \lambda^{'} N^{'}_{>}(t)} \mbox{             } c_{ {\bf{p}}, > }(t)
 c^{\dagger}_{ {\bf{p}}, > }(t^{'}) > \mbox{             }  n_F({\bf{k}})
\\
&\quad \times \left(  \theta (t-t^{'})+e^{ \beta (\epsilon_{ {\bf{p}} }-\epsilon_{ {\bf{k}} }) }\mbox{  } \theta (t^{'}-t)\right) - <T \mbox{   }e^{ - \lambda N_{>}(t)}  \mbox{             }e^{ - \lambda^{'} N^{'}_{>}(t)}
\mbox{        }  c_{ {\bf{k}}, < }(t)
c^{\dagger}_{ {\bf{k}},< }(t^{'}) >\mbox{             }
(1-  n_F( {\bf{p}} ) )\bigg]
\end{align*}

and

\begin{align*}
-\partial_{ \lambda } \mbox{             }<T  \mbox{   }e^{ - \lambda N_{>}(t)}  \mbox{             }&e^{ - \lambda^{'} N^{'}_{>}(t)} \mbox{             }  c_{ {\bf{k}}, > }(t)
c^{\dagger}_{ {\bf{k}},> }(t^{'}) >  \mbox{             }= \mbox{             }
\\
  &\sum_{ {\bf{p}} }\mbox{          }
 \frac{1}{e^{ \beta  (\epsilon_{ {\bf{k}} }-\epsilon_{ {\bf{p}} })  }e^{\lambda + \lambda^{'} }-1}
\bigg[e^{ \lambda+\lambda^{'} }\mbox{             } e^{ i   (t-t^{'}) (\epsilon_{ {\bf{p}} } - \epsilon_{ {\bf{k}} }) }
 \mbox{          }
  <T \mbox{   }e^{ - \lambda N_{>}(t)}  \mbox{             }e^{ - \lambda^{'} N^{'}_{>}(t)} \mbox{             } c_{ {\bf{p}}, < }(t)
 c^{\dagger}_{ {\bf{p}}  , < }(t^{'}) >   \mbox{             }(1- n_F({\bf{k}}) )\mbox{  }
 \\
 &\quad \times  \left(   \theta (t^{'}-t)+\theta (t-t^{'})e^{ \beta (\epsilon_{ {\bf{k}} }-\epsilon_{ {\bf{p}} }) } \right)
 -  <T \mbox{   }e^{ - \lambda N_{>}(t)}  \mbox{             }e^{ - \lambda^{'} N^{'}_{>}(t)} \mbox{             }
c_{ {\bf{k}}, > }(t)
c^{\dagger}_{ {\bf{k}},> }(t^{'}) >   \mbox{             } n_F({\bf{p}})\bigg]
\end{align*}
\\ \mbox{               } \\

\section{ Time evolution }
We may see from the above equations that the time evolution of the operators is a simple exponential.
\[
c_{ {\bf{p}} }(t) = {\tilde{c}}_{ {\bf{p}} }(t) \mbox{           } e^{ - i \epsilon_{ {\bf{p}} } \mbox{  } t }
\]
The piece-wise constant reduced correlation functions obey these equations,

\begin{align*}
-\partial_{ \lambda } \mbox{             }<T  \mbox{   }e^{ - \lambda N_{>}(t)}  &\mbox{             }e^{ - \lambda^{'} N^{'}_{>}(t)} \mbox{             } {\tilde{c}}_{ {\bf{k}}, > }(t)
{\tilde{c}}^{\dagger}_{ {\bf{k}},> }(t^{'}) >   \mbox{             }= \mbox{             }
\\
&
 \sum_{ {\bf{p}} }\mbox{          }
 \frac{1 }{e^{ \beta  (\epsilon_{ {\bf{k}} }-\epsilon_{ {\bf{p}} })  }e^{\lambda + \lambda^{'} }-1}
\bigg[ e^{ \lambda+\lambda^{'} }\mbox{             }
  <T \mbox{   }e^{ - \lambda N_{>}(t)}  \mbox{             }e^{ - \lambda^{'} N^{'}_{>}(t)} \mbox{             } {\tilde{c}}_{ {\bf{p}}, < }(t)
 {\tilde{c}}^{\dagger}_{ {\bf{p}}  , < }(t^{'}) >   \mbox{             }(1- n_F({\bf{k}}) )\mbox{  }
 \\
 & \quad \times   \left(   \theta (t^{'}-t)+\theta (t-t^{'})e^{ \beta (\epsilon_{ {\bf{k}} }-\epsilon_{ {\bf{p}} }) } \right)-  <T \mbox{   }e^{ - \lambda N_{>}(t)}  \mbox{             }e^{ - \lambda^{'} N^{'}_{>}(t)} \mbox{             }
{\tilde{c}}_{ {\bf{k}}, > }(t)
{\tilde{c}}^{\dagger}_{ {\bf{k}},> }(t^{'}) >   \mbox{             } n_F({\bf{p}})\bigg]
\end{align*}

and

\begin{align*}
-\partial_{ \lambda^{'} } \mbox{             }<T \mbox{   }e^{ - \lambda N_{>}(t)} & \mbox{             }e^{ - \lambda^{'} N^{'}_{>}(t)}
\mbox{        }  {\tilde{ c}}_{ {\bf{k}}, < }(t)
{\tilde{c}}^{\dagger}_{ {\bf{k}},< }(t^{'}) >
\mbox{             }= \mbox{             }
\\
&
  \sum_{ {\bf{p}} }\mbox{          }
  \frac{ 1}{ e^{ \beta (\epsilon_{ {\bf{p}} }-\epsilon_{ {\bf{k}} }) }e^{\lambda + \lambda^{'} }-1 }
 \bigg[<T\mbox{   }e^{ - \lambda N_{>}(t)}  \mbox{             }e^{ - \lambda^{'} N^{'}_{>}(t)} \mbox{             } {\tilde{c}}_{ {\bf{p}}, > }(t)
 {\tilde{c}}^{\dagger}_{ {\bf{p}}, > }(t^{'}) > \mbox{             }  n_F({\bf{k}})
 \\
& \quad \times \left(  \theta (t-t^{'})+e^{ \beta (\epsilon_{ {\bf{p}} }-\epsilon_{ {\bf{k}} }) }\mbox{  } \theta (t^{'}-t)\right) - <T \mbox{   }e^{ - \lambda N_{>}(t)}  \mbox{             }e^{ - \lambda^{'} N^{'}_{>}(t)}
\mbox{        }  {\tilde{c}}_{ {\bf{k}}, < }(t)
{\tilde{c}}^{\dagger}_{ {\bf{k}},< }(t^{'}) >\mbox{             }
(1-  n_F( {\bf{p}} ) )\bigg]
\end{align*}
\\ \mbox{          } \\

\section{ Separability condition }

\begin{align*}
e^{ \lambda+\lambda^{'} }\mbox{             }
  <T \mbox{   }e^{ - \lambda N_{>}(t)}  \mbox{             }e^{ - \lambda^{'} N^{'}_{>}(t)} &\mbox{             } {\tilde{c}}_{ {\bf{p}}, < }(t)
 {\tilde{c}}^{\dagger}_{ {\bf{p}}  , < }(t^{'}) >   \mbox{             }(1- n_F({\bf{k}}) )\mbox{  }
   \left(   \theta (t^{'}-t)+\theta (t-t^{'})e^{ \beta (\epsilon_{ {\bf{k}} }-\epsilon_{ {\bf{p}} }) } \right)
   \\
   -  <T \mbox{   }e^{ - \lambda N_{>}(t)}  \mbox{             }&e^{ - \lambda^{'} N^{'}_{>}(t)} \mbox{             }
{\tilde{c}}_{ {\bf{k}}, > }(t)
{\tilde{c}}^{\dagger}_{ {\bf{k}},> }(t^{'}) >   \mbox{             } n_F({\bf{p}}) \mbox{             }= \mbox{             }(e^{ \beta  (\epsilon_{ {\bf{k}} }-\epsilon_{ {\bf{p}} })  }e^{\lambda + \lambda^{'} }-1)\mbox{               }
L_{>}({\bf{k}}; \lambda, \lambda^{'};t-t^{'}) R_{<}({\bf{p}}; \lambda, \lambda^{'};t-t^{'})
\end{align*}

and

\begin{align*}
 <T\mbox{   }e^{ - \lambda N_{>}(t)}  \mbox{             }e^{ - \lambda^{'} N^{'}_{>}(t)} &\mbox{             } {\tilde{c}}_{ {\bf{p}}, > }(t)
 {\tilde{c}}^{\dagger}_{ {\bf{p}}, > }(t^{'}) > \mbox{             }  n_F({\bf{k}})
\left(  \theta (t-t^{'})+e^{ \beta (\epsilon_{ {\bf{p}} }-\epsilon_{ {\bf{k}} }) }\mbox{  } \theta (t^{'}-t)\right)
\\
 - <T \mbox{   }e^{ - \lambda N_{>}(t)}  \mbox{             }&e^{ - \lambda^{'} N^{'}_{>}(t)}
\mbox{        } {\tilde{ c}}_{ {\bf{k}}, < }(t)
{\tilde{c}}^{\dagger}_{ {\bf{k}},< }(t^{'}) >\mbox{             }
(1-  n_F( {\bf{p}} ) )
\mbox{             }= \mbox{             }
( e^{ \beta (\epsilon_{ {\bf{p}} }-\epsilon_{ {\bf{k}} }) }e^{\lambda + \lambda^{'} }-1 )
\mbox{               }
L_{<}({\bf{k}}; \lambda, \lambda^{'};t-t^{'}) R_{>}({\bf{p}}; \lambda, \lambda^{'};t-t^{'})
\end{align*}
============================================================

\section{ Compact equations to be solved }

\[
-\partial_{ \lambda } \mbox{             }<T  \mbox{   }e^{ - \lambda N_{>}(t)}  \mbox{             }e^{ - \lambda^{'} N^{'}_{>}(t)} \mbox{             } {\tilde{c}}_{ {\bf{k}}, > }(t)
{\tilde{c}}^{\dagger}_{ {\bf{k}},> }(t^{'}) >   \mbox{             }= \mbox{             } L_{>}({\bf{k}}; \lambda, \lambda^{'};t-t^{'}) \sum_{ {\bf{p}} }\mbox{          }
R_{<}({\bf{p}}; \lambda, \lambda^{'};t-t^{'})
\]

and

\[
-\partial_{ \lambda^{'} } \mbox{             }<T \mbox{   }e^{ - \lambda N_{>}(t)}  \mbox{             }e^{ - \lambda^{'} N^{'}_{>}(t)}
\mbox{        }   {\tilde{c}}_{ {\bf{k}}, < }(t)
{\tilde{c}}^{\dagger}_{ {\bf{k}},< }(t^{'}) >
\mbox{             }= \mbox{             }L_{<}({\bf{k}}; \lambda, \lambda^{'};t-t^{'})  \sum_{ {\bf{p}} }\mbox{          } R_{>}({\bf{p}}; \lambda, \lambda^{'};t-t^{'})
\]

\section{ Green function in terms of the separated functions }

\begin{align*}
e^{ \lambda+\lambda^{'} }\mbox{             }
  <T \mbox{   }e^{ - \lambda N_{>}(t)}  \mbox{             }e^{ - \lambda^{'} N^{'}_{>}(t)} &\mbox{             } {\tilde{c}}_{ {\bf{p}}, < }(t)
 {\tilde{c}}^{\dagger}_{ {\bf{p}}  , < }(t^{'}) >   \mbox{             }(1- n_F({\bf{k}}) )\mbox{  }
   \left(   \theta (t^{'}-t)+\theta (t-t^{'})e^{ \beta (\epsilon_{ {\bf{k}} }-\epsilon_{ {\bf{p}} }) } \right)
   \\
   -  <T \mbox{   }e^{ - \lambda N_{>}(t)}  \mbox{             }&e^{ - \lambda^{'} N^{'}_{>}(t)} \mbox{             }
{\tilde{c}}_{ {\bf{k}}, > }(t)
{\tilde{c}}^{\dagger}_{ {\bf{k}},> }(t^{'}) >   \mbox{             } n_F({\bf{p}}) \mbox{             }= \mbox{             }
(e^{ \beta  (\epsilon_{ {\bf{k}} }-\epsilon_{ {\bf{p}} })  }e^{\lambda + \lambda^{'} }-1)\mbox{               }
L_{>}({\bf{k}}; \lambda, \lambda^{'};t-t^{'}) R_{<}({\bf{p}}; \lambda, \lambda^{'};t-t^{'})
\end{align*}

and

\begin{align*}
 <T\mbox{   }e^{ - \lambda N_{>}(t)}  \mbox{             }e^{ - \lambda^{'} N^{'}_{>}(t)}& \mbox{             } {\tilde{c}}_{ {\bf{k}}, > }(t)
 {\tilde{c}}^{\dagger}_{ {\bf{k}}, > }(t^{'}) > \mbox{             }  n_F({\bf{p}})
\left(  \theta (t-t^{'})+e^{ \beta (\epsilon_{ {\bf{k}} }-\epsilon_{ {\bf{p}} }) }\mbox{  } \theta (t^{'}-t)\right)
\\
- <T \mbox{   }e^{ - \lambda N_{>}(t)}  \mbox{             }&e^{ - \lambda^{'} N^{'}_{>}(t)}
\mbox{        } {\tilde{ c}}_{ {\bf{p}}, < }(t)
{\tilde{c}}^{\dagger}_{ {\bf{p}},< }(t^{'}) >\mbox{             }
(1-  n_F( {\bf{k}} ) )
\mbox{             }= \mbox{             }
( e^{ \beta (\epsilon_{ {\bf{k}} }-\epsilon_{ {\bf{p}} }) }e^{\lambda + \lambda^{'} }-1 )
\mbox{               }
L_{<}({\bf{p}}; \lambda, \lambda^{'};t-t^{'}) R_{>}({\bf{k}}; \lambda, \lambda^{'};t-t^{'})
\end{align*}

=============================================================

\begin{align*}
<T \mbox{   }e^{ - \lambda N_{>}(t)}  &\mbox{             }e^{ - \lambda^{'} N^{'}_{>}(t)}
\mbox{     } {\tilde{ c}}_{ {\bf{p}}, < }(t)
{\tilde{c}}^{\dagger}_{ {\bf{p}},< }(t^{'}) >\mbox{      } (1-n_F({\bf{k}}))\mbox{            } =
\\
& \mbox{      }
L_{<}({\bf{p}}; \lambda, \lambda^{'};t-t^{'}) R_{>}({\bf{k}}; \lambda, \lambda^{'};t-t^{'})
+L_{>}({\bf{k}}; \lambda, \lambda^{'};t-t^{'}) R_{<}({\bf{p}}; \lambda, \lambda^{'};t-t^{'})
\mbox{      } (\theta (t^{'}-t) e^{ \beta (\epsilon_{ {\bf{k}} }- \epsilon_{ {\bf{p}} })  }+\theta (t-t^{'}) )
\end{align*}

and

\begin{align*}
 <T \mbox{   }e^{ - \lambda N_{>}(t)} & \mbox{             }e^{ - \lambda^{'} N^{'}_{>}(t)} \mbox{             }
{\tilde{c}}_{ {\bf{k}}, > }(t)
{\tilde{c}}^{\dagger}_{ {\bf{k}},> }(t^{'}) > \mbox{        } n_F({\bf{p}}) \mbox{            } =
\\
& \mbox{                }  L_{>}({\bf{k}}; \lambda, \lambda^{'};t-t^{'}) R_{<}({\bf{p}}; \lambda, \lambda^{'};t-t^{'}) +L_{<}({\bf{p}}; \lambda, \lambda^{'};t-t^{'}) R_{>}({\bf{k}}; \lambda, \lambda^{'};t-t^{'})
\mbox{           } (e^{\beta  (\epsilon_{ {\bf{k}} }-\epsilon_{ {\bf{p}} })} \theta (t-t^{'})+\theta (t^{'}-t) )\mbox{          }
 e^{\lambda + \lambda^{'} }
\end{align*}

OR,

\begin{align*}
<T \mbox{   }e^{ - \lambda N_{>}(t)}  \mbox{             }e^{ - \lambda^{'} N^{'}_{>}(t)}
\mbox{        } &{\tilde{ c}}_{ {\bf{p}}, < }(t)
{\tilde{c}}^{\dagger}_{ {\bf{p}},< }(t^{'}) >\mbox{      } (1-n_F({\bf{k}}))\mbox{            } =
\\
& \mbox{                }
L_{<}({\bf{p}}; \lambda, \lambda^{'};t-t^{'}) R_{>}({\bf{k}}; \lambda, \lambda^{'};t-t^{'})
+L_{>}({\bf{k}}; \lambda, \lambda^{'};t-t^{'}) R_{<}({\bf{p}}; \lambda, \lambda^{'};t-t^{'})
\mbox{             } e^{ \theta(t^{'}-t) \beta (\epsilon_{ {\bf{k}} }- \epsilon_{ {\bf{p}} })  }
\end{align*}

and

\begin{align*}
 <T \mbox{   }e^{ - \lambda N_{>}(t)}  \mbox{             }e^{ - \lambda^{'} N^{'}_{>}(t)} \mbox{             }
&{\tilde{c}}_{ {\bf{k}}, > }(t)
{\tilde{c}}^{\dagger}_{ {\bf{k}},> }(t^{'}) > \mbox{  } n_F({\bf{p}}) \mbox{            } =
\\
& \mbox{                }  L_{>}({\bf{k}}; \lambda, \lambda^{'};t-t^{'}) R_{<}({\bf{p}}; \lambda, \lambda^{'};t-t^{'}) +L_{<}({\bf{p}}; \lambda, \lambda^{'};t-t^{'}) R_{>}({\bf{k}}; \lambda, \lambda^{'};t-t^{'})
\mbox{           } e^{\theta (t-t^{'}) \beta  (\epsilon_{ {\bf{k}} }-\epsilon_{ {\bf{p}} })}  \mbox{          }
 e^{\lambda + \lambda^{'} }
\end{align*}

\section{ Reduction }

\begin{align*}
<T \mbox{   }e^{ - \lambda N_{>}(t)}  &\mbox{             }e^{ - \lambda^{'} N^{'}_{>}(t)}
\mbox{        } {\tilde{ c}}_{ {\bf{p}}, < }(t)
{\tilde{c}}^{\dagger}_{ {\bf{p}},< }(t^{'}) >\mbox{   } (1-n_F({\bf{k}}))\mbox{            } = 
\\
&\mbox{                }
L_{<}({\bf{p}}; \lambda, \lambda^{'};t-t^{'}) R_{>}({\bf{k}}; \lambda, \lambda^{'};t-t^{'})
+L_{>}({\bf{k}}; \lambda, \lambda^{'};t-t^{'})
\mbox{             } e^{ \theta(t^{'}-t) \beta \epsilon_{ {\bf{k}} }  } \mbox{             }  R_{<}({\bf{p}}; \lambda, \lambda^{'};t-t^{'})
\mbox{             } e^{ -\theta(t^{'}-t) \beta  \epsilon_{ {\bf{p}} } }
\end{align*}
and

\begin{align*}
 <T \mbox{   }e^{ - \lambda N_{>}(t)}&  \mbox{             }e^{ - \lambda^{'} N^{'}_{>}(t)} \mbox{             }
{\tilde{c}}_{ {\bf{k}}, > }(t)
{\tilde{c}}^{\dagger}_{ {\bf{k}},> }(t^{'}) > \mbox{         } n_F({\bf{p}}) \mbox{            } = 
\\
&\mbox{                }  L_{>}({\bf{k}}; \lambda, \lambda^{'};t-t^{'}) R_{<}({\bf{p}}; \lambda, \lambda^{'};t-t^{'}) +L_{<}({\bf{p}}; \lambda, \lambda^{'};t-t^{'})\mbox{           }
 e^{-\theta (t-t^{'}) \beta  \epsilon_{ {\bf{p}} } } R_{>}({\bf{k}}; \lambda, \lambda^{'};t-t^{'})
\mbox{           } e^{\theta (t-t^{'}) \beta  \epsilon_{ {\bf{k}} } }  \mbox{          }
 e^{\lambda + \lambda^{'} }
\end{align*}

===========================================================================
Set,

\[
 c_{<}(\lambda,\lambda^{'};t-t^{'})\mbox{                  } L_{<}({\bf{p}}; \lambda, \lambda^{'};t-t^{'}) \mbox{                 } = \mbox{                  }  R_{<}({\bf{p}}; \lambda, \lambda^{'};t-t^{'})
\mbox{             } e^{ -\theta(t^{'}-t) \beta  \epsilon_{ {\bf{p}} } }
\]

\[
c_{>}(\lambda,\lambda^{'};t-t^{'}) \mbox{          } L_{>}({\bf{k}}; \lambda, \lambda^{'};t-t^{'})  \mbox{                 } = \mbox{                  } R_{>}({\bf{k}}; \lambda, \lambda^{'};t-t^{'})
\mbox{           } e^{\theta (t-t^{'}) \beta  \epsilon_{ {\bf{k}} } }
\]

=============================================================================

\begin{align*}
<T \mbox{   }e^{ - \lambda N_{>}(t)}  \mbox{             }e^{ - \lambda^{'} N^{'}_{>}(t)}
\mbox{        } {\tilde{ c}}_{ {\bf{p}}, < }(t)
{\tilde{c}}^{\dagger}_{ {\bf{p}},< }(t^{'})& >\mbox{      } (1-n_F({\bf{k}}))\mbox{            } = 
\\
& \mbox{                }
L_{<}({\bf{p}}; \lambda, \lambda^{'};t-t^{'}) \mbox{             }R_{>}({\bf{k}}; \lambda, \lambda^{'};t-t^{'})   ( 1
+   e^{  \beta  \epsilon_{ {\bf{k}} } }   \mbox{             } c^{-1}_{>}(\lambda,\lambda^{'};t-t^{'}) \mbox{          }c_{<}(\lambda,\lambda^{'};t-t^{'}))
\end{align*}

and

\begin{align*}
 <T \mbox{   }e^{ - \lambda N_{>}(t)}  \mbox{             }e^{ - \lambda^{'} N^{'}_{>}(t)} \mbox{             }
{\tilde{c}}_{ {\bf{k}}, > }(t)
{\tilde{c}}^{\dagger}_{ {\bf{k}},> }&(t^{'}) >\mbox{   }n_F({\bf{p}})  \mbox{            } =
\\
& \mbox{                }  L_{>}({\bf{k}}; \lambda, \lambda^{'};t-t^{'})
R_{<}({\bf{p}}; \lambda, \lambda^{'};t-t^{'}) \mbox{ }( 1+
 e^{-  \beta  \epsilon_{ {\bf{p}} } }  c^{-1}_{<}(\lambda,\lambda^{'};t-t^{'})  c_{>}(\lambda,\lambda^{'};t-t^{'}) \mbox{          }
 e^{\lambda + \lambda^{'} })
\end{align*}

From this we may conclude,
\[
R_{<}({\bf{p}}; \lambda, \lambda^{'};t-t^{'})\mbox{        }  =  \mbox{     } g_{<}(\lambda,\lambda^{'};t-t^{'})\mbox{      } n_F({\bf{p}})\mbox{      }
\left( 1+
 e^{-  \beta  \epsilon_{ {\bf{p}} } }  c^{-1}_{<}(\lambda,\lambda^{'};t-t^{'})  c_{>}(\lambda,\lambda^{'};t-t^{'}) \mbox{          }
 e^{\lambda + \lambda^{'} } \right)^{-1}
\]
and
\[
R_{>}({\bf{k}}; \lambda, \lambda^{'};t-t^{'}) \mbox{             }=\mbox{     } g_{>}(\lambda,\lambda^{'};t-t^{'})\mbox{      } (1-n_F({\bf{k}}))\mbox{         }   \left( 1
+   e^{  \beta  \epsilon_{ {\bf{k}} } }   \mbox{             } c^{-1}_{>}(\lambda,\lambda^{'};t-t^{'}) \mbox{          }c_{<}(\lambda,\lambda^{'};t-t^{'})
\right)^{-1}
\]
and
\[
  L_{<}({\bf{p}}; \lambda, \lambda^{'};t-t^{'}) \mbox{                 } = \mbox{                  }c^{-1}_{<}(\lambda,\lambda^{'};t-t^{'}) \mbox{     } g_{<}(\lambda,\lambda^{'};t-t^{'})\mbox{      } n_F({\bf{p}})\mbox{      }
\left( 1+
 e^{-  \beta  \epsilon_{ {\bf{p}} } }  c^{-1}_{<}(\lambda,\lambda^{'};t-t^{'})  c_{>}(\lambda,\lambda^{'};t-t^{'}) \mbox{          }
 e^{\lambda + \lambda^{'} } \right)^{-1}
\mbox{             } e^{ -\theta(t^{'}-t) \beta  \epsilon_{ {\bf{p}} } }
\]
and
\[
 L_{>}({\bf{k}}; \lambda, \lambda^{'};t-t^{'})  \mbox{                 } = \mbox{                  } c^{-1}_{>}(\lambda,\lambda^{'};t-t^{'}) \mbox{          }\mbox{     } g_{>}(\lambda,\lambda^{'};t-t^{'})\mbox{      } (1-n_F({\bf{k}}))\mbox{         }   \left( 1
+   e^{  \beta  \epsilon_{ {\bf{k}} } }   \mbox{             } c^{-1}_{>}(\lambda,\lambda^{'};t-t^{'}) \mbox{          }c_{<}(\lambda,\lambda^{'};t-t^{'})
\right)^{-1}
\mbox{           } e^{\theta (t-t^{'}) \beta  \epsilon_{ {\bf{k}} } }
\]
or,

\begin{align*}
<T \mbox{   }e^{ - \lambda N_{>}(t)}  \mbox{             }e^{ - \lambda^{'} N^{'}_{>}(t)}
\mbox{        } {\tilde{ c}}_{ {\bf{p}}, < }(t)
{\tilde{c}}^{\dagger}_{ {\bf{p}},< }(t^{'}) >\mbox{            } = \mbox{                } c^{-1}_{<}&(\lambda,\lambda^{'};t-t^{'}) \mbox{     } g_{<}(\lambda,\lambda^{'};t-t^{'})  g_{>}(\lambda,\lambda^{'};t-t^{'})\mbox{      }  n_F({\bf{p}})\mbox{      }
\\
& \times \left( 1+
 e^{-  \beta  \epsilon_{ {\bf{p}} } }  c^{-1}_{<}(\lambda,\lambda^{'};t-t^{'})  c_{>}(\lambda,\lambda^{'};t-t^{'}) \mbox{          }
 e^{\lambda + \lambda^{'} } \right)^{-1}
\mbox{             } e^{ -\theta(t^{'}-t) \beta  \epsilon_{ {\bf{p}} } }
\end{align*}

and

\begin{align*}
 <T \mbox{   }e^{ - \lambda N_{>}(t)}  \mbox{             }e^{ - \lambda^{'} N^{'}_{>}(t)} \mbox{             }
{\tilde{c}}_{ {\bf{k}}, > }(t)
{\tilde{c}}^{\dagger}_{ {\bf{k}},> }(t^{'}) >   \mbox{            } = \mbox{                }
c^{-1}_{>}&(\lambda,\lambda^{'};t-t^{'}) \mbox{          }\mbox{     } g_{>}(\lambda,\lambda^{'};t-t^{'})
 g_{<}(\lambda,\lambda^{'};t-t^{'}) \mbox{      } (1-n_F({\bf{k}}))\mbox{         }  
 \\
 &\times \left( 1
+   e^{  \beta  \epsilon_{ {\bf{k}} } }   \mbox{             } c^{-1}_{>}(\lambda,\lambda^{'};t-t^{'}) \mbox{          }c_{<}(\lambda,\lambda^{'};t-t^{'})
\right)^{-1}
\mbox{           } e^{\theta (t-t^{'}) \beta  \epsilon_{ {\bf{k}} } }
\end{align*}

\newpage
\section{ Finding the remaining unknowns }
\[
-\partial_{ \lambda } \mbox{             }<T  \mbox{   }e^{ - \lambda N_{>}(t)}  \mbox{             }e^{ - \lambda^{'} N^{'}_{>}(t)} \mbox{             } {\tilde{c}}_{ {\bf{k}}, > }(t)
{\tilde{c}}^{\dagger}_{ {\bf{k}},> }(t^{'}) >   \mbox{             }= \mbox{             } L_{>}({\bf{k}}; \lambda, \lambda^{'};t-t^{'}) \sum_{ {\bf{p}} }\mbox{          }
R_{<}({\bf{p}}; \lambda, \lambda^{'};t-t^{'})
\]
and
\[
-\partial_{ \lambda^{'} } \mbox{             }<T \mbox{   }e^{ - \lambda N_{>}(t)}  \mbox{             }e^{ - \lambda^{'} N^{'}_{>}(t)}
\mbox{        }   {\tilde{c}}_{ {\bf{p}}, < }(t)
{\tilde{c}}^{\dagger}_{ {\bf{p}},< }(t^{'}) >
\mbox{             }= \mbox{             }L_{<}({\bf{p}}; \lambda, \lambda^{'};t-t^{'})  \sum_{ {\bf{k}} }\mbox{          } R_{>}({\bf{k}}; \lambda, \lambda^{'};t-t^{'})
\]

====================================================================================

\begin{align*}
-\partial_{ \lambda } \mbox{             }
c^{-1}_{>}(\lambda,\lambda^{'};t-t^{'}) \mbox{          }\mbox{     } g_{>}(\lambda,\lambda^{'};&t-t^{'})
 g_{<}(\lambda,\lambda^{'};t-t^{'}) \mbox{      } (1-n_F({\bf{k}}))\mbox{         }   \left( 1
+   e^{  \beta  \epsilon_{ {\bf{k}} } }   \mbox{             } c^{-1}_{>}(\lambda,\lambda^{'};t-t^{'}) \mbox{          }c_{<}(\lambda,\lambda^{'};t-t^{'})
\right)^{-1}
\mbox{           } e^{\theta (t-t^{'}) \beta  \epsilon_{ {\bf{k}} } }
\\
\mbox{             }= \mbox{             }  c^{-1}_{>}&(\lambda,\lambda^{'};t-t^{'}) \mbox{          }\mbox{     } g_{>}(\lambda,\lambda^{'};t-t^{'})\mbox{      } (1-n_F({\bf{k}}))\mbox{         }   \left( 1
+   e^{  \beta  \epsilon_{ {\bf{k}} } }   \mbox{             } c^{-1}_{>}(\lambda,\lambda^{'};t-t^{'}) \mbox{          }c_{<}(\lambda,\lambda^{'};t-t^{'})
\right)^{-1}
\\
& \times\mbox{           } e^{\theta (t-t^{'}) \beta  \epsilon_{ {\bf{k}} } }  \sum_{ {\bf{p}} }\mbox{          }
  g_{<}(\lambda,\lambda^{'};t-t^{'})\mbox{      } n_F({\bf{p}})\mbox{      }
\left( 1+
 e^{-  \beta  \epsilon_{ {\bf{p}} } }  c^{-1}_{<}(\lambda,\lambda^{'};t-t^{'})  c_{>}(\lambda,\lambda^{'};t-t^{'}) \mbox{          }
 e^{\lambda + \lambda^{'} } \right)^{-1}
\end{align*}

and

\begin{align*}
-\partial_{ \lambda^{'} } \mbox{             }
   c^{-1}_{<}(\lambda,\lambda^{'};t-t^{'}) \mbox{     } g_{<}(\lambda,\lambda^{'};&t-t^{'})  g_{>}(\lambda,\lambda^{'};t-t^{'})\mbox{      }  n_F({\bf{p}})\mbox{      }
\left( 1+
 e^{-  \beta  \epsilon_{ {\bf{p}} } }  c^{-1}_{<}(\lambda,\lambda^{'};t-t^{'})  c_{>}(\lambda,\lambda^{'};t-t^{'}) \mbox{          }
 e^{\lambda + \lambda^{'} } \right)^{-1}
\mbox{             } e^{ -\theta(t^{'}-t) \beta  \epsilon_{ {\bf{p}} } }
\\
\mbox{             }= \mbox{             }  c^{-1}_{<}&(\lambda,\lambda^{'};t-t^{'}) \mbox{     } g_{<}(\lambda,\lambda^{'};t-t^{'})\mbox{      } n_F({\bf{p}})\mbox{      }
\left( 1+
 e^{-  \beta  \epsilon_{ {\bf{p}} } }  c^{-1}_{<}(\lambda,\lambda^{'};t-t^{'})  c_{>}(\lambda,\lambda^{'};t-t^{'}) \mbox{          }
 e^{\lambda + \lambda^{'} } \right)^{-1}
 \\
 &
\mbox{             } e^{ -\theta(t^{'}-t) \beta  \epsilon_{ {\bf{p}} } }  \sum_{ {\bf{k}} }\mbox{          } g_{>}(\lambda,\lambda^{'};t-t^{'})\mbox{      } (1-n_F({\bf{k}}))\mbox{         }   \left( 1
+   e^{  \beta  \epsilon_{ {\bf{k}} } }   \mbox{             } c^{-1}_{>}(\lambda,\lambda^{'};t-t^{'}) \mbox{          }c_{<}(\lambda,\lambda^{'};t-t^{'})
\right)^{-1}
\end{align*}

================================================================================

\begin{align*}
-\partial_{ \lambda } \mbox{             }
c^{-1}_{>}(\lambda,\lambda^{'};t-t^{'}) \mbox{          }\mbox{     } g_{>}&(\lambda,\lambda^{'};t-t^{'})
 g_{<}(\lambda,\lambda^{'};t-t^{'}) \mbox{      }   \left( 1
+   e^{  \beta  \epsilon_{ {\bf{k}} } }   \mbox{             } c^{-1}_{>}(\lambda,\lambda^{'};t-t^{'}) \mbox{          }c_{<}(\lambda,\lambda^{'};t-t^{'})
\right)^{-1}
\\
\mbox{             }= \mbox{             }  c^{-1}_{>}&(\lambda,\lambda^{'};t-t^{'}) \mbox{          }\mbox{     } g_{>}(\lambda,\lambda^{'};t-t^{'})
  g_{<}(\lambda,\lambda^{'};t-t^{'}) \left( 1
+   e^{  \beta  \epsilon_{ {\bf{k}} } }   \mbox{             } c^{-1}_{>}(\lambda,\lambda^{'};t-t^{'}) \mbox{          }c_{<}(\lambda,\lambda^{'};t-t^{'})
\right)^{-1}
\mbox{           }
\\
& \times   \sum_{ {\bf{p}} }\mbox{      } n_F({\bf{p}})\mbox{      }
\left( 1+
 e^{-  \beta  \epsilon_{ {\bf{p}} } }  c^{-1}_{<}(\lambda,\lambda^{'};t-t^{'})  c_{>}(\lambda,\lambda^{'};t-t^{'}) \mbox{          }
 e^{\lambda + \lambda^{'} } \right)^{-1}
\end{align*}

and

\begin{align*}
-\partial_{ \lambda^{'} } \mbox{             }
   c^{-1}_{<}(\lambda,\lambda^{'}; t-t^{'}) \mbox{     } g_{<}&(\lambda,\lambda^{'}; t-t^{'})  g_{>}(\lambda,\lambda^{'}; t-t^{'})\mbox{      }
\left( 1+
 e^{-  \beta  \epsilon_{ {\bf{p}} } }  c^{-1}_{<}(\lambda,\lambda^{'}; t-t^{'})  c_{>}(\lambda,\lambda^{'}; t-t^{'}) \mbox{          }
 e^{\lambda + \lambda^{'} } \right)^{-1}
\\
\mbox{             }= \mbox{             }  c^{-1}_{<}&(\lambda,\lambda^{'}; t-t^{'}) \mbox{     } g_{<}(\lambda,\lambda^{'}; t-t^{'})g_{>}(\lambda,\lambda^{'}; t-t^{'})
\left( 1+
 e^{-  \beta  \epsilon_{ {\bf{p}} } }  c^{-1}_{<}(\lambda,\lambda^{'}; t-t^{'})  c_{>}(\lambda,\lambda^{'}; t-t^{'}) \mbox{          }
 e^{\lambda + \lambda^{'} } \right)^{-1}
\mbox{             } 
 \\
& \times \sum_{ {\bf{k}} }\mbox{      } (1-n_F({\bf{k}}))\mbox{         }   \left( 1
+   e^{  \beta  \epsilon_{ {\bf{k}} } }   \mbox{             } c^{-1}_{>}(\lambda,\lambda^{'}; t-t^{'}) \mbox{          }c_{<}(\lambda,\lambda^{'}; t-t^{'})
\right)^{-1}
\end{align*}

==================================================================================

\[
\Gamma_{>}(\lambda,\lambda^{'}; t-t^{'}) \mbox{      } = \mbox{            } c^{-1}_{>}(\lambda,\lambda^{'}; t-t^{'}) \mbox{          }\mbox{     } g_{>}(\lambda,\lambda^{'}; t-t^{'})
 g_{<}(\lambda,\lambda^{'}; t-t^{'})
\]
and
\[
\Gamma_{<}(\lambda,\lambda^{'}; t-t^{'}) \mbox{      } = \mbox{            }    c^{-1}_{<}(\lambda,\lambda^{'}; t-t^{'}) \mbox{     } g_{<}(\lambda,\lambda^{'}; t-t^{'})  g_{>}(\lambda,\lambda^{'}; t-t^{'})
\]
and
\[
\gamma(\lambda,\lambda^{'}; t-t^{'}) \mbox{      } = \mbox{            }  c^{-1}_{>}(\lambda,\lambda^{'}; t-t^{'}) \mbox{          }c_{<}(\lambda,\lambda^{'}; t-t^{'})
\]
==================================================================================
\begin{align*}
-\partial_{ \lambda } \mbox{             }
\Gamma_{>}(\lambda,\lambda^{'}; t-t^{'}) \mbox{      }   &\left( 1
+   e^{  \beta  \epsilon_{ {\bf{k}} } }   \mbox{             }\gamma(\lambda,\lambda^{'}; t-t^{'})
\right)^{-1} \mbox{             }=
\\
&
 \mbox{             } \Gamma_{>}(\lambda,\lambda^{'}; t-t^{'}) \mbox{      } \left( 1
+   e^{  \beta  \epsilon_{ {\bf{k}} } }   \mbox{             }
\gamma(\lambda,\lambda^{'}; t-t^{'}) \right)^{-1}
\mbox{           }
\sum_{ {\bf{p}} }\mbox{      } n_F({\bf{p}})\mbox{      }
\left( 1+
 e^{-  \beta  \epsilon_{ {\bf{p}} } }  \mbox{          }
\frac{  e^{\lambda + \lambda^{'} } }{ \gamma(\lambda,\lambda^{'}; t-t^{'}) } \right)^{-1}
\end{align*}

and

\begin{align*}
-\partial_{ \lambda^{'} } \mbox{             }
   \Gamma_{<}(\lambda,\lambda^{'}; t-t^{'}) \mbox{      }
&\left( 1+
 e^{-  \beta  \epsilon_{ {\bf{p}} } } \mbox{          }
 \frac{ e^{\lambda + \lambda^{'} } }{\gamma(\lambda,\lambda^{'}; t-t^{'}) } \right)^{-1} \mbox{             }=
 \\
 & \mbox{             } \Gamma_{<}(\lambda,\lambda^{'}; t-t^{'}) \mbox{      }
\left( 1+
 e^{-  \beta  \epsilon_{ {\bf{p}} } }  \mbox{          }
\frac{ e^{\lambda + \lambda^{'} } }{ \gamma(\lambda,\lambda^{'}; t-t^{'}) } \right)^{-1}
\mbox{             }  \sum_{ {\bf{k}} }\mbox{      } (1-n_F({\bf{k}}))\mbox{         }   \left( 1
+   e^{  \beta  \epsilon_{ {\bf{k}} } }   \mbox{             } \gamma(\lambda,\lambda^{'}; t-t^{'})
\right)^{-1}
\end{align*}
=====================================================================================

\[
 \Gamma_{>}(\lambda,\lambda^{'}; t-t^{'}) \mbox{      }   \left( 1
+   e^{  \beta  \epsilon_{ {\bf{k}} } }   \mbox{             } \gamma(\lambda,\lambda^{'}; t-t^{'})
\right)^{-1} \mbox{             }= \mbox{             } \Gamma_{>}(\lambda_0,\lambda^{'}; t-t^{'}) \mbox{      } \left( 1
+   e^{  \beta  \epsilon_{ {\bf{k}} } }   \mbox{             }
\gamma(\lambda_0,\lambda^{'}) \right)^{-1}
\mbox{           } e^{  - \int^{\lambda}_{\lambda_0} d s\mbox{             }   \sum_{ {\bf{p}} }\mbox{      } \frac{ n_F({\bf{p}})
 }{
\left( 1+
 e^{-  \beta  \epsilon_{ {\bf{p}} } }  \mbox{          }
\frac{  e^{s + \lambda^{'} } }{ \gamma(s,\lambda^{'}; t-t^{'}) } \right) } }
\]
and
\[
    \Gamma_{<}(\lambda,\lambda^{'}; t-t^{'}) \mbox{      }
\left( 1+
 e^{-  \beta  \epsilon_{ {\bf{p}} } } \mbox{          }
 \frac{ e^{\lambda + \lambda^{'} } }{\gamma(\lambda,\lambda^{'}; t-t^{'}) } \right)^{-1} \mbox{             }= \mbox{             } \Gamma_{<}(\lambda,\lambda_0; t-t^{'}) \mbox{      }
\left( 1+
 e^{-  \beta  \epsilon_{ {\bf{p}} } } \mbox{          }
 \frac{ e^{\lambda + \lambda_0 } }{\gamma(\lambda,\lambda_0; t-t^{'}) } \right)^{-1}
\mbox{             }  e^{ - \int^{ \lambda^{'} }_{ \lambda_0 } ds \sum_{ {\bf{k}} }\mbox{      } \frac{ (1-n_F({\bf{k}}))
 }{ \left( 1
+   e^{  \beta  \epsilon_{ {\bf{k}} } }   \mbox{             } \gamma(\lambda,s; t-t^{'})
\right) }   }
\]

=======================================================================================

\[
 \Gamma_{>}(\lambda,\lambda^{'}; t-t^{'}) \mbox{      }   \left( 1
+   e^{  \beta  \epsilon_{ {\bf{k}} } }   \mbox{             } \gamma(\lambda,\lambda^{'}; t-t^{'})
\right)^{-1} \mbox{             }= \mbox{             } \Gamma_{>}(0,\lambda^{'}; t-t^{'}) \mbox{      } \left( 1
+   e^{  \beta  \epsilon_{ {\bf{k}} } }   \mbox{             }
\gamma(0,\lambda^{'}; t-t^{'}) \right)^{-1}
\mbox{           } e^{  - \int^{\lambda}_{0} d s\mbox{             }   \sum_{ {\bf{p}} }\mbox{      } \frac{ n_F({\bf{p}})
 }{
\left( 1+
 e^{-  \beta  \epsilon_{ {\bf{p}} } }  \mbox{          }
\frac{  e^{s + \lambda^{'} } }{ \gamma(s,\lambda^{'}; t-t^{'}) } \right) } }
\]
and
\[
    \Gamma_{<}(\lambda,\lambda^{'}; t-t^{'}) \mbox{      }
\left( 1+
 e^{-  \beta  \epsilon_{ {\bf{p}} } } \mbox{          }
 \frac{ e^{\lambda + \lambda^{'} } }{\gamma(\lambda,\lambda^{'}; t-t^{'}) } \right)^{-1} \mbox{             }= \mbox{             } \Gamma_{<}(\lambda,0; t-t^{'}) \mbox{      }
\left( 1+
 e^{-  \beta  \epsilon_{ {\bf{p}} } } \mbox{          }
 \frac{ e^{\lambda  } }{\gamma(\lambda,0; t-t^{'}) } \right)^{-1}
\mbox{             }  e^{ - \int^{ \lambda^{'} }_{ 0 } ds \sum_{ {\bf{k}} }\mbox{      } \frac{ (1-n_F({\bf{k}}))
 }{ \left( 1
+   e^{  \beta  \epsilon_{ {\bf{k}} } }   \mbox{             } \gamma(\lambda,s; t-t^{'})
\right) }   }
\]

=======================================================================================

\newpage

=========================================================================================

\[
 \Gamma_{>}(\lambda,\lambda^{'}; t-t^{'}) \mbox{      } \left( 1
+   e^{  \beta  \epsilon_{ {\bf{k}} } }   \mbox{             }
\gamma(0,\lambda^{'}; t-t^{'}) \right) \mbox{             }= \mbox{             } \Gamma_{>}(0,\lambda^{'}; t-t^{'}) \mbox{      }  \left( 1
+   e^{  \beta  \epsilon_{ {\bf{k}} } }   \mbox{             } \gamma(\lambda,\lambda^{'}; t-t^{'})
\right)
\mbox{           } e^{  - \int^{\lambda}_{0} d s\mbox{             }   \sum_{ {\bf{p}} }\mbox{      } \frac{ n_F({\bf{p}})
 }{
\left( 1+
 e^{-  \beta  \epsilon_{ {\bf{p}} } }  \mbox{          }
\frac{  e^{s + \lambda^{'} } }{ \gamma(s,\lambda^{'}; t-t^{'}) } \right) } }
\]
and
\[
    \Gamma_{<}(\lambda,\lambda^{'}; t-t^{'}) \mbox{      }  \left( 1+
 e^{-  \beta  \epsilon_{ {\bf{p}} } } \mbox{          }
 \frac{ e^{\lambda  } }{\gamma(\lambda,0; t-t^{'}) } \right)
\mbox{             }= \mbox{             } \Gamma_{<}(\lambda,0; t-t^{'}) \mbox{      }
\left( 1+
 e^{-  \beta  \epsilon_{ {\bf{p}} } } \mbox{          }
 \frac{ e^{\lambda + \lambda^{'} } }{\gamma(\lambda,\lambda^{'}; t-t^{'}) } \right)
\mbox{             }  e^{ - \int^{ \lambda^{'} }_{ 0 } ds \sum_{ {\bf{k}} }\mbox{      } \frac{ (1-n_F({\bf{k}}))
 }{ \left( 1
+   e^{  \beta  \epsilon_{ {\bf{k}} } }   \mbox{             } \gamma(\lambda,s; t-t^{'})
\right) }   }
\]

============================================================================

============================================================================

From this we may conclude,

\[
\gamma(0,\lambda^{'}; t-t^{'}) \mbox{            } = \mbox{                 }  \gamma(\lambda,\lambda^{'}; t-t^{'})
\]
and,
\[
 \frac{ e^{\lambda  } }{\gamma(\lambda,0; t-t^{'}) }
\mbox{             }= \mbox{             }  \frac{ e^{\lambda + \lambda^{'} } }{\gamma(\lambda,\lambda^{'}; t-t^{'}) }
\]

=================================================================================================================

\[
 \Gamma_{>}(\lambda,\lambda^{'}; t-t^{'})   \mbox{             }= \mbox{             } \Gamma_{>}(0,\lambda^{'}; t-t^{'})
\mbox{           } e^{  - \int^{\lambda}_{0} d s\mbox{             }   \sum_{ {\bf{p}} }\mbox{      } \frac{ n_F({\bf{p}})
 }{
\left( 1+
 e^{-  \beta  \epsilon_{ {\bf{p}} } }  \mbox{          }
\frac{  e^{s + \lambda^{'} } }{ \gamma(s,\lambda^{'}; t-t^{'}) } \right) } }
\]
and
\[
    \Gamma_{<}(\lambda,\lambda^{'}; t-t^{'})
\mbox{             }= \mbox{             } \Gamma_{<}(\lambda,0; t-t^{'})
\mbox{             }  e^{ - \int^{ \lambda^{'} }_{ 0 } ds \sum_{ {\bf{k}} }\mbox{      } \frac{ (1-n_F({\bf{k}}))
 }{ \left( 1
+   e^{  \beta  \epsilon_{ {\bf{k}} } }   \mbox{             } \gamma(\lambda,s; t-t^{'})
\right) }   }
\]

==================================================================================================================

\[
\gamma(0,\lambda^{'}; t-t^{'}) \mbox{            } = \mbox{                 }  \gamma(\lambda,\lambda^{'}; t-t^{'})
\]
and,
\[
  \gamma(\lambda,\lambda^{'}; t-t^{'})
\mbox{             }= \mbox{             }    e^{  \lambda^{'} } \mbox{               } \gamma(\lambda,0; t-t^{'})
\]
or,
\[
\gamma(0,0; t-t^{'}) \mbox{            } = \mbox{                 }  \gamma(\lambda,0; t-t^{'})
\]
or,
\[
  \gamma(\lambda,\lambda^{'}; t-t^{'})
\mbox{             }= \mbox{             }    e^{  \lambda^{'} } \mbox{               } \gamma(0,0; t-t^{'}); \mbox{           } Set
\mbox{                         } \gamma(0,0; t-t^{'}) \equiv e^{ - \beta \mu }; \mbox{           } \gamma(\lambda,\lambda^{'}; t-t^{'})
\mbox{             }= \mbox{             }    e^{  \lambda^{'} } \mbox{               }e^{ - \beta \mu }
\]

=================================================================================================================

\[
 \Gamma_{>}(\lambda,\lambda^{'}; t-t^{'})   \mbox{             }= \mbox{             } \Gamma_{>}(0,\lambda^{'}; t-t^{'})
\mbox{           } e^{  - \int^{\lambda}_{0} d s\mbox{             }   \sum_{ {\bf{p}} }\mbox{      } \frac{ n_F({\bf{p}})
 }{
\left( 1+
 e^{-  \beta  (\epsilon_{ {\bf{p}} }-\mu) }  \mbox{          }
  e^{s  } \right) } }
\]
and
\[
    \Gamma_{<}(\lambda,\lambda^{'}; t-t^{'})
\mbox{             }= \mbox{             } \Gamma_{<}(\lambda,0; t-t^{'})
\mbox{             }  e^{ - \int^{ \lambda^{'} }_{ 0 } ds \sum_{ {\bf{k}} }\mbox{      } \frac{ (1-n_F({\bf{k}}))
 }{ \left( 1
+   e^{  \beta  (\epsilon_{ {\bf{k}} }-\mu) }   \mbox{             }  e^{  s }
\right) }   }
\]

==================================================================================================================

\[
\Gamma_{>}(\lambda,\lambda^{'}; t-t^{'}) \mbox{      } = \mbox{            } c^{-1}_{>}(\lambda,\lambda^{'}; t-t^{'}) \mbox{          }\mbox{     } g_{>}(\lambda,\lambda^{'}; t-t^{'})
 g_{<}(\lambda,\lambda^{'}; t-t^{'})
\]
and
\[
\Gamma_{<}(\lambda,\lambda^{'}; t-t^{'}) \mbox{      } = \mbox{            }    c^{-1}_{<}(\lambda,\lambda^{'}; t-t^{'}) \mbox{     } g_{<}(\lambda,\lambda^{'}; t-t^{'})  g_{>}(\lambda,\lambda^{'}; t-t^{'})
\]

====================================================================================================================

\[
<T \mbox{   }e^{ - \lambda N_{>}(t)}  \mbox{             }e^{ - \lambda^{'} N^{'}_{>}(t)}
\mbox{        } {\tilde{ c}}_{ {\bf{p}}, < }(t)
{\tilde{c}}^{\dagger}_{ {\bf{p}},< }(t^{'}) >\mbox{            } = \mbox{                }\Gamma_{<}(\lambda,\lambda^{'}; t-t^{'})\mbox{      }  \frac{ n_F({\bf{p}})\mbox{      }
  e^{ -\theta(t^{'}-t) \beta  \epsilon_{ {\bf{p}} } } }{\left( 1+
 e^{-  \beta  \epsilon_{ {\bf{p}} } }   \mbox{          }
\frac{ e^{\lambda + \lambda^{'} } }{\gamma(\lambda,\lambda^{'}; t-t^{'}) } \right)}
\]
and
\[
 <T \mbox{   }e^{ - \lambda N_{>}(t)}  \mbox{             }e^{ - \lambda^{'} N^{'}_{>}(t)} \mbox{             }
{\tilde{c}}_{ {\bf{k}}, > }(t)
{\tilde{c}}^{\dagger}_{ {\bf{k}},> }(t^{'}) >   \mbox{            } = \mbox{                }
\Gamma_{>}(\lambda,\lambda^{'}; t-t^{'})\mbox{      } \frac{ (1-n_F({\bf{k}})) e^{\theta (t-t^{'}) \beta  \epsilon_{ {\bf{k}} } } }{  \left( 1
+   e^{  \beta  \epsilon_{ {\bf{k}} } }   \mbox{             }
 \gamma(\lambda,\lambda^{'}; t-t^{'})
\right) }
\]

====================================================================================================================

\[
<T \mbox{   }e^{ - \lambda N_{>}(t)}  \mbox{             }e^{ - \lambda^{'} N^{'}_{>}(t)}
\mbox{        } {\tilde{ c}}_{ {\bf{p}}, < }(t)
{\tilde{c}}^{\dagger}_{ {\bf{p}},< }(t^{'}) >\mbox{            } = \mbox{                } \Gamma_{<}(\lambda,0; t-t^{'})
\mbox{      }  \frac{ n_F({\bf{p}})\mbox{      }
  e^{ -\theta(t^{'}-t) \beta  \epsilon_{ {\bf{p}} } } }{\left( 1+
 e^{-  \beta  (\epsilon_{ {\bf{p}} }-\mu) }   \mbox{          }
  e^{\lambda   }    \right)}\mbox{             }  e^{ - \int^{ \lambda^{'} }_{ 0 } ds \sum_{ {\bf{k}} }\mbox{      } \frac{ (1-n_F({\bf{k}}))
 }{ \left( 1
+   e^{  \beta  (\epsilon_{ {\bf{k}} }-\mu) }   \mbox{             }  e^{  s }
\right) }   }
\]
and
\[
 <T \mbox{   }e^{ - \lambda N_{>}(t)}  \mbox{             }e^{ - \lambda^{'} N^{'}_{>}(t)} \mbox{             }
{\tilde{c}}_{ {\bf{k}}, > }(t)
{\tilde{c}}^{\dagger}_{ {\bf{k}},> }(t^{'}) >   \mbox{            } = \mbox{                }
 \Gamma_{>}(0,\lambda^{'}; t-t^{'})
\mbox{      } \frac{ (1-n_F({\bf{k}})) e^{\theta (t-t^{'}) \beta  \epsilon_{ {\bf{k}} } } }{  \left( 1
+   e^{  \beta  (\epsilon_{ {\bf{k}} }-\mu) }   \mbox{             }
   e^{  \lambda^{'} }
\right) }\mbox{           } e^{  - \int^{\lambda}_{0} d s\mbox{             }   \sum_{ {\bf{p}} }\mbox{      } \frac{ n_F({\bf{p}})
 }{
\left( 1+
 e^{-  \beta  (\epsilon_{ {\bf{p}} }-\mu) }  \mbox{          }
  e^{s  } \right) } }
\]

\newpage

\section{ Solving for the remaining coefficients }

We now have to solve for the remaining coefficients viz. $  \Gamma_{>}(0,\lambda^{'}) $ and $ \Gamma_{<}(\lambda,0) $. For this we have to examine the complementary equations we have so far neglected. Consider,

\begin{align*}
-\partial_{ \lambda^{'} } \mbox{       } <T \mbox{   }e^{ - \lambda N_{>}(t)}  \mbox{             }e^{ - \lambda^{'} N^{'}_{>}(t)}
\mbox{        } &{\tilde{ c}}_{ {\bf{p}}, < }(t)
{\tilde{c}}^{\dagger}_{ {\bf{p}},< }(t^{'}) >
\\
&=\mbox{                } \Gamma_{<}(\lambda,0; t-t^{'})
\mbox{      }  \frac{ n_F({\bf{p}})\mbox{      }
  e^{ -\theta(t^{'}-t) \beta  \epsilon_{ {\bf{p}} } } }{\left( 1+
 e^{-  \beta  (\epsilon_{ {\bf{p}} }-\mu) }   \mbox{          }
  e^{\lambda   }    \right)}\mbox{             }  e^{ - \int^{ \lambda^{'} }_{ 0 } ds \sum_{ {\bf{k}} }\mbox{      } \frac{ (1-n_F({\bf{k}}))
 }{ \left( 1
+   e^{  \beta  (\epsilon_{ {\bf{k}} }-\mu) }   \mbox{             }  e^{  s }
\right) }   }\mbox{  } \mbox{      }\left(  \sum_{ {\bf{k}} }\mbox{      } \frac{ (1-n_F({\bf{k}}))
 }{ \left( 1
+   e^{  \beta  (\epsilon_{ {\bf{k}} }-\mu) }   \mbox{             }  e^{ \lambda^{'} }
\right) }  \right)
\end{align*}

and

\begin{align*}
-\partial_{ \lambda } \mbox{       }  <T \mbox{   }e^{ - \lambda N_{>}(t)}  \mbox{             }e^{ - \lambda^{'} N^{'}_{>}(t)} \mbox{             }
&{\tilde{c}}_{ {\bf{k}}, > }(t)
{\tilde{c}}^{\dagger}_{ {\bf{k}},> }(t^{'}) >  
\\
&=\mbox{                }
 \Gamma_{>}(0,\lambda^{'};t-t^{'})
\mbox{      } \frac{ (1-n_F({\bf{k}})) e^{\theta (t-t^{'}) \beta  \epsilon_{ {\bf{k}} } } }{  \left( 1
+   e^{  \beta  (\epsilon_{ {\bf{k}} }-\mu) }   \mbox{             }
   e^{  \lambda^{'} }
\right) }\mbox{           } e^{  - \int^{\lambda}_{0} d s\mbox{             }   \sum_{ {\bf{p}} }\mbox{      } \frac{ n_F({\bf{p}})
 }{
\left( 1+
 e^{-  \beta  (\epsilon_{ {\bf{p}} }-\mu) }  \mbox{          }
  e^{s  } \right) } }\mbox{          } \left(  \sum_{ {\bf{p}} }\mbox{      } \frac{ n_F({\bf{p}})
 }{
\left( 1+
 e^{-  \beta  (\epsilon_{ {\bf{p}} }-\mu) }  \mbox{          }
  e^{\lambda} \right) } \right)
\end{align*}

========================================================================================================================================

But on the other hand,

\begin{align*}
-\partial_{ \lambda^{'} } \mbox{       } <T \mbox{   }e^{ - \lambda N_{>}(t)}  \mbox{             }e^{ - \lambda^{'} N^{'}_{>}(t)}
\mbox{        } {\tilde{ c}}_{ {\bf{p}}, < }(t)
&{\tilde{c}}^{\dagger}_{ {\bf{p}},< }(t^{'}) >\mbox{            }\\
 &= \mbox{                }
  <T \mbox{   }e^{ - \lambda N_{>}(t)}  \mbox{             }e^{ - \lambda^{'} N^{'}_{>}(t)}
\mbox{        }N^{'}_{>}(t) {\tilde{ c}}_{ {\bf{p}}, < }(t)
{\tilde{c}}^{\dagger}_{ {\bf{p}},< }(t^{'}) >\mbox{            }\\
& = \mbox{                }
\sum_{ {\bf{k}} } \mbox{        }  <T \mbox{   }e^{ - \lambda N_{>}(t)}  \mbox{             }e^{ - \lambda^{'} N^{'}_{>}(t)}
\mbox{        }{\tilde{ c}}^{\dagger}_{ {\bf{k}}, > }(t){\tilde{ c}}_{ {\bf{k}}, > }(t) {\tilde{ c}}_{ {\bf{p}}, < }(t)
{\tilde{c}}^{\dagger}_{ {\bf{p}},< }(t^{'}) >
\end{align*}

Now assume $ t > t^{'} \rightarrow t $ then the above becomes,
\[
\sum_{ {\bf{k}} } \mbox{        }  <e^{ - \lambda N_{>}(t)}
\mbox{        }{\tilde{ c}}^{\dagger}_{ {\bf{k}}, > }(t) {\tilde{ c}}_{ {\bf{p}}, < }(t)
{\tilde{c}}^{\dagger}_{ {\bf{p}},< }(t){\tilde{ c}}_{ {\bf{k}}, > }(t) >   \mbox{            } = \mbox{                } \Gamma_{<}(\lambda,0;-i\epsilon)
\mbox{      }  \frac{ n_F({\bf{p}}) }{\left( 1+
 e^{-  \beta  (\epsilon_{ {\bf{p}} }-\mu) }   \mbox{          }
  e^{\lambda   }    \right)}\mbox{             }  \left(  \sum_{ {\bf{k}} }\mbox{      } \frac{ (1-n_F({\bf{k}}))
 }{ \left( 1
+   e^{  \beta  (\epsilon_{ {\bf{k}} }-\mu) }
\right) }  \right)
\]
and
\[
 \sum_{ {\bf{p}} }\mbox{          }
  < e^{ - \lambda^{'} N^{'}_{>}(t)} \mbox{             }
{\tilde{c}}_{ {\bf{p}}, < }(t){\tilde{c}}^{\dagger}_{ {\bf{p}}, < }(t)
{\tilde{c}}_{ {\bf{k}}, > }(t)
{\tilde{c}}^{\dagger}_{ {\bf{k}},> }(t) >   \mbox{            } = \mbox{                }
 \Gamma_{>}(0,\lambda^{'};-i\epsilon)
\mbox{      } \frac{ (1-n_F({\bf{k}})) e^{  \beta  \epsilon_{ {\bf{k}} } } }{  \left( 1
+   e^{  \beta  (\epsilon_{ {\bf{k}} }-\mu) }   \mbox{             }
   e^{  \lambda^{'} }
\right) }\mbox{           }   \left(  \sum_{ {\bf{p}} }\mbox{      } \frac{ n_F({\bf{p}})
 }{
\left( 1+
 e^{-  \beta  (\epsilon_{ {\bf{p}} }-\mu) }  \mbox{          }
 \right) } \right)
\]
We now want to evaluate the left hand side of the above equations in terms of bosonic algebra. Before we do this, we need a minor rearrangement. Even though this rearrangement uses Fermi algebra, the remainder of the evaluation is going to be done purely using bosonic algebra. We write,
\[
{\tilde{c}}_{ {\bf{p}}, < }(t){\tilde{c}}^{\dagger}_{ {\bf{p}}, < }(t)
 = - {\tilde{c}}^{\dagger}_{ {\bf{p}}, < }(t){\tilde{c}}_{ {\bf{p}}, < }(t) + n_F({\bf{p}})
\]
and
\[
{\tilde{c}}^{\dagger}_{ {\bf{p}}, < }{\tilde{c}}_{ {\bf{k}}, > }
\mbox{      } \equiv \mbox{           } a_{  \frac{1}{2}({\bf{p}}+{\bf{k}})  }({\bf{k}}-{\bf{p}})\mbox{      } ; \mbox{           }\mbox{           }\mbox{           }
{\tilde{c}}^{\dagger}_{ {\bf{k}}, > } {\tilde{c}}_{ {\bf{p}}, < }
\mbox{      } \equiv \mbox{           } a^{\dagger}_{  \frac{1}{2}({\bf{p}}+{\bf{k}})  }({\bf{k}}-{\bf{p}})
\]
or,
\[
\sum_{ {\bf{k}} } \mbox{        }  <e^{ - \lambda N_{>}(t)}
\mbox{        }a^{\dagger}_{  \frac{1}{2}({\bf{p}}+{\bf{k}})  }({\bf{k}}-{\bf{p}},t)
a_{  \frac{1}{2}({\bf{p}}+{\bf{k}})  }({\bf{k}}-{\bf{p}},t) >   \mbox{            } = \mbox{                } \Gamma_{<}(\lambda,0;-i\epsilon)
\mbox{      }  \frac{ n_F({\bf{p}}) }{\left( 1+
 e^{-  \beta  (\epsilon_{ {\bf{p}} }-\mu) }   \mbox{          }
  e^{\lambda   }    \right)}\mbox{             }  \left(  \sum_{ {\bf{k}} }\mbox{      } \frac{ (1-n_F({\bf{k}}))
 }{ \left( 1
+   e^{  \beta  (\epsilon_{ {\bf{k}} }-\mu) }
\right) }  \right)
\]

and

\begin{align*}
 N^0\mbox{          }
  < e^{ - \lambda^{'} N^{'}_{>}(t)} \mbox{             }
{\tilde{c}}_{ {\bf{k}}, > }(t)
{\tilde{c}}^{\dagger}_{ {\bf{k}},> }(t) > - \sum_{ {\bf{p}} }\mbox{          }
  < e^{ - \lambda^{'} N^{'}_{>}(t)} \mbox{             }
 & a_{  \frac{1}{2}({\bf{p}}+{\bf{k}})  }({\bf{k}}-{\bf{p}}) a^{\dagger}_{  \frac{1}{2}({\bf{p}}+{\bf{k}})  }({\bf{k}}-{\bf{p}})
>  \mbox{            }
\\
& = \mbox{                }
 \Gamma_{>}(0,\lambda^{'};-i\epsilon)
\mbox{      } \frac{ (1-n_F({\bf{k}})) e^{  \beta  \epsilon_{ {\bf{k}} } } }{  \left( 1
+   e^{  \beta  (\epsilon_{ {\bf{k}} }-\mu) }   \mbox{             }
   e^{  \lambda^{'} }
\right) }\mbox{           }   \left(  \sum_{ {\bf{p}} }\mbox{      } \frac{ n_F({\bf{p}})
 }{
\left( 1+
 e^{-  \beta  (\epsilon_{ {\bf{p}} }-\mu) }  \mbox{          }
 \right) } \right)
\end{align*}

\newpage

\[
\sum_{ {\bf{k}} } \mbox{        }  <e^{ - \lambda N_{>}(t)}
\mbox{        }a^{\dagger}_{  \frac{1}{2}({\bf{p}}+{\bf{k}})  }({\bf{k}}-{\bf{p}},t)
a_{  \frac{1}{2}({\bf{p}}+{\bf{k}})  }({\bf{k}}-{\bf{p}},t) >   \mbox{            } = \mbox{                } \Gamma_{<}(\lambda,0;-i\epsilon)
\mbox{      }  \frac{ n_F({\bf{p}}) }{\left( 1+
 e^{-  \beta  (\epsilon_{ {\bf{p}} }-\mu) }   \mbox{          }
  e^{\lambda   }    \right)}\mbox{             }  \left(  \sum_{ {\bf{k}} }\mbox{      } \frac{ (1-n_F({\bf{k}}))
 }{ \left( 1
+   e^{  \beta  (\epsilon_{ {\bf{k}} }-\mu) }
\right) }  \right)
\]
and
\begin{align*}
 N^0\mbox{          }
  < e^{ - \lambda^{'} N^{'}_{>}(t)} \mbox{             }
{\tilde{c}}_{ {\bf{k}}, > }(t)
{\tilde{c}}^{\dagger}_{ {\bf{k}},> }(t) > - \sum_{ {\bf{p}} }\mbox{          }
  < e^{ - \lambda^{'} N^{'}_{>}(t)} \mbox{             }
  &a_{  \frac{1}{2}({\bf{p}}+{\bf{k}})  }({\bf{k}}-{\bf{p}}) a^{\dagger}_{  \frac{1}{2}({\bf{p}}+{\bf{k}})  }({\bf{k}}-{\bf{p}})
>  \mbox{            }
\\
& = \mbox{                }
 \Gamma_{>}(0,\lambda^{'};-i\epsilon)
\mbox{      } \frac{ (1-n_F({\bf{k}})) e^{  \beta  \epsilon_{ {\bf{k}} } } }{  \left( 1
+   e^{  \beta  (\epsilon_{ {\bf{k}} }-\mu) }   \mbox{             }
   e^{  \lambda^{'} }
\right) }\mbox{           }   \left(  \sum_{ {\bf{p}} }\mbox{      } \frac{ n_F({\bf{p}})
 }{
\left( 1+
 e^{-  \beta  (\epsilon_{ {\bf{p}} }-\mu) }  \mbox{          }
 \right) } \right)
\end{align*}

but,
\[
 <T \mbox{   } e^{ - \lambda^{'} N^{'}_{>}(t)} \mbox{             }
{\tilde{c}}_{ {\bf{k}}, > }(t)
{\tilde{c}}^{\dagger}_{ {\bf{k}},> }(t^{'}) >   \mbox{            } = \mbox{                }
 \Gamma_{>}(0,\lambda^{'};t-t^{'})
\mbox{      } \frac{ (1-n_F({\bf{k}})) e^{\theta (t-t^{'}) \beta  \epsilon_{ {\bf{k}} } } }{  \left( 1
+   e^{  \beta  (\epsilon_{ {\bf{k}} }-\mu) }   \mbox{             }
   e^{  \lambda^{'} }
\right) }
\]
Thus we have to solve,
\[
\sum_{ {\bf{k}} } \mbox{        }  <e^{ - \lambda N_{>}(t)}
\mbox{        }a^{\dagger}_{  \frac{1}{2}({\bf{p}}+{\bf{k}})  }({\bf{k}}-{\bf{p}})
a_{  \frac{1}{2}({\bf{p}}+{\bf{k}})  }({\bf{k}}-{\bf{p}}) >   \mbox{            } = \mbox{                } \Gamma_{<}(\lambda,0;-i\epsilon)
\mbox{      }  \frac{ n_F({\bf{p}}) }{\left( 1+
 e^{-  \beta  (\epsilon_{ {\bf{p}} }-\mu) }   \mbox{          }
  e^{\lambda   }    \right)}\mbox{             }  \left(  \sum_{ {\bf{k}} }\mbox{      } \frac{ (1-n_F({\bf{k}}))
 }{ \left( 1
+   e^{  \beta  (\epsilon_{ {\bf{k}} }-\mu) }
\right) }  \right)
\]
and
\[
\sum_{ {\bf{p}} }\mbox{          }
  < e^{ - \lambda^{'} N^{'}_{>}(t)} \mbox{             }
  a_{  \frac{1}{2}({\bf{p}}+{\bf{k}})  }({\bf{k}}-{\bf{p}}) a^{\dagger}_{  \frac{1}{2}({\bf{p}}+{\bf{k}})  }({\bf{k}}-{\bf{p}})
>  \mbox{            } = \mbox{                }
\Gamma_{>}(0,\lambda^{'};-i\epsilon)
\mbox{      } \frac{ (1-n_F({\bf{k}})) e^{  \beta  \epsilon_{ {\bf{k}} } } }{  \left( 1
+   e^{  \beta  (\epsilon_{ {\bf{k}} }-\mu) }   \mbox{             }
   e^{  \lambda^{'} }
\right) }\mbox{           }   \left(N^0  -  \sum_{ {\bf{p}} }\mbox{      } \frac{ n_F({\bf{p}})
 }{
\left( 1+
 e^{-  \beta  (\epsilon_{ {\bf{p}} }-\mu) }  \mbox{          }
 \right) } \right)
\]
The left hand side of the above equations may be evaluated using boson-like algebra.

\[
<e^{ - \lambda N_{>}(t)}
\mbox{        }a^{\dagger}_{  \frac{1}{2}({\bf{p}}+{\bf{k}})  }({\bf{k}}-{\bf{p}})
a_{  \frac{1}{2}({\bf{p}}+{\bf{k}})  }({\bf{k}}-{\bf{p}}) >   \mbox{            } \equiv \mbox{                }\frac{Tr(e^{ - \beta (H-\mu N)}e^{ - \lambda N_{>}(t)}
\mbox{        }a^{\dagger}_{  \frac{1}{2}({\bf{p}}+{\bf{k}})  }({\bf{k}}-{\bf{p}})
a_{  \frac{1}{2}({\bf{p}}+{\bf{k}})  }({\bf{k}}-{\bf{p}}) )}{Tr(e^{ - \beta (H-\mu N)})}
\]
Using the cyclic property of trace we get,
\[
(e^{ \lambda N_{>}(t)}  e^{  \beta (H-\mu N)}a_{  \frac{1}{2}({\bf{p}}+{\bf{k}})  }({\bf{k}}-{\bf{p}})e^{ - \beta (H-\mu N)}e^{ - \lambda N_{>}(t)})
\mbox{        }= \mbox{           } a_{  \frac{1}{2}({\bf{p}}+{\bf{k}})  }({\bf{k}}-{\bf{p}})\mbox{          }
e^{ - \lambda }\mbox{     }
e^{ - \beta (\epsilon_{ {\bf{k}} } - \epsilon_{ {\bf{p}} }) }
\]
or,
\[
<e^{ - \lambda N_{>}(t)}
\mbox{        }a^{\dagger}_{  \frac{1}{2}({\bf{p}}+{\bf{k}})  }({\bf{k}}-{\bf{p}})
a_{  \frac{1}{2}({\bf{p}}+{\bf{k}})  }({\bf{k}}-{\bf{p}}) >   \mbox{            } \equiv \mbox{                }\mbox{          }
e^{ - \lambda }\mbox{     }
e^{ - \beta (\epsilon_{ {\bf{k}} } - \epsilon_{ {\bf{p}} }) }\frac{Tr(
e^{ - \beta (H-\mu N)}e^{ - \lambda N_{>}(t)}
\mbox{        }a_{  \frac{1}{2}({\bf{p}}+{\bf{k}})  }({\bf{k}}-{\bf{p}})
 a^{\dagger}_{  \frac{1}{2}({\bf{p}}+{\bf{k}})  }({\bf{k}}-{\bf{p}})
 )}{Tr(e^{ - \beta (H-\mu N)})}
\]
Now,
\[
a_{  \frac{1}{2}({\bf{p}}+{\bf{k}})  }({\bf{k}}-{\bf{p}})
 a^{\dagger}_{  \frac{1}{2}({\bf{p}}+{\bf{k}})  }({\bf{k}}-{\bf{p}})\mbox{        } = \mbox{        }
 a^{\dagger}_{  \frac{1}{2}({\bf{p}}+{\bf{k}})  }({\bf{k}}-{\bf{p}}) a_{  \frac{1}{2}({\bf{p}}+{\bf{k}})  }({\bf{k}}-{\bf{p}})
   +   [a_{  \frac{1}{2}({\bf{p}}+{\bf{k}})  }({\bf{k}}-{\bf{p}}),
 a^{\dagger}_{  \frac{1}{2}({\bf{p}}+{\bf{k}})  }({\bf{k}}-{\bf{p}})]
\]

\[
 [a_{  \frac{1}{2}({\bf{p}}+{\bf{k}})  }({\bf{k}}-{\bf{p}}),
 a^{\dagger}_{  \frac{1}{2}({\bf{p}}+{\bf{k}})  }({\bf{k}}-{\bf{p}})]\mbox{      } = \mbox{      }
 [ c^{\dagger}_{ {\bf{p}}, < }c_{ {\bf{k}}, >  } ,c^{\dagger}_{ {\bf{k}}, > }c_{ {\bf{p}}, < }]
\]

\[
 [a_{  \frac{1}{2}({\bf{p}}+{\bf{k}})  }({\bf{k}}-{\bf{p}}),
 a^{\dagger}_{  \frac{1}{2}({\bf{p}}+{\bf{k}})  }({\bf{k}}-{\bf{p}})]\mbox{      } = \mbox{      }
  (1-n_F({\bf{k}}))c^{\dagger}_{ {\bf{p}}, < }c_{ {\bf{p}}, < } - c^{\dagger}_{ {\bf{k}}, > }c_{ {\bf{k}}, >  }n_F({\bf{p}})
\]

\[
c^{\dagger}_{ {\bf{p}}, < }c_{ {\bf{p}}, < } = -c_{ {\bf{p}}, < } c^{\dagger}_{ {\bf{p}}, < } + n_F({\bf{p}})
\]

\[
c^{\dagger}_{ {\bf{k}}, > }c_{ {\bf{k}}, >  } =  - c_{ {\bf{k}}, >  }c^{\dagger}_{ {\bf{k}}, > } + (1-n_F({\bf{k}}))
\]

\[
 [a_{  \frac{1}{2}({\bf{p}}+{\bf{k}})  }({\bf{k}}-{\bf{p}}),
 a^{\dagger}_{  \frac{1}{2}({\bf{p}}+{\bf{k}})  }({\bf{k}}-{\bf{p}})]\mbox{      } = \mbox{      }
  -  (1-n_F({\bf{k}})) \mbox{      } c_{ {\bf{p}}, < } c^{\dagger}_{ {\bf{p}}, < }   + c_{ {\bf{k}}, >  }c^{\dagger}_{ {\bf{k}}, > } \mbox{  } n_F({\bf{p}})
\]

or,

\begin{align*}
<&e^{ - \lambda N_{>}(t)}
\mbox{        }a^{\dagger}_{  \frac{1}{2}({\bf{p}}+{\bf{k}})  }({\bf{k}}-{\bf{p}})
a_{  \frac{1}{2}({\bf{p}}+{\bf{k}})  }({\bf{k}}-{\bf{p}}) >   \mbox{            }
\\
& \qquad\equiv \mbox{                }\mbox{          }
e^{ - \lambda }\mbox{     }
e^{ - \beta (\epsilon_{ {\bf{k}} } - \epsilon_{ {\bf{p}} }) }\frac{Tr(
e^{ - \beta (H-\mu N)}e^{ - \lambda N_{>}(t)}
\mbox{        }
( a^{\dagger}_{  \frac{1}{2}({\bf{p}}+{\bf{k}})  }({\bf{k}}-{\bf{p}}) a_{  \frac{1}{2}({\bf{p}}+{\bf{k}})  }({\bf{k}}-{\bf{p}})
 -  (1-n_F({\bf{k}})) \mbox{      } c_{ {\bf{p}}, < } c^{\dagger}_{ {\bf{p}}, < }   + c_{ {\bf{k}}, >  }c^{\dagger}_{ {\bf{k}}, > } \mbox{  } n_F({\bf{p}}))
 )}{Tr(e^{ - \beta (H-\mu N)})}
\end{align*}

and
\[
 <e^{ - \lambda N_{>}(t)}
\mbox{        }a^{\dagger}_{  \frac{1}{2}({\bf{p}}+{\bf{k}})  }({\bf{k}}-{\bf{p}})
a_{  \frac{1}{2}({\bf{p}}+{\bf{k}})  }({\bf{k}}-{\bf{p}}) >   \mbox{            } \equiv \mbox{                }\mbox{          }
\frac{  < e^{ - \lambda N_{>}(t)}
\mbox{        }
(  -  (1-n_F({\bf{k}})) \mbox{      } c_{ {\bf{p}}, < } c^{\dagger}_{ {\bf{p}}, < }   + c_{ {\bf{k}}, >  }c^{\dagger}_{ {\bf{k}}, > } \mbox{  } n_F({\bf{p}})) >  }{ ( e^{   \lambda }\mbox{     }
e^{  \beta (\epsilon_{ {\bf{k}} } - \epsilon_{ {\bf{p}} }) } - 1) }
\]

==========================================================================================

\[
   < e^{ - \lambda^{'} N^{'}_{>}(t)} \mbox{             }
  a_{  \frac{1}{2}({\bf{p}}+{\bf{k}})  }({\bf{k}}-{\bf{p}}) a^{\dagger}_{  \frac{1}{2}({\bf{p}}+{\bf{k}})  }({\bf{k}}-{\bf{p}})
>  \mbox{            } = \mbox{                }
 \mbox{          }   e^{   \lambda^{'}    } \mbox{  }     e^{ \beta (\epsilon_{ {\bf{k}} }-\epsilon_{ {\bf{p}} } ) }
\mbox{       }
 <  e^{ - \lambda^{'} N^{'}_{>}(t)} \mbox{             }
       a^{\dagger}_{  \frac{1}{2}({\bf{p}}+{\bf{k}})  }({\bf{k}}-{\bf{p}})
  a_{  \frac{1}{2}({\bf{p}}+{\bf{k}})  }({\bf{k}}-{\bf{p}}) >
\]
 and,
\[
  a_{  \frac{1}{2}({\bf{p}}+{\bf{k}})  }({\bf{k}}-{\bf{p}}) a^{\dagger}_{  \frac{1}{2}({\bf{p}}+{\bf{k}})  }({\bf{k}}-{\bf{p}})
   +  (1-n_F({\bf{k}})) \mbox{      } c_{ {\bf{p}}, < } c^{\dagger}_{ {\bf{p}}, < }  -c_{ {\bf{k}}, >  }c^{\dagger}_{ {\bf{k}}, > } \mbox{  } n_F({\bf{p}})
   \mbox{            } = \mbox{                }      a^{\dagger}_{  \frac{1}{2}({\bf{p}}+{\bf{k}})  }({\bf{k}}-{\bf{p}})
  a_{  \frac{1}{2}({\bf{p}}+{\bf{k}})  }({\bf{k}}-{\bf{p}})
\]
 and,
 \[
 <  e^{ - \lambda^{'} N^{'}_{>}(t)} \mbox{             }
   a_{  \frac{1}{2}({\bf{p}}+{\bf{k}})  }({\bf{k}}-{\bf{p}}) a^{\dagger}_{  \frac{1}{2}({\bf{p}}+{\bf{k}})  }({\bf{k}}-{\bf{p}})
   >  \mbox{            } = \mbox{                }
\frac{ <  e^{ - \lambda^{'} N^{'}_{>}(t)} \mbox{             }
    ( (1-n_F({\bf{k}})) \mbox{      } c_{ {\bf{p}}, < } c^{\dagger}_{ {\bf{p}}, < } - c_{ {\bf{k}}, >  }c^{\dagger}_{ {\bf{k}}, > } \mbox{  } n_F({\bf{p}})) > }{(   e^{ -  \lambda^{'}    } \mbox{  }     e^{ \beta (\epsilon_{ {\bf{p}} }-\epsilon_{ {\bf{k}} } ) } - 1 )}
\]
 and,
\[
\sum_{ {\bf{p}} }\mbox{          }
  < e^{ - \lambda^{'} N^{'}_{>}(t)} \mbox{             }
  a_{  \frac{1}{2}({\bf{p}}+{\bf{k}})  }({\bf{k}}-{\bf{p}}) a^{\dagger}_{  \frac{1}{2}({\bf{p}}+{\bf{k}})  }({\bf{k}}-{\bf{p}})
>  \mbox{            } = \mbox{                }
\Gamma_{>}(0,\lambda^{'};-i\epsilon)
\mbox{      } \frac{ (1-n_F({\bf{k}})) e^{  \beta  \epsilon_{ {\bf{k}} } } }{  \left( 1
+   e^{  \beta  (\epsilon_{ {\bf{k}} }-\mu) }   \mbox{             }
   e^{  \lambda^{'} }
\right) }\mbox{           }   \left(N^0  -  \sum_{ {\bf{p}} }\mbox{      } \frac{ n_F({\bf{p}})
 }{
\left( 1+
 e^{-  \beta  (\epsilon_{ {\bf{p}} }-\mu) }  \mbox{          }
 \right) } \right)
\]

=============================================================================================

\newpage

=================================================================================================

\begin{align*}
 <e^{ - \lambda N_{>}(t)}
\mbox{        }a^{\dagger}_{  \frac{1}{2}({\bf{p}}+{\bf{k}})  }({\bf{k}}-{\bf{p}})
&a_{  \frac{1}{2}({\bf{p}}+{\bf{k}})  }({\bf{k}}-{\bf{p}}) >   \mbox{            }
\\
& \equiv \mbox{                }\mbox{          }
-\frac{
  (1-n_F({\bf{k}}))   }{ ( e^{   \lambda }\mbox{     }
e^{  \beta (\epsilon_{ {\bf{k}} } - \epsilon_{ {\bf{p}} }) } - 1) } \mbox{      }   < e^{ - \lambda N_{>}(t)}
\mbox{        }c_{ {\bf{p}}, < } c^{\dagger}_{ {\bf{p}}, < }>
 + \frac{
  n_F({\bf{p}})  }{ ( e^{   \lambda }\mbox{     }
e^{  \beta (\epsilon_{ {\bf{k}} } - \epsilon_{ {\bf{p}} }) } - 1) }   \mbox{  }
 < e^{ - \lambda N_{>}(t)}
\mbox{        }c_{ {\bf{k}}, >  }c^{\dagger}_{ {\bf{k}}, > }>
\end{align*}

 and,
 
 \begin{align*}
 <  e^{ - \lambda^{'} N^{'}_{>}(t)} \mbox{             }
   a_{  \frac{1}{2}({\bf{p}}+{\bf{k}})  }({\bf{k}}-{\bf{p}}) &a^{\dagger}_{  \frac{1}{2}({\bf{p}}+{\bf{k}})  }({\bf{k}}-{\bf{p}})
   >  \mbox{            } 
   \\
   &= \mbox{                }
\frac{
     (1-n_F({\bf{k}})) }{(   e^{ -  \lambda^{'}    } \mbox{  }     e^{ \beta (\epsilon_{ {\bf{p}} }-\epsilon_{ {\bf{k}} } ) } - 1 )} \mbox{      }<  e^{ - \lambda^{'} N^{'}_{>}(t)} \mbox{             }
     c_{ {\bf{p}}, < } c^{\dagger}_{ {\bf{p}}, < }  >
     - \frac{ n_F({\bf{p}}) }{(   e^{ -  \lambda^{'}    } \mbox{  }     e^{ \beta (\epsilon_{ {\bf{p}} }-\epsilon_{ {\bf{k}} } ) } - 1 )}
      \mbox{  }
      <  e^{ - \lambda^{'} N^{'}_{>}(t)} \mbox{             }
    c_{ {\bf{k}}, >  }c^{\dagger}_{ {\bf{k}}, > }  >
 \end{align*}

==================================================================================================

\[
\sum_{ {\bf{k}} } \mbox{        }  <e^{ - \lambda N_{>}(t)}
\mbox{        }a^{\dagger}_{  \frac{1}{2}({\bf{p}}+{\bf{k}})  }({\bf{k}}-{\bf{p}})
a_{  \frac{1}{2}({\bf{p}}+{\bf{k}})  }({\bf{k}}-{\bf{p}}) >   \mbox{            } = \mbox{                } \Gamma_{<}(\lambda,0; -i\epsilon)
\mbox{      }  \frac{ n_F({\bf{p}}) }{\left( 1+
 e^{-  \beta  (\epsilon_{ {\bf{p}} }-\mu) }   \mbox{          }
  e^{\lambda   }    \right)}\mbox{             }  \sum_{ {\bf{k}} }\mbox{      } \frac{ (1-n_F({\bf{k}}))
 }{ \left( 1
+   e^{  \beta  (\epsilon_{ {\bf{k}} }-\mu) }
\right) }
\]
and
\[
\sum_{ {\bf{p}} }\mbox{          }
  < e^{ - \lambda^{'} N^{'}_{>}(t)} \mbox{             }
  a_{  \frac{1}{2}({\bf{p}}+{\bf{k}})  }({\bf{k}}-{\bf{p}}) a^{\dagger}_{  \frac{1}{2}({\bf{p}}+{\bf{k}})  }({\bf{k}}-{\bf{p}})
>  \mbox{            } = \mbox{                }
\Gamma_{>}(0,\lambda^{'}; -i \epsilon)
\mbox{      } \frac{ (1-n_F({\bf{k}})) e^{  \beta  \epsilon_{ {\bf{k}} } } }{  \left( 1
+   e^{  \beta  (\epsilon_{ {\bf{k}} }-\mu) }   \mbox{             }
   e^{  \lambda^{'} }
\right) }\mbox{           } \sum_{ {\bf{p}} }\mbox{      }    \frac{  n_F({\bf{p}})
 }{ e^{ \beta  (\epsilon_{ {\bf{p}} }-\mu) }    + 1
  }
\]

======================================================================================

======================================================================================

\[
<T \mbox{   }e^{ - \lambda N_{>}(t)}
\mbox{        } {\tilde{ c}}_{ {\bf{p}}, < }(t)
{\tilde{c}}^{\dagger}_{ {\bf{p}},< }(t) >\mbox{            } = \mbox{                } \Gamma_{<}(\lambda,0; -i \epsilon)
\mbox{      }  \frac{ n_F({\bf{p}})  }{\left( 1+
 e^{-  \beta  (\epsilon_{ {\bf{p}} }-\mu) }   \mbox{          }
  e^{\lambda   }    \right)}
\]
and
\[
 <T \mbox{   }e^{ - \lambda N_{>}(t)}
{\tilde{c}}_{ {\bf{k}}, > }(t)
{\tilde{c}}^{\dagger}_{ {\bf{k}},> }(t) >   \mbox{            } = \mbox{                }
 \Gamma_{>}(0,0; -i \epsilon)
\mbox{      } \frac{ (1-n_F({\bf{k}})) e^{  \beta  \epsilon_{ {\bf{k}} } } }{  \left( 1
+   e^{  \beta  (\epsilon_{ {\bf{k}} }-\mu) }
\right) }\mbox{           } e^{  - \int^{\lambda}_{0} d s\mbox{             }   \sum_{ {\bf{p}} }\mbox{      } \frac{ n_F({\bf{p}})
 }{
\left( 1+
 e^{-  \beta  (\epsilon_{ {\bf{p}} }-\mu) }  \mbox{          }
  e^{s  } \right) } }
\]

====================================================================================

\[
<T \mbox{   } e^{ - \lambda^{'} N^{'}_{>}(t)}
\mbox{        } {\tilde{ c}}_{ {\bf{p}}, < }(t)
{\tilde{c}}^{\dagger}_{ {\bf{p}},< }(t) >\mbox{            } = \mbox{                } \Gamma_{<}(0,0; -i\epsilon)
\mbox{      }  \frac{ n_F({\bf{p}}) }{\left( 1+
 e^{-  \beta  (\epsilon_{ {\bf{p}} }-\mu) }   \right)}\mbox{             }  e^{ - \int^{ \lambda^{'} }_{ 0 } ds \sum_{ {\bf{k}} }\mbox{      } \frac{ (1-n_F({\bf{k}}))
 }{ \left( 1
+   e^{  \beta  (\epsilon_{ {\bf{k}} }-\mu) }   \mbox{             }  e^{  s }
\right) }   }
\]
and
\[
 <T \mbox{   } e^{ - \lambda^{'} N^{'}_{>}(t)} \mbox{             }
{\tilde{c}}_{ {\bf{k}}, > }(t)
{\tilde{c}}^{\dagger}_{ {\bf{k}},> }(t) >   \mbox{            } = \mbox{                }
 \Gamma_{>}(0,\lambda^{'}; -i \epsilon)
\mbox{      } \frac{ (1-n_F({\bf{k}})) e^{  \beta  \epsilon_{ {\bf{k}} } } }{  \left( 1
+   e^{  \beta  (\epsilon_{ {\bf{k}} }-\mu) }   \mbox{             }
   e^{  \lambda^{'} }
\right) }
\]

\section{  Reduced equations for $  \Gamma_{>}(0,\lambda^{'}; t-t^{'}) $ and $ \Gamma_{>}(\lambda,0; t-t^{'})  $  }

==================================================================================================

\begin{align*}
\sum_{ {\bf{k}} } \mbox{        } (-\frac{
  (1-n_F({\bf{k}}))   }{ ( e^{   \lambda }\mbox{     }
e^{  \beta (\epsilon_{ {\bf{k}} } - \epsilon_{ {\bf{p}} }) } - 1) } \mbox{      }   < e^{ - \lambda N_{>}(t)}
\mbox{        }c_{ {\bf{p}}, < } c^{\dagger}_{ {\bf{p}}, < }>
 + \frac{
  n_F({\bf{p}})  }{ ( e^{   \lambda }\mbox{     }
e^{  \beta (\epsilon_{ {\bf{k}} } - \epsilon_{ {\bf{p}} }) } - 1) }   \mbox{  }
 &< e^{ - \lambda N_{>}(t)}
\mbox{        }c_{ {\bf{k}}, >  }c^{\dagger}_{ {\bf{k}}, > }>)   \mbox{            } 
\\
&= \mbox{                } \Gamma_{<}(\lambda,0; -i\epsilon)
\mbox{      }  \frac{ n_F({\bf{p}}) }{\left( 1+
 e^{-  \beta  (\epsilon_{ {\bf{p}} }-\mu) }   \mbox{          }
  e^{\lambda   }    \right)}\mbox{             }  \sum_{ {\bf{k}} }\mbox{      } \frac{ (1-n_F({\bf{k}}))
 }{ \left( 1
+   e^{  \beta  (\epsilon_{ {\bf{k}} }-\mu) }
\right) }
\end{align*}

and

\begin{align*}
\sum_{ {\bf{p}} }\mbox{          }
 (\frac{
     (1-n_F({\bf{k}})) }{(   e^{ -  \lambda^{'}    } \mbox{  }     e^{ \beta (\epsilon_{ {\bf{p}} }-\epsilon_{ {\bf{k}} } ) } - 1 )} \mbox{      }<  e^{ - \lambda^{'} N^{'}_{>}(t)} \mbox{             }
     c_{ {\bf{p}}, < } c^{\dagger}_{ {\bf{p}}, < }  >
     - \frac{ n_F({\bf{p}}) }{(   e^{ -  \lambda^{'}    } \mbox{  }     e^{ \beta (\epsilon_{ {\bf{p}} }-\epsilon_{ {\bf{k}} } ) } - 1 )}
      \mbox{  }
      &<  e^{ - \lambda^{'} N^{'}_{>}(t)} \mbox{             }
    c_{ {\bf{k}}, >  }c^{\dagger}_{ {\bf{k}}, > }  >)  \mbox{            }
    \\
    & = \mbox{                }
\Gamma_{>}(0,\lambda^{'};-i\epsilon)
\mbox{      } \frac{ (1-n_F({\bf{k}})) e^{  \beta  \epsilon_{ {\bf{k}} } } }{  \left( 1
+   e^{  \beta  (\epsilon_{ {\bf{k}} }-\mu) }   \mbox{             }
   e^{  \lambda^{'} }
\right) }\mbox{           } \sum_{ {\bf{p}} }\mbox{      }    \frac{  n_F({\bf{p}})
 }{ e^{ \beta  (\epsilon_{ {\bf{p}} }-\mu) }    + 1
  }
\end{align*}

==================================================================================================

\begin{align*}
\sum_{ {\bf{k}} } \mbox{        }
  &\frac{
  n_F({\bf{p}})  }{ ( e^{   \lambda }\mbox{     }
e^{  \beta (\epsilon_{ {\bf{k}} } - \epsilon_{ {\bf{p}} }) } - 1) }    \mbox{                }
 \Gamma_{>}(0,0; -i \epsilon)
\mbox{      } \frac{ (1-n_F({\bf{k}})) e^{  \beta  \epsilon_{ {\bf{k}} } } }{  \left( 1
+   e^{  \beta  (\epsilon_{ {\bf{k}} }-\mu) }
\right) }\mbox{           } e^{  - \int^{\lambda}_{0} d s\mbox{             }   \sum_{ {\bf{p}} }\mbox{      } \frac{ n_F({\bf{p}})
 }{
\left( 1+
 e^{-  \beta  (\epsilon_{ {\bf{p}} }-\mu) }  \mbox{          }
  e^{s  } \right) } }
\\
&\mbox{            } = \mbox{                } \sum_{ {\bf{k}} }\mbox{      } \left( \Gamma_{<}(\lambda,0; -i\epsilon)
\mbox{      }  \frac{ n_F({\bf{p}}) }{\left( 1+
 e^{-  \beta  (\epsilon_{ {\bf{p}} }-\mu) }   \mbox{          }
  e^{\lambda   }    \right)}\mbox{             }  \frac{ (1-n_F({\bf{k}}))
 }{ \left( 1
+   e^{  \beta  (\epsilon_{ {\bf{k}} }-\mu) }
\right) }  +  \frac{
  (1-n_F({\bf{k}}))   }{ ( e^{   \lambda }\mbox{     }
e^{  \beta (\epsilon_{ {\bf{k}} } - \epsilon_{ {\bf{p}} }) } - 1) }  \mbox{                } \Gamma_{<}(\lambda,0; -i\epsilon)
\mbox{      }  \frac{ n_F({\bf{p}})  }{\left( 1+
 e^{-  \beta  (\epsilon_{ {\bf{p}} }-\mu) }   \mbox{          }
  e^{\lambda   }    \right)}
\right)
\end{align*}

and

\begin{align*}
 \Gamma_{<}(0,0; -i\epsilon)&
\mbox{      } \sum_{ {\bf{p}} }\mbox{          }
 \frac{
     (1-n_F({\bf{k}})) }{(   e^{ -  \lambda^{'}    } \mbox{  }     e^{ \beta (\epsilon_{ {\bf{p}} }-\epsilon_{ {\bf{k}} } ) } - 1 )} \mbox{      }
 \mbox{                } \frac{ n_F({\bf{p}}) }{\left( 1+
 e^{-  \beta  (\epsilon_{ {\bf{p}} }-\mu) }   \right)}\mbox{             }  e^{ - \int^{ \lambda^{'} }_{ 0 } ds \sum_{ {\bf{k}} }\mbox{      } \frac{ (1-n_F({\bf{k}}))
 }{ \left( 1
+   e^{  \beta  (\epsilon_{ {\bf{k}} }-\mu) }   \mbox{             }  e^{  s }
\right) }   }  \mbox{            }  
\\
&
= \Gamma_{>}(0,\lambda^{'}; -i\epsilon)
\mbox{      } \sum_{ {\bf{p}} }\mbox{      }\left( \frac{ (1-n_F({\bf{k}})) e^{  \beta  \epsilon_{ {\bf{k}} } } }{  \left( 1
+   e^{  \beta  (\epsilon_{ {\bf{k}} }-\mu) }   \mbox{             }
   e^{  \lambda^{'} }
\right) }\mbox{           }     \frac{  n_F({\bf{p}})
 }{ e^{ \beta  (\epsilon_{ {\bf{p}} }-\mu) }    + 1
  }  + \frac{ n_F({\bf{p}}) }{(   e^{ -  \lambda^{'}    } \mbox{  }     e^{ \beta (\epsilon_{ {\bf{p}} }-\epsilon_{ {\bf{k}} } ) } - 1 )}
      \mbox{  }\mbox{                } \frac{ (1-n_F({\bf{k}})) e^{  \beta  \epsilon_{ {\bf{k}} } } }{  \left( 1
+   e^{  \beta  (\epsilon_{ {\bf{k}} }-\mu) }   \mbox{             }
   e^{  \lambda^{'} }
\right) }\right)
\end{align*}

\newpage

==================================================================================================

\begin{align*}
\sum_{ {\bf{k}} } \mbox{        }
  &\frac{
  n_F({\bf{p}})  }{ ( e^{   \lambda }\mbox{     }
e^{  \beta (\epsilon_{ {\bf{k}} } - \epsilon_{ {\bf{p}} }) } - 1) }    \mbox{                }
 \Gamma_{>}(0,0; -i \epsilon)
\mbox{      } \frac{ (1-n_F({\bf{k}})) e^{  \beta  \epsilon_{ {\bf{k}} } } }{  \left( 1
+   e^{  \beta  (\epsilon_{ {\bf{k}} }-\mu) }
\right) }\mbox{           } e^{  - \int^{\lambda}_{0} d s\mbox{             }   \sum_{ {\bf{p}} }\mbox{      } \frac{ n_F({\bf{p}})
 }{
\left( 1+
 e^{-  \beta  (\epsilon_{ {\bf{p}} }-\mu) }  \mbox{          }
  e^{s  } \right) } }
\\
&
\mbox{            } = \mbox{                } \sum_{ {\bf{k}} }\mbox{      } \left( \Gamma_{<}(\lambda,0; -i \epsilon)
\mbox{      }  \frac{ n_F({\bf{p}}) }{\left( 1+
 e^{-  \beta  (\epsilon_{ {\bf{p}} }-\mu) }   \mbox{          }
  e^{\lambda   }    \right)}\mbox{             }  \frac{ (1-n_F({\bf{k}}))
 }{ \left( 1
+   e^{  \beta  (\epsilon_{ {\bf{k}} }-\mu) }
\right) }  +  \frac{
  (1-n_F({\bf{k}}))   }{ ( e^{   \lambda }\mbox{     }
e^{  \beta (\epsilon_{ {\bf{k}} } - \epsilon_{ {\bf{p}} }) } - 1) }  \mbox{                } \Gamma_{<}(\lambda,0; -i \epsilon)
\mbox{      }  \frac{ n_F({\bf{p}})  }{\left( 1+
 e^{-  \beta  (\epsilon_{ {\bf{p}} }-\mu) }   \mbox{          }
  e^{\lambda   }    \right)}
\right)
\end{align*}

and

\begin{align*}
 \Gamma_{<}(0,0; -i \epsilon)
&\mbox{      } \sum_{ {\bf{p}} }\mbox{          }
 \frac{
     (1-n_F({\bf{k}})) }{(   e^{ -  \lambda^{'}    } \mbox{  }     e^{ \beta (\epsilon_{ {\bf{p}} }-\epsilon_{ {\bf{k}} } ) } - 1 )} \mbox{      }
 \mbox{                } \frac{ n_F({\bf{p}}) }{\left( 1+
 e^{-  \beta  (\epsilon_{ {\bf{p}} }-\mu) }   \right)}\mbox{             }  e^{ - \int^{ \lambda^{'} }_{ 0 } ds \sum_{ {\bf{k}} }\mbox{      } \frac{ (1-n_F({\bf{k}}))
 }{ \left( 1
+   e^{  \beta  (\epsilon_{ {\bf{k}} }-\mu) }   \mbox{             }  e^{  s }
\right) }   }  
\\
&
=\Gamma_{>}(0,\lambda^{'}; -i \epsilon)
\mbox{      } \sum_{ {\bf{p}} }\mbox{      }\left( \frac{ (1-n_F({\bf{k}})) e^{  \beta  \epsilon_{ {\bf{k}} } } }{  \left( 1
+   e^{  \beta  (\epsilon_{ {\bf{k}} }-\mu) }   \mbox{             }
   e^{  \lambda^{'} }
\right) }\mbox{           }     \frac{  n_F({\bf{p}})
 }{ e^{ \beta  (\epsilon_{ {\bf{p}} }-\mu) }    + 1
  }  + \frac{ n_F({\bf{p}}) }{(   e^{ -  \lambda^{'}    } \mbox{  }     e^{ \beta (\epsilon_{ {\bf{p}} }-\epsilon_{ {\bf{k}} } ) } - 1 )}
      \mbox{  }\mbox{                } \frac{ (1-n_F({\bf{k}})) e^{  \beta  \epsilon_{ {\bf{k}} } } }{  \left( 1
+   e^{  \beta  (\epsilon_{ {\bf{k}} }-\mu) }   \mbox{             }
   e^{  \lambda^{'} }
\right) }\right)
\end{align*}

This means,

==================================================================================================

\begin{align*}
 \Gamma_{>}(0,0; -i \epsilon)\mbox{           } e^{  - \int^{\lambda}_{0} d s\mbox{             }   \sum_{ {\bf{p}} }\mbox{      } \frac{ n_F({\bf{p}})
 }{
\left( 1+
 e^{-  \beta  (\epsilon_{ {\bf{p}} }-\mu) }  \mbox{          }
  e^{s  } \right) } }
\mbox{      } \sum_{ {\bf{k}} } \mbox{        }
 & \frac{
  n_F({\bf{p}})  }{ ( e^{   \lambda }\mbox{     }
e^{  \beta (\epsilon_{ {\bf{k}} } - \epsilon_{ {\bf{p}} }) } - 1) }    \mbox{                }
\frac{ (1-n_F({\bf{k}})) e^{  \beta  \epsilon_{ {\bf{k}} } } }{  \left( 1
+   e^{  \beta  (\epsilon_{ {\bf{k}} }-\mu) }
\right) }
\mbox{            }
\\
& = \mbox{                }  \Gamma_{<}(\lambda,0; -i \epsilon)e^{-\beta  \mu }
\mbox{      }\sum_{ {\bf{k}} }\mbox{      }   \frac{
n_F({\bf{p}})}{  \left(e^{\beta  ( \epsilon_{ {\bf{k}} }- \epsilon_{ {\bf{p}} })+\lambda }-1\right)}
\mbox{      }   \frac{ (1-n_F({\bf{k}})) e^{\beta \epsilon_{ {\bf{k}} } }
 }{\left(1+e^{\beta  (\epsilon_{ {\bf{k}} }-\mu) }\right) }
\end{align*}

and

\begin{align*}
 \Gamma_{<}(0,0; -i \epsilon) e^{ - \int^{ \lambda^{'} }_{ 0 } ds \sum_{ {\bf{k}} }\mbox{      } \frac{ (1-n_F({\bf{k}}))
 }{ \left( 1
+   e^{  \beta  (\epsilon_{ {\bf{k}} }-\mu) }   \mbox{             }  e^{  s }
\right) }   }  \mbox{            }
\mbox{      } \sum_{ {\bf{p}} }\mbox{          }
 &\frac{
     (1-n_F({\bf{k}})) }{(   e^{ -  \lambda^{'}    } \mbox{  }     e^{ \beta (\epsilon_{ {\bf{p}} }-\epsilon_{ {\bf{k}} } ) } - 1 )} \mbox{      }
 \mbox{                } \frac{ n_F({\bf{p}}) }{\left( 1+
 e^{-  \beta  (\epsilon_{ {\bf{p}} }-\mu) }   \right)}\mbox{             }
 \\
 & = \mbox{                }e^{-\lambda^{'} } e^{\beta  \mu } \mbox{          }
\Gamma_{>}(0,\lambda^{'}; -i \epsilon)
\mbox{      } \sum_{ {\bf{p}} }\mbox{      }  \frac{(1-n_F({\bf{k}}))
}{  \left(e^{- \lambda^{'} }
e^{\beta  ( \epsilon_{ {\bf{p}} }- \epsilon_{ {\bf{k}} })}-1\right)} \mbox{      }  \frac{
n_F({\bf{p}})}{\left( 1+e^{-\beta (\epsilon_{ {\bf{p}} }-\mu) }\right) }
\end{align*}

This means,

\subsection{ The solution to the coefficients }

\[
\Gamma_{<}(\lambda,0; -i \epsilon)
\mbox{            } = \mbox{                }  e^{ \beta  \mu } \mbox{           }\Gamma_{>}(0,0; -i \epsilon)\mbox{           } e^{  - \int^{\lambda}_{0} d s\mbox{             }   \sum_{ {\bf{p}} }\mbox{      } \frac{ n_F({\bf{p}})
 }{
\left( 1+
 e^{-  \beta  (\epsilon_{ {\bf{p}} }-\mu) }  \mbox{          }
  e^{s  } \right) } }
\]
and
\[
\Gamma_{>}(0,\lambda^{'}; -i \epsilon)
 \mbox{             } = \mbox{                }e^{ \lambda^{'} } e^{ - \beta  \mu } \mbox{          }  \Gamma_{<}(0,0; -i \epsilon) \mbox{          } \mbox{          } e^{ - \int^{ \lambda^{'} }_{ 0 } ds \sum_{ {\bf{k}} }\mbox{      } \frac{ (1-n_F({\bf{k}}))
 }{ \left( 1
+   e^{  \beta  (\epsilon_{ {\bf{k}} }-\mu) }   \mbox{             }  e^{  s }
\right) }   }
\]
This also means,
\[
\Gamma_{<}(0,0; -i \epsilon)
\mbox{            } = \mbox{                }  e^{ \beta  \mu } \mbox{           }\Gamma_{>}(0,0; -i \epsilon)
\]
We make the assertion that this is also valid for general time differences. This means,
\[
\Gamma_{<}(\lambda,0; t-t^{'})
\mbox{            } = \mbox{                }  e^{ \beta  \mu } \mbox{           }\Gamma_{>}(0,0; t-t^{'})\mbox{           } e^{  - \int^{\lambda}_{0} d s\mbox{             }   \sum_{ {\bf{p}} }\mbox{      } \frac{ n_F({\bf{p}})
 }{
\left( 1+
 e^{-  \beta  (\epsilon_{ {\bf{p}} }-\mu) }  \mbox{          }
  e^{s  } \right) } }
\]
and
\[
\Gamma_{>}(0,\lambda^{'}; t-t^{'})
 \mbox{             } = \mbox{                }e^{ \lambda^{'} } e^{ - \beta  \mu } \mbox{          }  \Gamma_{<}(0,0; t-t^{'}) \mbox{          } \mbox{          } e^{ - \int^{ \lambda^{'} }_{ 0 } ds \sum_{ {\bf{k}} }\mbox{      } \frac{ (1-n_F({\bf{k}}))
 }{ \left( 1
+   e^{  \beta  (\epsilon_{ {\bf{k}} }-\mu) }   \mbox{             }  e^{  s }
\right) }   }
\]
This also means,
\[
\Gamma_{<}(0,0; t-t^{'})
\mbox{            } = \mbox{                }  e^{ \beta  \mu } \mbox{           }\Gamma_{>}(0,0; t-t^{'})
\]
This means,

\begin{align*}
<T \mbox{   }e^{ - \lambda N_{>}(t)}  \mbox{             }e^{ - \lambda^{'} N^{'}_{>}(t)}
\mbox{        } &{\tilde{ c}}_{ {\bf{p}}, < }(t)
{\tilde{c}}^{\dagger}_{ {\bf{p}},< }(t^{'}) >\mbox{            } 
\\
&= \mbox{                }   e^{ \beta  \mu } \mbox{           }\Gamma_{>}(0,0; t-t^{'})
\mbox{           } e^{  - \int^{\lambda}_{0} d s\mbox{             }   \sum_{ {\bf{p}} }\mbox{      } \frac{ n_F({\bf{p}})
 }{
\left( 1+
 e^{-  \beta  (\epsilon_{ {\bf{p}} }-\mu) }  \mbox{          }
  e^{s  } \right) } }
\mbox{      }  \frac{ n_F({\bf{p}})\mbox{      }
  e^{ -\theta(t^{'}-t) \beta  \epsilon_{ {\bf{p}} } } }{\left( 1+
 e^{-  \beta  (\epsilon_{ {\bf{p}} }-\mu) }   \mbox{          }
  e^{\lambda   }    \right)}\mbox{             }  e^{ - \int^{ \lambda^{'} }_{ 0 } ds \sum_{ {\bf{k}} }\mbox{      } \frac{ (1-n_F({\bf{k}}))
 }{ \left( 1
+   e^{  \beta  (\epsilon_{ {\bf{k}} }-\mu) }   \mbox{             }  e^{  s }
\right) }   }
\end{align*}

and

\begin{align*}
 <T \mbox{   }e^{ - \lambda N_{>}(t)}  \mbox{             }e^{ - \lambda^{'} N^{'}_{>}(t)} \mbox{             }
&{\tilde{c}}_{ {\bf{k}}, > }(t)
{\tilde{c}}^{\dagger}_{ {\bf{k}},> }(t^{'}) >   \mbox{            } 
\\
&= \mbox{                }
 e^{ \lambda^{'} } e^{ - \beta  \mu } \mbox{          }  \Gamma_{<}(0,0; t-t^{'})  \mbox{          } e^{ - \int^{ \lambda^{'} }_{ 0 } ds \sum_{ {\bf{k}} }\mbox{      } \frac{ (1-n_F({\bf{k}}))
 }{ \left( 1
+   e^{  \beta  (\epsilon_{ {\bf{k}} }-\mu) }   \mbox{             }  e^{  s }
\right) }   }
\mbox{      } \frac{ (1-n_F({\bf{k}})) e^{\theta (t-t^{'}) \beta  \epsilon_{ {\bf{k}} } } }{  \left( 1
+   e^{  \beta  (\epsilon_{ {\bf{k}} }-\mu) }   \mbox{             }
   e^{  \lambda^{'} }
\right) }\mbox{           } e^{  - \int^{\lambda}_{0} d s\mbox{             }   \sum_{ {\bf{p}} }\mbox{      } \frac{ n_F({\bf{p}})
 }{
\left( 1+
 e^{-  \beta  (\epsilon_{ {\bf{p}} }-\mu) }  \mbox{          }
  e^{s  } \right) } }
\end{align*}

The coefficient $ \Gamma_{>}(0,0;t-t^{'}) $ (or $ \Gamma_{<}(0,0;t-t^{'}) $) is the only that remains to be fixed. We now set $ \lambda, \lambda^{'} = 0 $ to get,
\[
<T \mbox{   } {\tilde{ c}}_{ {\bf{p}}, < }(t)
{\tilde{c}}^{\dagger}_{ {\bf{p}},< }(t^{'}) >\mbox{            } = \mbox{                }   e^{ \beta  \mu } \mbox{           }\Gamma_{>}(0,0;t-t^{'})\mbox{           }
   \frac{
  e^{ -\theta(t^{'}-t) \beta  \epsilon_{ {\bf{p}} } } }{\left( 1+
 e^{-  \beta  (\epsilon_{ {\bf{p}} }-\mu) }    \right)} \mbox{                }   n_F({\bf{p}})
\]
and
\[
 <T \mbox{   } {\tilde{c}}_{ {\bf{k}}, > }(t)
{\tilde{c}}^{\dagger}_{ {\bf{k}},> }(t^{'}) >   \mbox{            } = \mbox{                }
   \Gamma_{>}(0,0;t-t^{'})  \mbox{          }
\mbox{      } \frac{  e^{\theta (t-t^{'}) \beta  \epsilon_{ {\bf{k}} } } }{  \left( 1
+   e^{  \beta  (\epsilon_{ {\bf{k}} }-\mu) }
\right) }  \mbox{          }  (1-n_F({\bf{k}}))
\]

========================================================================================================

\[
<T \mbox{   } {\tilde{ c}}_{ {\bf{p}}, < }(t)
{\tilde{c}}^{\dagger}_{ {\bf{p}},< }(t^{'}) >\mbox{            } = \mbox{                } \Gamma_{>}(0,0;t-t^{'})\mbox{           }
   \frac{
 e^{ \theta(t-t^{'}) \beta  \epsilon_{ {\bf{p}} } } }{\left( 1+
 e^{  \beta  (\epsilon_{ {\bf{p}} }-\mu) }    \right)} \mbox{                }   n_F({\bf{p}})
\]
and
\[
 <T \mbox{   } {\tilde{c}}_{ {\bf{k}}, > }(t)
{\tilde{c}}^{\dagger}_{ {\bf{k}},> }(t^{'}) >   \mbox{            } = \mbox{                }
   \Gamma_{>}(0,0;t-t^{'})  \mbox{          }
\mbox{      } \frac{  e^{\theta (t-t^{'}) \beta  \epsilon_{ {\bf{k}} } } }{  \left( 1
+   e^{  \beta  (\epsilon_{ {\bf{k}} }-\mu) }
\right) }  \mbox{          }  (1-n_F({\bf{k}}))
\]
Now,
\[
<T \mbox{   } {\tilde{ c}}_{ {\bf{p}}, < }(t)
{\tilde{c}}^{\dagger}_{ {\bf{p}},< }(t^{'}) >\mbox{            } + \mbox{                }  <T \mbox{   } {\tilde{ c}}_{ {\bf{p}}, > }(t)
{\tilde{c}}^{\dagger}_{ {\bf{p}},> }(t^{'}) >\mbox{            } =\mbox{                } <T \mbox{   } {\tilde{ c}}_{ {\bf{p}} }(t)
{\tilde{c}}^{\dagger}_{ {\bf{p}} }(t^{'}) >
\]
Thus,
\[
 <T \mbox{   } {\tilde{ c}}_{ {\bf{p}} }(t)
{\tilde{c}}^{\dagger}_{ {\bf{p}} }(t^{'}) >\mbox{            } =\mbox{                }\Gamma_{>}(0,0;t-t^{'})\mbox{           }
   \frac{
 e^{ \theta(t-t^{'}) \beta  \epsilon_{ {\bf{p}} } } }{\left( 1+
 e^{  \beta  (\epsilon_{ {\bf{p}} }-\mu) }    \right)}
\]

\[
 < {\tilde{ c}}_{ {\bf{p}} }(t)
{\tilde{c}}^{\dagger}_{ {\bf{p}} }(t) >\mbox{            } =\mbox{                }\Gamma_{>}(0,0;-i\epsilon)\mbox{           }
   \frac{
 e^{   \beta  \epsilon_{ {\bf{p}} } } }{\left( 1+
 e^{  \beta  (\epsilon_{ {\bf{p}} }-\mu) }    \right)}
\]

\[
- <{\tilde{c}}^{\dagger}_{ {\bf{p}} }(t) {\tilde{ c}}_{ {\bf{p}} }(t) >\mbox{            } =\mbox{                }\Gamma_{>}(0,0;i\epsilon)\mbox{           }
   \frac{ 1}{\left( 1+
 e^{  \beta  (\epsilon_{ {\bf{p}} }-\mu) }    \right)}
\]
A choice,
\[
\Gamma_{>}(0,0;t-t^{'})\mbox{           }= \mbox{                   }sgn(t-t^{'})\mbox{                }  e^{ -\theta(t-t^{'}) \beta  \mu }
\]
and
\[
\Gamma_{<}(0,0; t-t^{'})
\mbox{            } = \mbox{                }  e^{ \beta  \mu } \mbox{           }\Gamma_{>}(0,0; t-t^{'})
\]
ensures
\[
 <  {\tilde{ c}}_{ {\bf{p}} }(t)
{\tilde{c}}^{\dagger}_{ {\bf{p}} }(t) >\mbox{            } + \mbox{                }  < {\tilde{c}}^{\dagger}_{ {\bf{p}} }(t) {\tilde{ c}}_{ {\bf{p}} }(t)
 >\mbox{            } = \mbox{                } 1
\]
Lastly, the meaning of the constant $ \mu $ that was introduced as a proxy for $ \gamma(0,0; t-t^{'}) $ through $ \gamma(0,0; t-t^{'})  \equiv e^{ - \beta \mu } $, is identical to the usual chemical potential since we must ensure that the average total number of fermions is conserved.
\[
\sum_{ {\bf{p}} } \frac{1}{e^{ \beta (\epsilon_{ {\bf{p}} } - \mu ) } + 1} \mbox{               } = \mbox{             } N^0 \mbox{               } = \mbox{             }
\sum_{ {\bf{p}} } n_F({\bf{p}})
\]

\section{ Final answer for the fermion correlation function derived using boson algebra }

\begin{align*}
<T \mbox{   }e^{ - \lambda N_{>}(t)}  \mbox{             }e^{ - \lambda^{'} N^{'}_{>}(t)}
\mbox{        } &{\tilde{ c}}_{ {\bf{p}}, < }(t)
{\tilde{c}}^{\dagger}_{ {\bf{p}},< }(t^{'}) >\mbox{            }
\\
& = \mbox{                }   sgn(t-t^{'})\mbox{                }
  \frac{ n_F({\bf{p}})\mbox{      }
  e^{ -\theta(t^{'}-t) \beta  (\epsilon_{ {\bf{p}} }-\mu) } }{\left( 1+
 e^{-  \beta  (\epsilon_{ {\bf{p}} }-\mu) }   \mbox{          }
  e^{\lambda   }    \right)}\mbox{           } e^{  - \int^{\lambda}_{0} d s\mbox{             }   \sum_{ {\bf{p}} }\mbox{      } \frac{ n_F({\bf{p}})
 }{
\left( 1+
 e^{-  \beta  (\epsilon_{ {\bf{p}} }-\mu) }  \mbox{          }
  e^{s  } \right) } }\mbox{             }  e^{ - \int^{ \lambda^{'} }_{ 0 } ds \sum_{ {\bf{k}} }\mbox{      } \frac{ (1-n_F({\bf{k}}))
 }{ \left( 1
+   e^{  \beta  (\epsilon_{ {\bf{k}} }-\mu) }   \mbox{             }  e^{  s }
\right) }   }
\end{align*}

and

\begin{align*}
 <T \mbox{   }e^{ - \lambda N_{>}(t)}  \mbox{             }e^{ - \lambda^{'} N^{'}_{>}(t)} \mbox{             }
&{\tilde{c}}_{ {\bf{k}}, > }(t)
{\tilde{c}}^{\dagger}_{ {\bf{k}},> }(t^{'}) >   \mbox{            } 
\\
&= \mbox{                }
 e^{ \lambda^{'} }  \mbox{           }
sgn(t-t^{'})\mbox{                }   \frac{ (1-n_F({\bf{k}})) e^{\theta (t-t^{'}) \beta  (\epsilon_{ {\bf{k}} }-\mu) } }{  \left( 1
+   e^{  \beta  (\epsilon_{ {\bf{k}} }-\mu) }   \mbox{             }
   e^{  \lambda^{'} }
\right) }\mbox{           } e^{  - \int^{\lambda}_{0} d s\mbox{             }   \sum_{ {\bf{p}} }\mbox{      } \frac{ n_F({\bf{p}})
 }{
\left( 1+
 e^{-  \beta  (\epsilon_{ {\bf{p}} }-\mu) }  \mbox{          }
  e^{s  } \right) } }  \mbox{          } e^{ - \int^{ \lambda^{'} }_{ 0 } ds \sum_{ {\bf{k}} }\mbox{      } \frac{ (1-n_F({\bf{k}}))
 }{ \left( 1
+   e^{  \beta  (\epsilon_{ {\bf{k}} }-\mu) }   \mbox{             }  e^{  s }
\right) }   }
\end{align*}

\section{ Result using direct trace over fermion states }

\begin{align*}
<T \mbox{   }e^{ - \lambda N_{>}(t)}  \mbox{             }&e^{ - \lambda^{'} N^{'}_{>}(t)}
\mbox{        } {\tilde{ c}}_{ {\bf{p}}, < }(t)
{\tilde{c}}^{\dagger}_{ {\bf{p}},< }(t^{'}) >
\\
&
 =\theta(t-t^{'}) \mbox{             } < e^{ - \lambda \sum_{ {\bf{k}} \neq {\bf{p}}  }\mbox{      }{\tilde{c}}_{ {\bf{k}},< } {\tilde{c}}^{\dagger}_{ {\bf{k}},< }} >
\mbox{        }
< e^{ - \lambda \mbox{      }{\tilde{c}}_{ {\bf{p}},< } {\tilde{c}}^{\dagger}_{ {\bf{p}},< }}
\mbox{        } {\tilde{ c}}_{ {\bf{p}}, < }
{\tilde{c}}^{\dagger}_{ {\bf{p}},< } > < \mbox{             }e^{ - \lambda^{'} \sum_{ {\bf{k}}  }\mbox{      } {\tilde{c}}^{\dagger}_{ {\bf{k}},> }{\tilde{c}}_{ {\bf{k}},> } }>
\\
&\quad - \theta(t^{'}-t) \mbox{             }
< {\tilde{c}}^{\dagger}_{ {\bf{p}},< }\mbox{        }e^{ - \lambda \mbox{      }{\tilde{c}}_{ {\bf{p}},< } {\tilde{c}}^{\dagger}_{ {\bf{p}},< }}
\mbox{        } {\tilde{ c}}_{ {\bf{p}}, < }
> <e^{ - \lambda \sum_{ {\bf{k}} \neq {\bf{p}}  }\mbox{      }{\tilde{c}}_{ {\bf{k}},< } {\tilde{c}}^{\dagger}_{ {\bf{k}},< }} >
\mbox{        }  \mbox{             }<e^{ - \lambda^{'} \sum_{ {\bf{k}}  }\mbox{      } {\tilde{c}}^{\dagger}_{ {\bf{k}},> }{\tilde{c}}_{ {\bf{k}},> } }>
\\
&= \mbox{                } \theta(t-t^{'}) \mbox{             } < e^{ - \lambda \sum_{ {\bf{k}} \neq {\bf{p}}  }\mbox{      }{\tilde{c}}_{ {\bf{k}},< } {\tilde{c}}^{\dagger}_{ {\bf{k}},< }} >
\mbox{        }
< e^{ - \lambda \mbox{      }{\tilde{c}}_{ {\bf{p}},< } {\tilde{c}}^{\dagger}_{ {\bf{p}},< }}
\mbox{        } {\tilde{ c}}_{ {\bf{p}}, < }
{\tilde{c}}^{\dagger}_{ {\bf{p}},< } > < \mbox{             }e^{ - \lambda^{'} \sum_{ {\bf{k}}  }\mbox{      } {\tilde{c}}^{\dagger}_{ {\bf{k}},> }{\tilde{c}}_{ {\bf{k}},> } }>
\\
&
\quad - \theta(t^{'}-t) \mbox{             }
< {\tilde{c}}^{\dagger}_{ {\bf{p}},< }\mbox{        }e^{ - \lambda \mbox{      }{\tilde{c}}_{ {\bf{p}},< } {\tilde{c}}^{\dagger}_{ {\bf{p}},< }}
\mbox{        } {\tilde{ c}}_{ {\bf{p}}, < }
> <e^{ - \lambda \sum_{ {\bf{k}} \neq {\bf{p}}  }\mbox{      }{\tilde{c}}_{ {\bf{k}},< } {\tilde{c}}^{\dagger}_{ {\bf{k}},< }} >
\mbox{        }  \mbox{             }<e^{ - \lambda^{'} \sum_{ {\bf{k}}  }\mbox{      } {\tilde{c}}^{\dagger}_{ {\bf{k}},> }{\tilde{c}}_{ {\bf{k}},> } }>
\end{align*}
\begin{align*}
\mbox{            } = \mbox{                } \theta(t-t^{'}) \mbox{             }
( \prod_{ {\bf{k}} \neq {\bf{p}} } \mbox{          }
\frac{ Tr ( e^{ - \beta (\epsilon_{ {\bf{k}} }- \mu)  {\tilde{c}}^{\dagger}_{ {\bf{k}},< }
  {\tilde{c}}_{ {\bf{k}},< }  }
 e^{ - \lambda  \mbox{      }{\tilde{c}}_{ {\bf{k}},< } {\tilde{c}}^{\dagger}_{ {\bf{k}},< }} ) }{Tr ( e^{ - \beta (\epsilon_{ {\bf{k}} }- \mu)  {\tilde{c}}^{\dagger}_{ {\bf{k}},< }
  {\tilde{c}}_{ {\bf{k}},< }  })} )
\mbox{        }
&\frac{ Tr (e^{ - \beta (\epsilon_{ {\bf{p}} }-\mu){\tilde{c}}^{\dagger}_{ {\bf{p}},< }{\tilde{c}}_{ {\bf{p}},< } }  e^{ - \lambda \mbox{      }{\tilde{c}}_{ {\bf{p}},< } {\tilde{c}}^{\dagger}_{ {\bf{p}},< }}
\mbox{        } {\tilde{ c}}_{ {\bf{p}}, < }
{\tilde{c}}^{\dagger}_{ {\bf{p}},< } ) }{ Tr( e^{ - \beta (\epsilon_{ {\bf{p}} }-\mu){\tilde{c}}^{\dagger}_{ {\bf{p}},< }{\tilde{c}}_{ {\bf{p}},< } }) }
\mbox{             }
\\
&\times( \prod_{ {\bf{k}}  } \mbox{             }\frac{ Tr( e^{ - \beta (\epsilon_{ {\bf{k}} }-\mu)\mbox{      } {\tilde{c}}^{\dagger}_{ {\bf{k}},> }{\tilde{c}}_{ {\bf{k}},> }  }
e^{ - \lambda^{'}  \mbox{      } {\tilde{c}}^{\dagger}_{ {\bf{k}},> }{\tilde{c}}_{ {\bf{k}},> } } ) }{
 Tr( e^{ - \beta (\epsilon_{ {\bf{k}} }-\mu)\mbox{      } {\tilde{c}}^{\dagger}_{ {\bf{k}},> }{\tilde{c}}_{ {\bf{k}},> }  }
  ) }  )
\\
- \theta(t^{'}-t) \mbox{             }
\frac{ Tr( e^{ - \beta ( \epsilon_{ {\bf{p}} }-\mu) {\tilde{c}}^{\dagger}_{ {\bf{p}},< } {\tilde{c}}_{ {\bf{p}},< }  }
{\tilde{c}}^{\dagger}_{ {\bf{p}},< }\mbox{        }e^{ - \lambda \mbox{      }{\tilde{c}}_{ {\bf{p}},< } {\tilde{c}}^{\dagger}_{ {\bf{p}},< }}
\mbox{        } {\tilde{ c}}_{ {\bf{p}}, < } ) }{ Tr( e^{ - \beta ( \epsilon_{ {\bf{p}} }-\mu) {\tilde{c}}^{\dagger}_{ {\bf{p}},< } {\tilde{c}}_{ {\bf{p}},< }  } ) }
 & ( \prod_{ {\bf{k}} \neq {\bf{p}}  }\mbox{      }\frac{ Tr( e^{ -\beta ( \epsilon_{ {\bf{k}} }-\mu ){\tilde{c}}^{\dagger}_{ {\bf{k}},< }{\tilde{c}}_{ {\bf{k}},< } } e^{ - \lambda{\tilde{c}}_{ {\bf{k}},< } {\tilde{c}}^{\dagger}_{ {\bf{k}},< }}  ) }{  Tr( e^{ -\beta ( \epsilon_{ {\bf{k}} }-\mu ){\tilde{c}}^{\dagger}_{ {\bf{k}},< }{\tilde{c}}_{ {\bf{k}},< } } ) })
  \\
  & \times\prod_{ {\bf{k}} }
  \mbox{             }\frac{  Tr ( e^{ - \beta (\epsilon_{ {\bf{k}} }-\mu){\tilde{c}}^{\dagger}_{ {\bf{k}},> }{\tilde{c}}_{ {\bf{k}},> }   }e^{ - \lambda^{'} \mbox{      } {\tilde{c}}^{\dagger}_{ {\bf{k}},> }{\tilde{c}}_{ {\bf{k}},> } }  ) }{ Tr ( e^{ - \beta (\epsilon_{ {\bf{k}} }-\mu){\tilde{c}}^{\dagger}_{ {\bf{k}},> }{\tilde{c}}_{ {\bf{k}},> }   }   ) }
\end{align*}

\[
\mbox{            } = \mbox{                } \theta(t-t^{'}) \frac{ (   e^{ - \lambda  }
\mbox{        } n_F({\bf{p}})  ) }{ (
 e^{ - \lambda    } + e^{ - \beta (\epsilon_{ {\bf{p}} }- \mu) }  ) }\mbox{             }
( \prod_{ {\bf{k}}   } \mbox{          }
\frac{ (
 e^{ - \lambda n_F({\bf{k}})   } + e^{ - \beta (\epsilon_{ {\bf{k}} }- \mu)n_F({\bf{k}})  }  ) }
 {  (1 +  e^{ - \beta (\epsilon_{ {\bf{k}} }- \mu)n_F({\bf{k}})  })} )
\mbox{             }
( \prod_{ {\bf{k}}  } \mbox{             }
\frac{  (1 +  e^{ - \beta (\epsilon_{ {\bf{k}} }-\mu)\mbox{      }(1-n_F({\bf{k}})) }
e^{ - \lambda^{'}\mbox{      }(1-n_F({\bf{k}}))   } ) }{
  ( 1 +  e^{ - \beta (\epsilon_{ {\bf{k}} }-\mu)\mbox{      }(1-n_F({\bf{k}})) }
  ) }  )
\]
\[
- \theta(t^{'}-t) \mbox{             } e^{ -\beta ( \epsilon_{ {\bf{p}} }-\mu)    } \mbox{            }
\frac{  n_F({\bf{p}}) \mbox{        }e^{ - \lambda  }
 }{ (  e^{ - \lambda  }
+e^{ -\beta ( \epsilon_{ {\bf{p}} }-\mu )   } ) }
\mbox{             }
  ( \prod_{ {\bf{k}}    }\mbox{      }\frac{ (  e^{ - \lambda n_F({\bf{k}}) }
+e^{ -\beta ( \epsilon_{ {\bf{k}} }-\mu ) n_F({\bf{k}}) } ) }{  (1 +
e^{ -\beta ( \epsilon_{ {\bf{k}} }-\mu ) n_F({\bf{k}}) } ) })
\mbox{        }\prod_{ {\bf{k}} }
  \mbox{             }\frac{   ( 1  + e^{ - \beta (\epsilon_{ {\bf{k}} }-\mu) (1-n_F({\bf{k}}))   }
e^{ - \lambda^{'} \mbox{      } (1-n_F({\bf{k}})) } ) }{
( 1+e^{ - \beta (\epsilon_{ {\bf{k}} }-\mu) (1-n_F({\bf{k}})) }   ) }
\]

\newpage

\begin{align*}
<T \mbox{   }e^{ - \lambda N_{>}(t)}  \mbox{             }e^{ - \lambda^{'} N^{'}_{>}(t)}
\mbox{        } {\tilde{ c}}_{ {\bf{p}}, < }(t)
&{\tilde{c}}^{\dagger}_{ {\bf{p}},< }(t^{'}) >\mbox{            } = \mbox{                }
\\
sgn(t-t^{'}) \mbox{             } e^{ -\theta(t^{'}-t) \beta ( \epsilon_{ {\bf{p}} }-\mu)    } \mbox{            }
\frac{  n_F({\bf{p}}) \mbox{        }e^{ - \lambda  }
 }{ (  e^{ - \lambda  }
+e^{ -\beta ( \epsilon_{ {\bf{p}} }-\mu )   } ) }
\mbox{             }
  &( \prod_{ {\bf{k}}    }\mbox{      }\frac{ (  e^{ - \lambda n_F({\bf{k}}) }
+e^{ -\beta ( \epsilon_{ {\bf{k}} }-\mu ) n_F({\bf{k}}) } ) }{  (1 +
e^{ -\beta ( \epsilon_{ {\bf{k}} }-\mu ) n_F({\bf{k}}) } ) })
\mbox{        }\prod_{ {\bf{k}} }
  \mbox{             }\frac{   ( 1  + e^{ - \beta (\epsilon_{ {\bf{k}} }-\mu) (1-n_F({\bf{k}}))   }
e^{ - \lambda^{'} \mbox{      } (1-n_F({\bf{k}})) } ) }{
( 1+e^{ - \beta (\epsilon_{ {\bf{k}} }-\mu) (1-n_F({\bf{k}})) }   ) }
\end{align*}

which agrees with the above result from bosonic algebra. \\ \mbox{        } \\
Similarly,

\begin{align*}
<T \mbox{   }e^{ - \lambda N_{>}(t)}  \mbox{             }&e^{ - \lambda^{'} N^{'}_{>}(t)}
\mbox{        } {\tilde{ c}}_{ {\bf{k}}, > }(t)
{\tilde{c}}^{\dagger}_{ {\bf{k}},> }(t^{'}) >\mbox{  }=
\\
& \mbox{                }
\frac{Tr(e^{ - \beta \sum_{ {\bf{p}} } (\epsilon_{ {\bf{p}} }-\mu)c^{\dagger}_{ {\bf{p}} }c_{ {\bf{p}} } }
[\theta(t-t^{'}) \mbox{ }e^{ - \lambda N_{>}(t)}  \mbox{             }e^{ - \lambda^{'} N^{'}_{>}(t)}
\mbox{        } {\tilde{ c}}_{ {\bf{k}}, > }(t)
{\tilde{c}}^{\dagger}_{ {\bf{k}},> }(t^{'})
-  \theta(t^{'}-t) \mbox{ }{\tilde{c}}^{\dagger}_{ {\bf{k}},> }(t^{'})e^{ - \lambda N_{>}(t)}  \mbox{             }e^{ - \lambda^{'} N^{'}_{>}(t)}
\mbox{        } {\tilde{ c}}_{ {\bf{k}}, > }(t)
])}{Tr(e^{ - \beta \sum_{ {\bf{p}} } (\epsilon_{ {\bf{p}} }-\mu)c^{\dagger}_{ {\bf{p}} }c_{ {\bf{p}} } })}
\end{align*}

\[
 = \!
\frac{Tr(e^{ - \beta \sum_{ {\bf{p}} } (\epsilon_{ {\bf{p}} }-\mu)c^{\dagger}_{ {\bf{p}} }c_{ {\bf{p}} } }
[\theta(t-t^{'}) \mbox{ }e^{ - \sum_{ {\bf{p}} }( \lambda \mbox{  }  c_{ {\bf{p}}, < }c^{\dagger}_{ {\bf{p}}, < } 
 + \lambda^{'} \mbox{  } c^{\dagger}_{ {\bf{p}}, > } c_{ {\bf{p}}, > }  ) }
\mbox{        } {\tilde{ c}}_{ {\bf{k}}, > }(t)
{\tilde{c}}^{\dagger}_{ {\bf{k}},> }(t^{'})
-  \theta(t^{'}-t) \mbox{ }{\tilde{c}}^{\dagger}_{ {\bf{k}},> }(t^{'})\mbox{ }e^{ - \sum_{ {\bf{p}} }( \lambda \mbox{  }  c_{ {\bf{p}}, < }c^{\dagger}_{ {\bf{p}}, < }
 + \lambda^{'} \mbox{  } c^{\dagger}_{ {\bf{p}}, > } c_{ {\bf{p}}, > }  ) }
\mbox{        } {\tilde{ c}}_{ {\bf{k}}, > }(t)
])}{Tr(e^{ - \beta \sum_{ {\bf{p}} } (\epsilon_{ {\bf{p}} }-\mu)c^{\dagger}_{ {\bf{p}} }c_{ {\bf{p}} } })}
\]

\begin{align*}
 \mbox{            } = \mbox{                }\theta(t-t^{'}) \mbox{ }&
\frac{Tr(e^{ - \beta \sum_{ {\bf{p}} } (\epsilon_{ {\bf{p}} }-\mu)c^{\dagger}_{ {\bf{p}} }c_{ {\bf{p}} } }
e^{ - \sum_{ {\bf{p}} }( \lambda \mbox{  }  c_{ {\bf{p}}, < }c^{\dagger}_{ {\bf{p}}, < }
 + \lambda^{'} \mbox{  } c^{\dagger}_{ {\bf{p}}, > } c_{ {\bf{p}}, > }  ) }
\mbox{        } {\tilde{ c}}_{ {\bf{k}}, > }(t)
{\tilde{c}}^{\dagger}_{ {\bf{k}},> }(t^{'})
)}{Tr(e^{ - \beta \sum_{ {\bf{p}} } (\epsilon_{ {\bf{p}} }-\mu)c^{\dagger}_{ {\bf{p}} }c_{ {\bf{p}} } })}
\\
 &\qquad \mbox{            } -\theta(t^{'}-t)  \mbox{                }
\frac{Tr(e^{ - \beta \sum_{ {\bf{p}} } (\epsilon_{ {\bf{p}} }-\mu)c^{\dagger}_{ {\bf{p}} }c_{ {\bf{p}} } }
  \mbox{ }{\tilde{c}}^{\dagger}_{ {\bf{k}},> }(t^{'})\mbox{ }e^{ - \sum_{ {\bf{p}} }( \lambda \mbox{  }  c_{ {\bf{p}}, < }c^{\dagger}_{ {\bf{p}}, < }
 + \lambda^{'} \mbox{  } c^{\dagger}_{ {\bf{p}}, > } c_{ {\bf{p}}, > }  ) }
\mbox{        } {\tilde{ c}}_{ {\bf{k}}, > }(t)
 )}{Tr(e^{ - \beta \sum_{ {\bf{p}} } (\epsilon_{ {\bf{p}} }-\mu)c^{\dagger}_{ {\bf{p}} }c_{ {\bf{p}} } })}
\\
 \mbox{            } = \mbox{                }\theta(t-t^{'}) \mbox{ }&
\frac{Tr(e^{ - \beta \sum_{ {\bf{p}} } (\epsilon_{ {\bf{p}} }-\mu)c^{\dagger}_{ {\bf{p}} }c_{ {\bf{p}} } }
e^{ - \sum_{ {\bf{p}} }( \lambda \mbox{  }  c_{ {\bf{p}}, < }c^{\dagger}_{ {\bf{p}}, < }
 + \lambda^{'} \mbox{  } c^{\dagger}_{ {\bf{p}}, > } c_{ {\bf{p}}, > }  ) }
\mbox{        } {\tilde{ c}}_{ {\bf{k}}, > }(t)
{\tilde{c}}^{\dagger}_{ {\bf{k}},> }(t^{'})
)}{Tr(e^{ - \beta \sum_{ {\bf{p}} } (\epsilon_{ {\bf{p}} }-\mu)c^{\dagger}_{ {\bf{p}} }c_{ {\bf{p}} } })}
\\
 &\qquad \mbox{            } -\theta(t^{'}-t)  \mbox{                }
\frac{Tr(e^{ - \beta \sum_{ {\bf{p}} } (\epsilon_{ {\bf{p}} }-\mu)c^{\dagger}_{ {\bf{p}} }c_{ {\bf{p}} } }
  \mbox{ }e^{ - \sum_{ {\bf{p}} }( \lambda \mbox{  }  c_{ {\bf{p}}, < }c^{\dagger}_{ {\bf{p}}, < }
 + \lambda^{'} \mbox{  } c^{\dagger}_{ {\bf{p}}, > } c_{ {\bf{p}}, > }  ) }
\mbox{        }
  e^{ \lambda^{'} }
\mbox{        } {\tilde{c}}^{\dagger}_{ {\bf{k}},> }(t^{'})\mbox{ } 
\mbox{        } {\tilde{ c}}_{ {\bf{k}}, > }(t)
 )}{Tr(e^{ - \beta \sum_{ {\bf{p}} } (\epsilon_{ {\bf{p}} }-\mu)c^{\dagger}_{ {\bf{p}} }c_{ {\bf{p}} } })}
\end{align*}
\begin{align*}
 \mbox{            } = \mbox{                }\bigg[ \theta(t&-t^{'}) \mbox{ }
\frac{Tr(e^{ - \beta (\epsilon_{ {\bf{k}} }-\mu)c^{\dagger}_{ {\bf{k}} }c_{ {\bf{k}} } }
e^{ -  ( \lambda \mbox{  }  c_{ {\bf{k}}, < }c^{\dagger}_{ {\bf{k}}, < }
 + \lambda^{'} \mbox{  } c^{\dagger}_{ {\bf{k}}, > } c_{ {\bf{k}}, > }  ) }
\mbox{        } {\tilde{ c}}_{ {\bf{k}}, > }(t)
{\tilde{c}}^{\dagger}_{ {\bf{k}},> }(t^{'})
)}{Tr(e^{ - \beta   (\epsilon_{ {\bf{k}} }-\mu)c^{\dagger}_{ {\bf{k}} }c_{ {\bf{k}} } })}
 \mbox{            } 
 -\theta(t^{'}-t) 
 \\ 
&\times \frac{Tr(e^{ - \beta  (\epsilon_{ {\bf{k}} }-\mu)c^{\dagger}_{ {\bf{k}} }c_{ {\bf{k}} } }
  \mbox{ }e^{ -  ( \lambda \mbox{  }  c_{ {\bf{k}}, < }c^{\dagger}_{ {\bf{k}}, < }
 + \lambda^{'} \mbox{  } c^{\dagger}_{ {\bf{k}}, > } c_{ {\bf{k}}, > }  ) }
\mbox{        }
  e^{ \lambda^{'} }
\mbox{        } {\tilde{c}}^{\dagger}_{ {\bf{k}},> }(t^{'})\mbox{ }
\mbox{        } {\tilde{ c}}_{ {\bf{k}}, > }(t)
 )}{Tr(e^{ - \beta   (\epsilon_{ {\bf{k}} }-\mu)c^{\dagger}_{ {\bf{k}} }c_{ {\bf{k}} } })}\bigg]
\mbox{        }
\prod_{ {\bf{p}} \neq {\bf{k}} } \mbox{  } \frac{Tr(e^{ - \beta  (\epsilon_{ {\bf{p}} }-\mu)c^{\dagger}_{ {\bf{p}} }c_{ {\bf{p}} } }
e^{ - ( \lambda \mbox{  }  c_{ {\bf{p}}, < }c^{\dagger}_{ {\bf{p}}, < }
 + \lambda^{'} \mbox{  } c^{\dagger}_{ {\bf{p}}, > } c_{ {\bf{p}}, > }  ) }
)}{Tr(e^{ - \beta (\epsilon_{ {\bf{p}} }-\mu)c^{\dagger}_{ {\bf{p}} }c_{ {\bf{p}} } })}
\end{align*}

\begin{align*}
 \mbox{            } = \mbox{                } (1-n_F({\bf{k}})) \mbox{                } \bigg[ 
\frac{ \theta(t-t^{'}) }{ (1  + e^{ - \beta  (\epsilon_{ {\bf{k}} }-\mu) }
e^{ -   \lambda^{'}    }
) }
 \mbox{            } -  \mbox{                }
\frac{ \theta(t^{'}-t)
  }{( e^{   \beta  (\epsilon_{ {\bf{k}} }-\mu)  }  +  
e^{ -   \lambda^{'}    }
) }\bigg] 
 \mbox{        }
\prod_{ {\bf{p}}   } \mbox{  } &\frac{ (e^{ - \beta n_F({\bf{p}}) (\epsilon_{ {\bf{p}} }-\mu)  }
  + 
e^{ -  \lambda \mbox{  }  n_F({\bf{p}})
  }
)}{ (1 + e^{ - \beta \mbox{  }  n_F({\bf{p}}) (\epsilon_{ {\bf{p}} }-\mu)  })}
  \mbox{        }
  \\
&\times\prod_{ {\bf{p}}   } \mbox{  } \frac{ (e^{ - \beta (1-n_F({\bf{p}})) (\epsilon_{ {\bf{p}} }-\mu)  }
e^{ -    \lambda^{'} \mbox{  } (1-n_F({\bf{p}}))   } + 1
)}{ (1 + e^{ - \beta (\epsilon_{ {\bf{p}} }-\mu) \mbox{  } (1-n_F({\bf{p}})) })}
\end{align*}

But,
\[
  (1-n_F({\bf{k}})) \mbox{                } \bigg[ 
\frac{ \theta(t-t^{'}) }{ (1  + e^{ - \beta  (\epsilon_{ {\bf{k}} }-\mu) }
e^{ -   \lambda^{'}    }
) }
 \mbox{            } -  \mbox{                }
\frac{ \theta(t^{'}-t)
  }{( e^{   \beta  (\epsilon_{ {\bf{k}} }-\mu)  }  +  
e^{ -   \lambda^{'}    }
) }\bigg] \mbox{           }  = \mbox{                 }
 e^{ \lambda^{'} }  \mbox{           }
sgn(t-t^{'})\mbox{                }   \frac{ (1-n_F({\bf{k}})) e^{\theta (t-t^{'}) \beta  (\epsilon_{ {\bf{k}} }-\mu) } }{  \left( 1
+   e^{  \beta  (\epsilon_{ {\bf{k}} }-\mu) }   \mbox{             }
   e^{  \lambda^{'} }
\right) }
\]
This means,
\[
<T \mbox{   }e^{ - \lambda N_{>}(t)}  \mbox{             }e^{ - \lambda^{'} N^{'}_{>}(t)}
\mbox{        } {\tilde{ c}}_{ {\bf{k}}, > }(t)
{\tilde{c}}^{\dagger}_{ {\bf{k}},> }(t^{'}) >\mbox{            } = \mbox{                }
\]
\[
 e^{ \lambda^{'} }  \mbox{           }
sgn(t-t^{'})\mbox{                }   \frac{ (1-n_F({\bf{k}})) e^{\theta (t-t^{'}) \beta  (\epsilon_{ {\bf{k}} }-\mu) } }{  \left( 1
+   e^{  \beta  (\epsilon_{ {\bf{k}} }-\mu) }   \mbox{             }
   e^{  \lambda^{'} }
\right) }
 \mbox{        }
\prod_{ {\bf{p}}   } \mbox{  } \frac{ (e^{ - \beta n_F({\bf{p}}) (\epsilon_{ {\bf{p}} }-\mu)  }
  + 
e^{ -  \lambda \mbox{  }  n_F({\bf{p}})
  }
)}{ (1 + e^{ - \beta \mbox{  }  n_F({\bf{p}}) (\epsilon_{ {\bf{p}} }-\mu)  })}
  \mbox{        }
\prod_{ {\bf{p}}   } \mbox{  } \frac{ (e^{ - \beta (1-n_F({\bf{p}})) (\epsilon_{ {\bf{p}} }-\mu)  }
e^{ -    \lambda^{'} \mbox{  } (1-n_F({\bf{p}}))   } + 1
)}{ (1 + e^{ - \beta (\epsilon_{ {\bf{p}} }-\mu) \mbox{  } (1-n_F({\bf{p}})) })}
\]
which agrees with the above result from bosonic algebra.